%% file: main.tex
\documentclass[11pt,ruled,lined,linesnumbered,noresetcount]{article}

\usepackage{palatino}
\usepackage{wrapfig,graphicx,psfrag,xspace,color,latexsym,amsmath,amssymb,booktabs}
\usepackage{comment}
\usepackage{setspace}
\usepackage[dvips]{epsfig}      
\usepackage{longtable}          
\usepackage{xurl,hyperref}       
\usepackage{multirow}           
\usepackage{color}
\usepackage{algorithm2e}
\usepackage{algorithmic}
\usepackage{graphicx}
\usepackage{tikz}
\usepackage{xcolor}
\usepackage{euscript,enumerate,float,afterpage}
\usepackage{rotate}
\usepackage{verbatim}
\usepackage{epstopdf}
\usepackage{hhline}
\usepackage{subcaption}
\usepackage[font=footnotesize,labelfont=bf]{caption}
\usepackage[utf8]{inputenc}
\usepackage{listings}
\usepackage{xcolor}

\definecolor{codegreen}{rgb}{0,0.5,0}
\definecolor{codegray}{rgb}{0.5,0.5,0.5}
\definecolor{codepurple}{rgb}{0.58,0,0.82}
\definecolor{backcolour}{rgb}{0.95,0.95,0.92}
\lstdefinestyle{mystyle}{
  backgroundcolor=\color{backcolour},   commentstyle=\color{codegreen},
  keywordstyle=\color{magenta},
  numberstyle=\tiny\color{codegray},
  stringstyle=\color{codepurple},
  basicstyle=\ttfamily\large,
  breakatwhitespace=false,         
  breaklines=true,                 
  captionpos=b,                    
  keepspaces=true,                 
  numbers=left,                    
  numbersep=5pt,                  
  showspaces=false,                
  showstringspaces=false,
  showtabs=false,                  
  tabsize=2,morekeywords={predict,detect},
    keywordstyle=\bfseries,
    showstringspaces=false
}

\newcommand{\squishlist}{
   \begin{list}{$\bullet$}
    { \setlength{\itemsep}{0pt}      \setlength{\parsep}{0pt}
      \setlength{\topsep}{3pt}       \setlength{\partopsep}{0pt}
      \setlength{\listparindent}{-2pt}
      \setlength{\itemindent}{-5pt}
      \setlength{\leftmargin}{1em} \setlength{\labelwidth}{0em}
      \setlength{\labelsep}{0.5em} } }

\newcommand{\squishend}{
    \end{list}  }

\usepackage{lipsum}
\definecolor{ccred}{RGB}{227, 0, 34}
\definecolor{cyan}{RGB}{0, 150, 200}

\newcommand\bu[1]{{\color{black}{#1}}}
\newcommand\vc[1]{{\color{black}{#1}}}
\newcommand\sz[1]{{\color{black}{#1}}}

\usepackage{graphicx}
\usepackage{enumitem}
\usepackage{mdframed}
\usepackage{url}
\usepackage{colortbl}
\usepackage[font=footnotesize,labelfont=bf]{caption}
\usepackage{subcaption}
\usepackage{multirow}

\newcommand{\legostore}[0]{L{\small EGO\-S}tore}
\newcommand{\lego}[0]{\legostore}

\newcommand{\ul}{\underline}

\begin{document}
\title{LEGOStore: A Linearizable Geo-Distributed Store Combining Replication and Erasure Coding}

\author{Hamidreza Zare, Viveck R. Cadambe, Bhuvan Urgaonkar, \\
Chetan Sharma, Praneet Soni, Nader Alfares, and Arif Merchant$\diamondsuit$\\
{\em The Pennsylvania State University, Google$\diamondsuit$}}

\date{}
\setcounter{page}{1}


\maketitle
 
\noindent {\bf Abstract---}We design and implement LEGOStore, an erasure coding (EC) based linearizable data store over geo-distributed public cloud data centers (DCs). 
For such a data store, the confluence of the  following factors opens up opportunities for EC to be latency-competitive with replication: (a) the necessity of communicating with remote DCs to tolerate entire DC failures and implement linearizability; and (b) the emergence of DCs near most large population centers. 
LEGOStore employs an  optimization framework that, for a given object, carefully chooses among replication and EC, as well as among various DC placements to minimize overall costs. To handle workload dynamism, LEGOStore employs a novel agile reconfiguration protocol. Our evaluation using a LEGOStore prototype spanning 9 Google Cloud Platform DCs demonstrates the efficacy of our ideas. We observe cost savings ranging from moderate (5-20\%) to significant (60\%)  
over baselines representing the state of the art while meeting tail latency SLOs. Our reconfiguration protocol is able to transition key placements in 3 to 4 inter-DC RTTs ($<$ 1s in our experiments), allowing for agile adaptation to dynamic conditions.

\begin{NoHyper}
\renewcommand\thefootnote{}\footnote{Praneet Soni is currently with Apple, and Chetan Sharma is currently with Microsoft.}
\addtocounter{footnote}{-1}
\end{NoHyper}

\input{1introduction}
\input{2background}
\input{3design}
\input{4opt}
\input{5dyn}
\input{7evaluation}

\input{8related_work}

\input{9conclusion}

\clearpage

\bibliographystyle{plain}
\bibliography{bibs/all}

\clearpage
\pagebreak
\input{appendix/appendix}

\end{document}

%% file: 1introduction.tex
\section{Introduction}


Consistent geo-distributed key-value (KV) stores are crucial building blocks of 
modern Internet-scale services including databases and web applications. 
Strong consistency (i.e., linearizability) is especially 
preferred by users for the ease of development and testing it offers. 
As a case in point, the  hugely popular S3 store from Amazon Web Services (AWS), in essence a KV store with a GET/PUT interface,  was recently re-designed to switch its consistency model from eventual to linearizable~\cite{s3-blog}. However, because of the inherent lower bound in  \cite{attiya1994sequential}, linearizable KV stores inevitably incur significant latency and cost overheads compared to weaker consistency models such as causal and eventual consistency. These drawbacks are particularly pronounced in the geo-distributed setting because of the  high inter-data center networking costs, and large network latencies (see Table~\ref{tab:net+rtt}). To further exacerbate the problem, dynamic phenomena such as shifts in arrival rates, appearance of clients in new locations far from where data is stored,  increase in network delays, etc., can lead to gaps between predicted and actual performance both in terms of costs and tail latencies. 

\begin{table*}[ht]\centering
\footnotesize
\caption{Storage and VM prices for the 9 GCP data centers that our prototype spans. We use custom VMs with 1 core and 1 GB of RAM from General-purpose machine type family to run the \lego's servers and Standard provisioned space for its storage~\cite{gcp-pricing}.}
\resizebox{0.9\textwidth}{!}{
\begin{tabular}{lrccccccccc}\toprule
&\multicolumn{9}{c}{\cellcolor[HTML]{fff2cc}GCP data center location} \\\cmidrule{2-10}
&\cellcolor[HTML]{d9d9ff}Tokyo &\cellcolor[HTML]{d9d9ff}Sydney &\cellcolor[HTML]{d9d9ff}Singapore &\cellcolor[HTML]{d9d9ff}Frankfurt &\cellcolor[HTML]{d9d9ff}London &\cellcolor[HTML]{d9d9ff}Virginia &\cellcolor[HTML]{d9d9ff}São Paulo &\cellcolor[HTML]{d9d9ff}Los Angeles &\cellcolor[HTML]{d9d9ff}Oregon \\\cmidrule{2-10}
\cellcolor[HTML]{d9ead3}Storage (\$/GB.Month) &0.052 &0.054 &0.044 &0.048 &0.048 &0.044 &0.06 &0.048 &0.04 \\
\cellcolor[HTML]{d9ead3}Virtual machine (\$/hour) &0.0261 &0.0283 &0.0253 &0.0262 &0.0262 &0.0226 &0.0310 &0.0248 &0.0215 \\
\bottomrule
\end{tabular}}
\label{tab:dcs}
\end{table*}

\begin{table*}[ht]\centering
\footnotesize
\caption{Diverse RTTs and network pricing for 9 chosen GCP data centers. 
The RTTs are measured between VMs placed within the DCs; they would be higher if one of the VMs were external to GCP but not by enough to change the outcome of our optimizer. For the same general recipient location, the outbound network prices are sometimes higher if the recipient is located outside of GCP (egress pricing) but these prices exhibit a similar geographical diversity ~\cite{gcp-network-pricing}.}
\resizebox{0.9\textwidth}{!}{
\begin{tabular}{llccccccccccccccccccc}\toprule
& &\multicolumn{18}{c}{\cellcolor[HTML]{fff2cc}User location} \\\cmidrule{3-20}
& &\multicolumn{2}{c}{\cellcolor[HTML]{d9d9ff}Tokyo} &\multicolumn{2}{c}{\cellcolor[HTML]{d9d9ff}Sydney} &\multicolumn{2}{c}{\cellcolor[HTML]{d9d9ff}Singapore} &\multicolumn{2}{c}{\cellcolor[HTML]{d9d9ff}Frankfurt} &\multicolumn{2}{c}{\cellcolor[HTML]{d9d9ff}London} &\multicolumn{2}{c}{\cellcolor[HTML]{d9d9ff}Virginia} &\multicolumn{2}{c}{\cellcolor[HTML]{d9d9ff}São Paulo} &\multicolumn{2}{c}{\cellcolor[HTML]{d9d9ff}Los Angeles} &\multicolumn{2}{c}{\cellcolor[HTML]{d9d9ff}Oregon} \\\cmidrule{3-20}
& &\cellcolor[HTML]{b6d7a8}\begin{tabular}{@{}c@{}}n/w\\ price\\(\$/GB)\end{tabular} &\cellcolor[HTML]{b6d7a8}\begin{tabular}{@{}c@{}}\rotatebox[origin=c]{90}{Latency (ms)}\end{tabular}
&\cellcolor[HTML]{b6d7a8}\begin{tabular}{@{}c@{}}n/w\\ price\\(\$/GB)\end{tabular} &\cellcolor[HTML]{b6d7a8}\begin{tabular}{@{}c@{}}\rotatebox[origin=c]{90}{Latency (ms)}\end{tabular}
&\cellcolor[HTML]{b6d7a8}\begin{tabular}{@{}c@{}}n/w\\ price\\(\$/GB)\end{tabular} &\cellcolor[HTML]{b6d7a8}\begin{tabular}{@{}c@{}}\rotatebox[origin=c]{90}{Latency (ms)}\end{tabular}
&\cellcolor[HTML]{b6d7a8}\begin{tabular}{@{}c@{}}n/w\\ price\\(\$/GB)\end{tabular} &\cellcolor[HTML]{b6d7a8}\begin{tabular}{@{}c@{}}\rotatebox[origin=c]{90}{Latency (ms)}\end{tabular}
&\cellcolor[HTML]{b6d7a8}\begin{tabular}{@{}c@{}}n/w\\ price\\(\$/GB)\end{tabular} &\cellcolor[HTML]{b6d7a8}\begin{tabular}{@{}c@{}}\rotatebox[origin=c]{90}{Latency (ms)}\end{tabular}
&\cellcolor[HTML]{b6d7a8}\begin{tabular}{@{}c@{}}n/w\\ price\\(\$/GB)\end{tabular} &\cellcolor[HTML]{b6d7a8}\begin{tabular}{@{}c@{}}\rotatebox[origin=c]{90}{Latency (ms)}\end{tabular}
&\cellcolor[HTML]{b6d7a8}\begin{tabular}{@{}c@{}}n/w\\ price\\(\$/GB)\end{tabular} &\cellcolor[HTML]{b6d7a8}\begin{tabular}{@{}c@{}}\rotatebox[origin=c]{90}{Latency (ms)}\end{tabular}
&\cellcolor[HTML]{b6d7a8}\begin{tabular}{@{}c@{}}n/w\\ price\\(\$/GB)\end{tabular} &\cellcolor[HTML]{b6d7a8}\begin{tabular}{@{}c@{}}\rotatebox[origin=c]{90}{Latency (ms)}\end{tabular}
&\cellcolor[HTML]{b6d7a8}\begin{tabular}{@{}c@{}}n/w\\ price\\(\$/GB)\end{tabular} &\cellcolor[HTML]{b6d7a8}\begin{tabular}{@{}c@{}}\rotatebox[origin=c]{90}{Latency (ms)}\end{tabular}
 \\\midrule
\cellcolor[HTML]{fff2cc}&\cellcolor[HTML]{d9d9ff}Tokyo & - &2 &0.15 &115 &0.12 &70 &0.12 &226 &0.12 &218 &0.12 &148 &0.12 &253 &0.12 &100 &0.12 &90 \\
\cellcolor[HTML]{fff2cc}&\cellcolor[HTML]{d9d9ff}Sydney &\cellcolor[HTML]{efefef}0.15 &\cellcolor[HTML]{efefef}115 &\cellcolor[HTML]{efefef}- &\cellcolor[HTML]{efefef}2 &\cellcolor[HTML]{efefef}0.15 &\cellcolor[HTML]{efefef}94 &\cellcolor[HTML]{efefef}0.15 &\cellcolor[HTML]{efefef}289 &\cellcolor[HTML]{efefef}0.15 &\cellcolor[HTML]{efefef}277 &\cellcolor[HTML]{efefef}0.15 &\cellcolor[HTML]{efefef}204 &\cellcolor[HTML]{efefef}0.15 &\cellcolor[HTML]{efefef}291 &\cellcolor[HTML]{efefef}0.15 &\cellcolor[HTML]{efefef}139 &\cellcolor[HTML]{efefef}0.15 &\cellcolor[HTML]{efefef}162 \\
\cellcolor[HTML]{fff2cc}&\cellcolor[HTML]{d9d9ff}Singapore &0.09 &72 &0.15 &94 &- &2 &0.09 &202 &0.09 &203 &0.09 &214 &0.09 &319 &0.09 &165 &0.09 &166 \\
\cellcolor[HTML]{fff2cc}&\cellcolor[HTML]{d9d9ff}Frankfurt &\cellcolor[HTML]{efefef}0.08 &\cellcolor[HTML]{efefef}229 &\cellcolor[HTML]{efefef}0.15 &\cellcolor[HTML]{efefef}289 &\cellcolor[HTML]{efefef}0.08 &\cellcolor[HTML]{efefef}201 &\cellcolor[HTML]{efefef}- &\cellcolor[HTML]{efefef}2 &\cellcolor[HTML]{efefef}0.08 &\cellcolor[HTML]{efefef}15 &\cellcolor[HTML]{efefef}0.08 &\cellcolor[HTML]{efefef}89 &\cellcolor[HTML]{efefef}0.08 &\cellcolor[HTML]{efefef}202 &\cellcolor[HTML]{efefef}0.08 &\cellcolor[HTML]{efefef}153 &\cellcolor[HTML]{efefef}0.08 &\cellcolor[HTML]{efefef}139 \\
\cellcolor[HTML]{fff2cc}&\cellcolor[HTML]{d9d9ff}London &0.08 &222 &0.15 &280 &0.08 &204 &0.08 &15 &- &2 &0.08 &79 &0.08 &192 &0.08 &141 &0.08 &131 \\
\cellcolor[HTML]{fff2cc}&\cellcolor[HTML]{d9d9ff}Virginia &\cellcolor[HTML]{efefef}0.08 &\cellcolor[HTML]{efefef}146 &\cellcolor[HTML]{efefef}0.15 &\cellcolor[HTML]{efefef}204 &\cellcolor[HTML]{efefef}0.08 &\cellcolor[HTML]{efefef}214 &\cellcolor[HTML]{efefef}0.08 &\cellcolor[HTML]{efefef}90 &\cellcolor[HTML]{efefef}0.08 &\cellcolor[HTML]{efefef}79 &\cellcolor[HTML]{efefef}- &\cellcolor[HTML]{efefef}2 &\cellcolor[HTML]{efefef}0.08 &\cellcolor[HTML]{efefef}116 &\cellcolor[HTML]{efefef}0.08 &\cellcolor[HTML]{efefef}68 &\cellcolor[HTML]{efefef}0.08 &\cellcolor[HTML]{efefef}58 \\
\cellcolor[HTML]{fff2cc}&\cellcolor[HTML]{d9d9ff}São Paulo &0.08 &252 &0.15 &292 &0.08 &317 &0.08 &202 &0.08 &192 &0.08 &117 &- &1 &0.08 &155 &0.08 &172 \\
\cellcolor[HTML]{fff2cc}&\cellcolor[HTML]{d9d9ff}Los Angeles &\cellcolor[HTML]{efefef}0.08 &\cellcolor[HTML]{efefef}101 &\cellcolor[HTML]{efefef}0.15 &\cellcolor[HTML]{efefef}139 &\cellcolor[HTML]{efefef}0.08 &\cellcolor[HTML]{efefef}180 &\cellcolor[HTML]{efefef}0.08 &\cellcolor[HTML]{efefef}153 &\cellcolor[HTML]{efefef}0.08 &\cellcolor[HTML]{efefef}142 &\cellcolor[HTML]{efefef}0.08 &\cellcolor[HTML]{efefef}67 &\cellcolor[HTML]{efefef}0.08 &\cellcolor[HTML]{efefef}155 &\cellcolor[HTML]{efefef}- &\cellcolor[HTML]{efefef}2 &\cellcolor[HTML]{efefef}0.08 &\cellcolor[HTML]{efefef}26 \\
\multirow{-9}{*}{\cellcolor[HTML]{fff2cc}\rotatebox[origin=c]{90}{GCP data center}}&\cellcolor[HTML]{d9d9ff}Oregon &0.08 &95 &0.15 &164 &0.08 &165 &0.08 &142 &0.08 &131 &0.08 &58 &0.08 &173 &0.08 &26 &- &2 \\
\bottomrule
\end{tabular}}
\label{tab:net+rtt}
\end{table*}

We design LEGOStore, a geo-distributed linearizable KV store (with the familiar GET/PUT or read/write API), meant for a global user-base. LEGOStore procures its resources from a cloud provider's fleet of data centers (DCs) like many  storage service providers~\cite{evernote,overleaf,aws-outage-github}. Since an entire DC may become unavailable~\cite{aws-outage-github,azure-outage1,azure-outage2,Veeraraghavan:2018:MMD:3291168.3291196}, LEGOStore employs  redundancy across geo-distributed DCs to operate despite such events. LEGOStore's goal is to {\em  offer  tail  latency  service-level objectives (SLOs) that are predictable and robust in the face of myriad sources of dynamism at a low cost}. For achieving these properties, LEGOStore is built upon the following three pillars:

\textbf{1. Erasure Coding (EC)} is a generalization of replication that is more storage-efficient than replication for a given fault tolerance. A long line of research  
has helped establish EC's efficacy {\em within a DC} or for weaker consistency needs. 
However, EC's efficacy in 
the linearizable geo-distributed setting is relatively 
less well-understood.  Two recent works Giza~\cite{giza} and Pando~\cite{pando} demonstrate some aspects of EC's promise in the geo-\-distributed context. In particular, because EC allows fragmenting the data and storing it in a fault-tolerant manner, it leads to smaller storage costs and (more importantly) smaller inter-DC networking costs. However, the smaller-sized fragments inevitably require contacting more DCs and,  therefore, are thought to imply higher latencies in a first-order estimation. By comprehensively exploring a wide gamut of workload features and SLOs,  and careful design of the data placement and EC parameters through an optimization framework,  LEGOStore brings out the full potential of EC in the geo-distributed setting.

\textbf{2. 
} 
LEGOStore adapts {\bf non blocking, leaderless, quorum based linearizability protocols} for both EC~\cite{cadambe2017coded} and replication~\cite{ABD_conference,ABD}. When used with carefully optimized quorums, these protocols help realize LEGOStore's goal of predictable performance (i.e.,  meeting latency SLOs with a very high likelihood as long as workload features match their predicted values considered by our optimization). A second important reason is that, when used in conjunction with a well-designed resource autoscaling strategy, the  latency resulting from non-blocking protocols depends primarily on its inter-DC latency and data transfer time components. 
 This is in contrast with the  leader-based consensus used in Giza and Pando (see Figure~\ref{fig:concurrency} ). 
While these works 
may be able to offer lower latencies for certain workloads, they are susceptible to severe performance degradation under concurrency-induced contention. 

\textbf{3. 
} To offer robust SLOs in the face of dynamism, LEGOStore implements an {\bf agile reconfiguration mechanism}. LEGOStore continually weighs the pros and cons of changing the {\em configuration}\footnote{By the configuration of a key, we mean the following  aspects of its placement: (i) whether EC or replication is being used ; (ii) the EC parameters or replication degree being used; and (iii) the DCs that comprise various quorums. 
} of a key or a group of keys via a cost-benefit analysis rooted in its cost/performance modeling.  
If prompted by this  analysis, it uses a novel reconfiguration protocol to safely switch the configurations of concerned keys   without breaking linearizability. We have designed this reconfiguration protocol carefully and specifically to work alongside our GET/PUT protocols to keep its execution time small which,  in turn, allows us to limit the performance degradation experienced by user requests issued during a reconfiguration. 

\noindent 
{\bf Contributions:} 
We design \lego, a cost-effective KV store with predictable tail latency  that adapts to dynamism. We develop an optimization framework that, for a group of keys with similar suitable workload features,  takes as input these features, and public cloud characteristics and determines configurations that satisfy SLOs at minimal cost. The configuration choice involves selecting one from a  family of linearizability protocols, and the protocol parameters that can include the degree of redundancy, DC placement, quorum sizes, and EC parameters. In this manner it selectively uses EC for its better storage/networking costs when allowed by latency goals, and uses replication otherwise. A key’s storage method may change over time based on changes in workload features. 
We develop a 
safe reconfiguration protocol, and an accompanying  heuristic cost/benefit analysis that allows \lego~ to control costs by dynamically adapting configurations. 

We build a prototype LEGOStore system.  We carry out extensive evaluations using our optimizer and prototype spanning 9 Google Cloud Platform (GCP) DCs. We get insights from our evaluation to make suitable modeling and protocol design choices. Because of our design, our prototype has close match with the performance predicted by the optimizer. Potential cost savings over baselines based on state-of-the-art works such as SPANStore~\cite{spanstore} and  Pando~\cite{pando} range from moderate (5-20\%) to significant (up to 60\%). The most significant cost savings emerge from carefully avoiding the use of DCs with high outbound network prices. Our work offers a number of general trends and insights relating workload and infrastructure properties to cost-effective realization of linearizability. Some of our findings are perhaps non-intuitive: (i) smaller EC-fragments do not always lead to lower costs (Section~\ref{subsubsec:opt-k-dep}); (ii) there exists a significant asymmetry between GETs and PUTs in terms of costs, and read-heavy workloads lead to different choices than write-heavy workloads (Section~\ref{subsec:asymmetry}); (iii) we find scenarios where the optimizer is able to exploit the lower costs of EC without a latency penalty (Section~\ref{subsec:ECvsrep}); and (iv) even when a majority of the requests to a key arise at a particular location, the DC near that location may not necessarily be used for this key in our optimizer's solution. 
    

%% file: 2background.tex
\section{Background}
\label{sec:back}

\noindent {\bf Public Cloud Latencies and Pricing:} The lower bound of~\cite{attiya1994sequential,Attiya_Welch} implies that {\em both} GET and PUT operations in LEGOStore necessarily involve inter-DC latencies and data transfers 
unlike with weaker consistency models. 
The latencies between users and various DCs of a public cloud provider span a large range. In Table~\ref{tab:net+rtt} we depict our measurements  of round-trip times (RTTs) between pairs of DCs out of a set of 9 Google Cloud Platform (GCP) data centers we use. The smallest RTTs are 15-20 msec while the largest exceed 300 msec; RTTs between nodes within the same DC are 1-2 msec and pale in comparison. Similarly, the prices for storage, computational, and network resources across DCs also exhibit geographical diversity as shown in Tables~\ref{tab:dcs} and~\ref{tab:net+rtt}. This diversity is the most prominent for data transfers---the cheapest per-byte transfer is  \$0.08/GB (e.g., London to Tokyo), the costliest is \$0.15/GB (e.g., Sydney to Tokyo). LEGOStore's design must carefully navigate these sources of diversity to meet latency SLOs at minimum cost.

\noindent {\bf Our Choice of Consistent Storage Algorithms:} 
Due to the lower bound mentioned above, in leader-based protocols (e.g., Raft~\cite{raft}), for the geo-distributed scenario of interest to us, one round trip time to the leader is inevitable. Using such leader-based protocols can drive up latency when the workload is distributed over a wide geographical area, and there is no leader node that is sufficiently close to all DCs so as to satisfy the SLO requirements. Such a design can also place excessive load on DCs that are more centrally located. 
Therefore, as such, we choose algorithms with leaderless, quorum-based protocols in our design and implementation. We describe these protocols (ABD for replication and CAS for EC) next. 


\noindent{\bf The ABD Algorithm:}\footnote{The name ``ABD'' comes from the authors, Attya, Bar-Noy and Dolev ~\cite{ABD_conference,ABD}.} 
Let $N$ denote the degree of replication (specifically, spanning $N$ separate DCs) being used for the key under discussion. In the ABD algorithm, for a given key, each of the nodes (i.e., DCs) stores a (tag, value) pair, where the tag is a (logical timestamp, client ID) pair. The node replaces this tuple when it receives a value with a higher tag from a client operation.
The PUT operation consists of two phases. The first phase, which involves a logical-time query, requires responses from a quorum of $q_1$ nodes, and the second phase, which involves sending the  new (tag, value) pair requires acknowledgements from a quorum of $q_2$ nodes. The GET operation also consists of two phases. The first phase again involves logical-time queries from a quorum of $q_1$ nodes and determining the highest of these tags.  The second ``write-back'' phase 
sends the  (tag, value) pair chosen from the first-phase responses to a quorum of $q_2$ nodes. 
If $q_1+q_2 > N$, then ABD guarantees linearizability. If $q_1, q_2 \leq N-f,$ then operations terminate so long as the number of node failures is at most $f.$  Note that this is a stronger liveness guarantee as compared to Paxos; ABD circumvents FLP impossibility \cite{FLP} because it implements a data type (read/write memory) that is weaker than consensus \vc{(see also  \cite{ABD_Djisktra} and Theorems 17.5, 17.9 in \cite{Lynch1996})}.  This liveness property translates into excellent robustness of operation latency (see Section~\ref{sec:concurrency}). See formal description of ABD in Appendix~\ref{app:ABD} of ~\cite{legostore-arxiv}.

Whereas the above vanilla ABD requires two phases for all its GET operations, a slight enhancement allows some (potentially many) GET operations to complete in only one phase; we refer to such a scenario as an ``Optimized GET,'' see details in \cite{legostore-arxiv}.

\begin{table}[t]\centering
\footnotesize
\caption{Coarse cost comparison of replication (ABD) versus erasure coding (CAS). Costs reported are per GET/PUT operation, and the per-server storage cost. We assume that each value has $B$ bits and the metadata size is negligible. All quorum sizes are assumed to be $(N+k)/2$ for CAS and $(N+1)/2$ for ABD;   $N,k$ are assumed to be odd to ignore integer rounding. Latency is reported as the number of round trips of the protocol.}
\begin{minipage}{\columnwidth}
\begin{center}
\begin{tabular}{lcccccc}\toprule
&\cellcolor[HTML]{d9d9ff}PUT cost &\cellcolor[HTML]{d9d9ff}PUT latency &\cellcolor[HTML]{d9d9ff}GET cost &\cellcolor[HTML]{d9d9ff}GET latency &\cellcolor[HTML]{d9d9ff}Storage cost \\\cmidrule{2-6}
\cellcolor[HTML]{d9ead3}CAS\footnote{With efficient garbage collection, $\delta$ can be kept small; it is equal to $1$ for keys with sufficiently low arrival rates.} &$\frac{NB}{k}$ & $3$ rounds & $\frac{(N-K) B}{2K}$ & 2 rounds & $\delta\frac{B}{k}$ \\
\cellcolor[HTML]{d9ead3}ABD\footnote{ABD has a higher communication cost vs. CAS for GETs, even if $k=1$, since it propagates values in the write-back phase, whereas CAS only propagates metadata.} &$NB$ & 2 rounds& $(N-1) B$ & $2$ rounds & $1$ \\
\bottomrule
\end{tabular}
\label{table:comparison}
\end{center}
\end{minipage}
\end{table}

\noindent \textbf{Erasure Coding:} Erasure coding (EC) is a generalization of replication that is attractive for modern storage systems because of its potential cost savings over replication. An $(N,K)$ Reed Solomon Code stores a value over $N$ nodes, with each node storing a codeword symbol 
of size $1/K$ of the original value, unlike replication where each node stores the entire value.  The value can be decoded from \emph{any} $K$ of the $N$ nodes, so the code tolerates  up to $f=N-K$ failures. On the other hand, replication duplicates the data $N=f+1$ times to tolerate $f$ failures. For a fixed value of $N$, EC leads to a $K$-fold reduction in storage cost compared to $N$-way replication for the same fault-tolerance. It also leads to a $K$-fold reduction in communication cost for PUTs, which  can be significant because of the inter-DC network transfer pricing. While this suggests that costs reduce with increasing $K$, we will see that the actual dependence of costs on $K$ in LEGOStore is far more complex (see Section~\ref{subsubsec:opt-k-dep}).

In EC-based protocols, GET operations require $K > 1$ nodes to respond with codeword symbols for 
the \emph{same} version of the key. However, due to asynchrony, different nodes may store different versions at a given time. Reconciling the different versions incurs additional communication overheads for EC-based algorithms. 

\noindent \textbf{The CAS Algorithm:}
We use the  \emph{coded atomic storage} (CAS) algorithm\footnote{{The algorithm in Appendix \ref{app:cas} of~\cite{legostore-arxiv} is a modification the algorithm in~\cite{CadambeCoded_NCA, cadambe2017coded} to allow for flexible quorum sizes, which in turn this exposes more cost-saving opportunities.} 
} of~\cite{CadambeCoded_NCA, cadambe2017coded}, described 
in Appendix \ref{app:cas} of~\cite{legostore-arxiv}.   
In CAS, servers store a list of triples, each consisting of a tag, a codeword symbol, and a label that can be `pre' or `fin'.  The GET protocol operates in two phases like ABD; however, PUT operates in {\em three} phases. Similar to ABD, the first phase of PUT  acquires the latest tag. The second phase sends an encoded value to servers, and servers store this symbol with a `pre' label. The third phase propagates the `fin' label to  servers, and servers which receive it update the label for that tag. 
The three phases of PUT require quorums of  $q_1,q_2,q_3$, resp., responses to complete.  Servers respond to queries from GETs/PUTs only with  latest tag labeled `fin' in their lists. A GET operates in two phases, the first phase to acquire the highest tag labeled `fin' and the second to acquire the chunks for that tag, decode the value and do a write-back. The two phases of GET require responses from quorums of size $q_1,q_4$, resp. In the write-back phase, CAS only sends a `fin' label with the tag, unlike ABD which sends the entire value. 
\vc{In fact, the structural differences between ABD and CAS protocols translates to lower communication costs for CAS even if $k=1$ (i.e., replication) is used as compared to ABD, with the penalty of incurring higher PUT latency due to the additional phase (\vc{See Table \ref{table:comparison}}). This variation between ABD and CAS offers LEGOStore further flexibility in adapting to workloads as demonstrated in Section~\ref{sec:opt-ins}.}
Similar to ABD, we also employ an "Optimized GET" for CAS that enables some (potentially many) GET operations to complete in only one phase. This optimized GET is based on a client-side cache (recall a client is different from a user, cf. Section~\ref{sec:arch}) for the value  computed in second phase of GET.  LEGOStore exploits these  differences to reduce costs based on if the workload is read- or write-intensive. On the server-side protocol, unlike ABD where a server simply replaces a value with a higher tagged value, CAS  requires servers to store a history of the codeword symbols corresponding to multiple versions, and then garbage collect (GC) older versions at a later point. 
However, in practice, the overhead is negligible for the workloads we study (see Appendix \ref{app:gc} in \cite{legostore-arxiv}, \vc{also remarks in Table \ref{table:comparison}.)}. \vc{The preliminary cost comparison of ABD and CAS in Table \ref{table:comparison} ignores several important aspects of practical key-value stores, in particular the spatial diversity of pricing and latency, flexibility of choosing quorum sizes and locations, and the impact of arrival rates. Our paper refines the insights of Table \ref{table:comparison} in the context of LEGOStore (See Section~\ref{sec:opt} for details).}

%% file: 3design.tex
\section{LEGOStore System Design}
\label{sec:arch}


\subsection{Interface and Components}
\label{sec:interface}

LEGOStore is a linearizable key-value store spanning a set $\mathcal{D}$ of $D$ DCs of a public cloud provider.
Applications using LEGOStore ("users")  link the LEGOStore library that offers them an API comprising the following linearizable operations:

\squishlist
    \item CREATE(k,v): creates the key {\tt k} using default configuration {\tt c} (we will  define a configuration momentarily) if it doesn't already exist and stores {\tt (k,c)} in the local meta-data server (MDS); returns an error if the key already exists.\footnote{A default configuration uses the nearest DCs for various quorums in terms of their RTTs from the client.}
    \item GET(k): returns value for {\tt k} if {\tt k} exists; else  returns an error.
    \item PUT(k,v): sets value of {\tt k} to {\tt v}; returns error if {\tt k} doesn't exist. 
    \item DELETE(k): removes {\tt k};  returns error if {\tt k} doesn't exist.
\squishend

To service these operations, the library issues RPCs to a LEGOStore "client" within a DC in $\mathcal{D}$. A LEGOStore client implements the client-end of LEGOStore's consistency protocols. A user resident within a DC  in $\mathcal{D}$ would be assisted by a client within the same DC. For users outside of $\mathcal{D}$, a natural choice would be a client in the nearest DC. 
The client assisting a user may change over time (e.g., due to user movement) but only across operations. Since the user-client delay is negligible compared to other RTTs involved in request servicing (recall Table~\ref{tab:net+rtt}), we will ignore it in our modeling.  

In order to service a {\tt GET} or a {\tt PUT} request for a key {\tt k}, a client first determines the "configuration" for {\tt k} which consists of the following elements: (i) replication or erasure coding to be used (and, correspondingly,  ABD or CAS); (ii) the DCs that comprise relevant quorums; and (iii) the identities of the LEGOStore "proxies" within each of these DCs. 
Having obtained the configuration, a client issues protocol-specific Remote Procedure Calls (RPCs) to proxies in relevant quorums to service the user request. Each DC's proxy serves as the intermediary between the client and the compute/storage servers that (a) implement the server-end of our consistency protocols and (b) store actual data (replicas for ABD, EC chunks for CAS) along with appropriate tags. 

%% file: 4opt.tex
\subsection{Finding Cost-Effective Configurations}
\label{sec:opt}
\allowdisplaybreaks{
We develop an optimization 
that determines cost-effective configurations assuming perfect knowledge of workload and system properties. 
Since our protocols operate at a per-key granularity due to the composability of linearizability~\cite{herlihy90}---notice how the ABD and CAS algorithms in  Appendix~\ref{app:ABD} and~\ref{app:cas} of~\cite{legostore-arxiv} are described for a generic key---we can decompose our datastore-wide optimization into smaller optimization problems, one per key.\footnote{Although we design our optimizer for  individual keys, aggregating keys with similar workload features and considering such a "group" of keys in the optimizer may be useful (perhaps even necessary) for LEGOStore to scale to large number of keys.}

\noindent {\bf Inputs (See Table \ref{tbl:inputsandvars}):} We assume that \legostore~spans $D$ geo-distributed DCs numbered $1, ..., D$. We assume that the following are available at a per-key granularity: (i) overall request arrival rate;  (ii) geographical distribution of requests (specifically, fractions of the overall arrival rate emerging in/near each DC); (iii) fraction of requests that are GET operations; (iv) average object size  and meta-data\footnote{Meta-data transferred over the network can have non-negligible cost/latency implications and that is what we explicitly capture. On the other hand, the storage of meta-data  contributes relatively negligibly to costs and we do not consider those costs.} size; (v) GET and PUT latency SLOs. 
We assume that SLOs are in terms of 
$99^{th}$ percentile latencies. 
We assume that the availability requirement is expressed via the single parameter $f$> 0; \legostore~must continue servicing requests despite up to $f$ DC failures. The system properties considered in our formulation are: (i) inter-DC latencies and prices for network traffic between 
clients and servers; (ii) storage; (iii) computational resources in the form of  virtual machines (VMs). 

\begin{table}[!t]
\footnotesize
\caption{Input and decision variables used by \legostore's optimization. 
}
\centering
\begin{tabular}{|p{0.7cm}|p{12cm}|p{1cm}|}
  \hline
 \cellcolor[HTML]{D9D9FF}{\bf \small Input} & \cellcolor[HTML]{D9D9FF}{\bf \small Interpretation} & \cellcolor[HTML]{D9D9FF}{\bf \small Type}\\ 
\hline\hline
$D $ & Number of data centers & integer\\\hline
$\mathcal{D} $ & Set of data centers numbered $1, ..., D$& set\\\hline
$l_{ij}$ & Latency from DC $i$ to DC $j$ (RTT/2) & real\\\hline
$B_{ij}$ & Bandwidth between DC $i$ and DC $j$  
& real\\\hline
$\mathcal{G}$ & Set of keys & set\\\hline
$\lambda_g$ & Aggregate request arrival rate for key $g\in \mathcal{G}$ & integer\\\hline
$\rho_g$ & Read-write ratio for $g$ & real [0,1]\\\hline
$\alpha_{ig}$ & Fraction of requests originating at/near DC $i$ for key $g$ & real \\\hline
$o_g$ & Average object size, {\em including} protocol-specific metadata exchanged between a client and a server 
& integer\\\hline
$o_m$ & Average protocol-specific metadata exchanged between a client and a server for each phase & integer \\\hline
$l_{get}$ & GET latency SLO & real\\\hline
$l_{put}$ & PUT latency SLO & real\\\hline
$f$ & Availability requirement (i.e., number of failed DCs  to tolerate) & integer\\\hline
$p_i^s$ & Storage price (per byte per unit time) for DC $i \in \mathcal{D}$ & real \\\hline
$p^n_{ij}$ & Network price per byte from location $i$ to location $j$& real\\\hline
$p^{v}_i$ & VM price at DC $i$ (simplifying assumption: all VMs of a single size) & real\\\hline
$\theta^v$ & This quantity multiplied by the request arrival rate at DC $i$ captures the VM capacity required at $i$ & real\\
   \hline\hline
\cellcolor[HTML]{D9D9FF}{\bf \small Var.} & \cellcolor[HTML]{D9D9FF}{\bf \small Interpretation} & \cellcolor[HTML]{D9D9FF}{\bf \small Type}\\ 
\hline\hline
$e_g$ & Protocol ($0$ for ABD, $1$ for CAS) for key $g$ & boolean\\\hline
$m_g$ & Length of code (replication factor for ABD) & integer\\\hline
$k_g$ & Dimension of code (equals 1 for replication) & integer\\\hline
$q_{i,g}$ & Quorum size for $i^{\text{th}}$ quorum of key $g$ & integer\\\hline
$v_i$ & Capacity of VMs at DC $i$ & real\\\hline
$\mathbf{iq}^k_g$ & Indicator for data placement for $k^{\text{th}}$ quorum of key $g$. $iq_{ijg}^k=1$ iff DC $j$ in $k^{\text{th}}$ quorum of clients in/near DC $i$ & boolean\\ 
\hline
\end{tabular}
\label{tbl:inputsandvars}
\end{table}

\noindent {\bf Decision Variables:} Our decision variables, as described at the bottom of Table~\ref{tbl:inputsandvars}, help capture all aspects of a valid configuration. These include: (i) boolean variable $e_{g}$ whether this key would be served using ABD, and (ii) variables $\mathbf{iq}_g^{k}$ which DCs constitute various quorums that the chosen algorithm (2 and 4 quorums, resp.,  for ABD and  CAS) requires (see variable $iq_{g}^{k}$ in Table \ref{tbl:inputsandvars}). 


\noindent {\bf Optimization:} Our optimization tries to minimize the cost of operating key $g\in {G}$ in the next \emph{epoch} -- a period of relative stability in workload features. Our objective for key $g\in{G}$, which is cost per unit time during the epoch being considered, is expressed as: 


\begin{equation}
\begin{split}
  \text{minimize}~~~\big(C_{g, get} + C_{g, put} + C_{g, Storage} + C_{g, VM}\big)\\
  \text{s.t.}~~~(\ref{eq:put-cost})-(\ref{eq:get-cas-cost})~~~\text{in Appendix~\ref{app:optimization} of~\cite{legostore-arxiv}}.
\end{split}
\label{eq:opt}
\end{equation}

The first two components of the objective with  $put$ and $get$ in their subscripts denote the networking costs per unit time of PUT and GET operations, resp.,  for key $g$ while the last two denote costs per unit time spent towards storage and computation, resp. \vc{The details of the optimization are in Appendix \ref{app:optimization} of \cite{legostore-arxiv}, we reproduce some representative constraints and equations here. 

The networking cost per unit time of PUTs for key $g$ must be represented differently based on whether ABD or CAS is used: $   C_{g,put} = \underbrace{e_g \cdot C_{g,put,CAS}}_{\textrm{n/w cost if CAS chosen}} + \underbrace{(1-e_g) \cdot C_{g,put,ABD}}_{\textrm{n/w cost if ABD chosen}},$ where,
\allowdisplaybreaks{
\begin{equation*}
\begin{split}
C_{g,put,ABD} = (1-\rho_g) \cdot \lambda_g \sum\limits_{i=1}^D \alpha_{ig}\Big(\underbrace{o_m\sum\limits_{j=1}^D {p^n_{ji} \cdot iq^1_{ijg}}}_{\textrm{n/w cost for phase 1}} + \\
        \underbrace{o_g\sum\limits_{k=1}^D {p^n_{ik} \cdot iq^2_{ikg}}}_{\textrm{n/w cost for phase 2}}\Big).
\end{split}
\end{equation*}
\begin{equation}
\begin{split}
 C_{g,put,CAS} = (1-\rho_g) \cdot \lambda_g  \sum\limits_{i=1}^D
 \alpha_{ig}\Big(o_m\Big(\underbrace{\sum\limits_{j=1}^D{p^n_{ji}} \cdot iq^1_{ijg}}_{\textrm{phase 1}} + \\ 
  \underbrace{\sum\limits_{k=1}^D{p^n_{ik} \cdot iq^3_{ikg}}}_{phase 3}\Big) +  \underbrace{\frac{o_g}{k_g}\sum\limits_{m=1}^D{p^n_{im} \cdot iq^2_{img}}}_{phase 2}\Big).
\end{split}
\end{equation}

}

Note the role played by the key boolean decision variable $iq^k_{ijg}$ whose interpretation is: $iq^k_{ijg}$=1 iff data center $j$ is in the $k^{th}$ quorum for clients in/near data center $i$. 
In the above expressions, $(1-\rho_g) \cdot \lambda_g \cdot \alpha_{ig}$ captures the PUT request rate arising at/near data center $i$ and the $o_m$ and $o_g$ multipliers convert this into bytes per unit time. The terms within the braces model the per-byte network transfer prices. The first term represents network transfer prices that apply to the first phase of the ABD PUT protocol whereas the second term does the same for ABD PUT's second phase. The  term $p^n_{ji} \cdot iq^1_{ijg}$ should be understood as follows: since ABD's first phase involves clients in/near data center $i$ sending relatively small-sized {\em write-query} messages to all servers in their quorum (i.e., quorum 1, hence the 1 in the superscript of $iq$) followed by these servers responding with their (tag, value) pairs, the subscript in $p^n_{ji}$ is selected to denote the price of data transfer from $j$ (for the server at data center $j$) to $i$ (for clients located in/near data center $i$).
The network cost per unit time for CAS is similar, with the number of phases being $3$ and the network cost savings offered by CAS reflected in phase $3$, where the value size $o_g$ is divided by the parameter $k_g.$ The networking costs for GET are presented in Appendix \ref{app:optimization} in \cite{legostore-arxiv}.

 The storage cost is modeled as: $C_{g,storage} = p^s \cdot \big(e_g \cdot m_g \cdot \frac{o_g}{k_g} + (1 - e_g) \cdot m_g \cdot o_g \big),$ see \cite{legostore-arxiv} for explanations. Finally, we consider the  VM costs per unit time for key $g$. Our assumptions on modeling VM costs include: ability of procurement of VMs at fine granularity (see, e.g., \cite{small-vm}) and VM autoscaling \cite{burscale, guo-autoscale} to ensure satisfactory provisioning of VM capacity at each DC.  We assume that this suitable VM capacity chosen by such an autoscaling policy is proportional to the total request arrival rate at data center $i$ for key $g$. With these assumptions, the VM cost for key $g$ at data center $i$ is:   
$C_{g,VM} = \theta^v \cdot \sum\limits_{j=1}^D p_j^v \cdot \lambda_g +\sum\limits_{i=1}^D \alpha_{ig} +\sum\limits_{k=1}^4 iq^k_{ijg},$ where $\theta^v$ is an empirically determined multiplier that estimates VM capacity needed to serve the the request rate arriving at data center $j$ for $g$.  
}

\vc{\noindent {\bf Constraints:} Our optimization needs to capture the 3 types of constraints related to: (i) ensuring linearizability; (ii) meeting availability guarantees corresponding to the parameter $f$; and (iii) meeting latency SLOs.  
The key modeling choices we make are: (i) to use worst-case latency as a "proxy" for tail latency; and (ii) ignore latency contributors within a data center other than data transfer time (e.g., queuing delays, encoding/ decoding time). For PUT operations in CAS, the constraints are $\forall {i, j, k \in \mathcal{D},}$:

\allowdisplaybreaks{
\begin{equation}
\begin{split}
&\underbrace{iq^1_{ijg} \cdot \Big(l_{ij} + l_{ji} + \frac{o_m}{B_{ji}}\Big)}_{\textrm{Latency of first phase of PUT}} + \underbrace{iq^2_{img}\cdot \Big(l_{im} + {\frac{o_g/k_g}{B_{im}} + l_{mi}\Big)}}_{\textrm{Latency of second phase of PUT}} ~~~~+ \\
&\underbrace{iq^3_{ikg}\cdot \Big(l_{ik} + \frac{o_m}{B_{ik}} + l_{ki}\Big)}_{\textrm{Latency of third phase of PUT}} \leq l_{put}.
\end{split}
\end{equation}}
See Appendix \ref{app:optimization} in \cite{legostore-arxiv} for explanations, and for constraints for PUT operations of ABD, and for GETs.  Linearizability and availability targets manifest as constraints on quorum sizes.}

}

%% file: 5dyn.tex
\subsection{How to Reconfigure?}
\label{sec:reconfig-proto}

LEGOStore uses a 
reconfiguration protocol that transitions chosen keys from their old configurations to their new configurations without violating linearizability. \vc{Unlike our approach, consensus-based protocols such as Raft and Viewstamped Replication~\cite{Ousterhout:2013:SDL:2517349.2522716,liskov2012viewstamped}, implement the key-value store as a log of commands applied sequentially to a replicated state machine. These solutions implement reconfiguration by adding it as a special command to this log.} Thus, when a reconfiguration request is issued, the commands that are issued before the reconfiguration request are first applied to the state machine before executing the reconfiguration. To execute the reconfiguration, the leader transfers the state to the new configuration. After the transfer, it resumes handling of client  commands that are serialized after the reconfiguration request, but replicating the state machine in the new configuration. 
While our approach 
does not involve a replicated log\footnote{Rather than a replicated log, we simply have a replicated single read/write variable per key, which is updated on receiving new values.}, it is possible to develop an approach that inherits the essential idea of consensus-based reconfiguration as follows:  (i) On receiving a reconfiguration request, wait for all ongoing operations to complete, and pause all new operations; 
(ii) Perform the reconfiguration by transferring state from the old to the new configurations; (iii) Resume all operations.


Under the reasonable assumption that reconfigurations {\em of a given key are performed relatively infrequently},\footnote{{More precisely, we assume that reconfigurations to a key are separated in time by periods that are much longer (several minutes to hours or even longer)  than the time it takes to reconfigure (sub-second to a second, see measurements in Section~\ref{sec:reconf-load}).}} our design goal is to ensure that user performance is not degraded in the common case where the configurations remain static. For this, it is crucial that user operations that are not concurrent with reconfigurations  follow the baseline static protocols without requiring additional steps/phases (such as contacting a controller). 
Our protocol does not assume any special relation between the old and new configurations. It can handle all types transitions, including changing of the replication factor, EC parameters, quorum structure, and the protocol itself. 

\vc{We wish to keep the number of communication phases as well as the number of operations affected as small as possible. Towards this, \legostore's reconfiguration protocol improves upon steps (i)-(iii) above.} Reconfigurations are conducted by a controller that reads data from the old configuration and transfers it to the new configuration. {On detecting a workload or system change (See Section~\ref{sec:reconfig-whenwhat} for details), the controller immediately performs the reconfiguration without having to wait for all ongoing operations to complete, i.e., without having to perform step (i).} This enables \legostore~to adapt quickly to  workload changes. Furthermore, steps (ii),(iii) are conducted jointly through a single round of messaging. \vc{In particular, \legostore's protocol  integrates with the underlying protocols of CAS and ABD and piggybacks the reconfiguration requests from the controller along with the actions that read or transfer  the data. \legostore's algorithm is provably linearizable (Appendix \ref{app:reconfig} of~\cite{legostore-arxiv}), and therefore, preserves the correctness of the overall data store.} The algorithm blocks certain concurrent operations and then resumes them on completing the reconfiguration.

The reconfiguration protocol is described formally 
in Appendix \ref{app:reconfig} of ~\cite{legostore-arxiv}. 
We assume that reconfigurations are applied sequentially by the reconfiguration controller (or simply controller). 
The controller sends a \texttt{reconfig\_query} message to all the servers in the old configuration. \vc{The \texttt{reconfig\_query} message serves to both signal a change in configuration, as well as an internal `get' request for the controller to read a consistent value in order to transfer it to the new configuration.}  On receiving this message, the servers pause all the ongoing operations and respond with the latest value if the old configuration is performing ABD, or the highest tag labeled `fin' if it is performing CAS. The controller waits for a quorum to respond from the old configuration. If the old configuration is performing CAS, then the reconfiguration controller sends a \texttt{reconfiguration\_get} message to servers in the old configuration with the highest tag among the messages received from the quorum \vc{(the quorum size is $N-q_2+1$ if the old configuration is performing ABD and $N-\min(q_3,q_4)+1$ if it is performing CAS).} \vc{A server that receives the \texttt{reconfiguration\_get} message with tag $t$ sends a codeword symbol corresponding to that tag, if it is available locally; else it responds with an acknowledgement. The reconfiguration controller then obtains responses from a quorum of $q_4$ messages and decodes the value from the responses.}
 The controller then proceeds to write the $(tag,value)$ pair to the new configuration, performing encoding if the new configuration involves CAS. On completing writing the value to the new configuration, the controller sends a $\texttt{finish\_reconfig}$ message to servers in the old configuration. On receiving these messages, the servers complete all operations with tag less than or equal to $t_{highest}$ and send $\texttt{operation\_fail}$ messages along with information of new configuration to other pending operations that were paused. On receiving $\texttt{operation\_fail}$ messages, the clients restart the operation in the new configuration.


\subsection{When and What to Reconfigure?}
\label{sec:reconfig-whenwhat}

At the heart of our strategy are these questions: (i) is a key configured poorly for its current or upcoming workload? (ii) if yes, should it be stored using a different configuration?
While our discussion focuses on workload dynamism, our ideas also apply to changes in system/infrastructure properties. 

\noindent {\bf Is a Key Configured Sub-Optimally?} Some workload\- changes can be predicted (e.g., cyclical temporal patterns or domain-specific insights from users) while others can only be determined after they have occurred. Generally speaking, a system such as \legostore~would employ a combination of predictive and reactive approaches for detecting workload changes~\cite{wilkespaper,icac05,burscale}. In this paper, we pursue a purely reactive approach. We employ two types of reactive rules that are indicative of a key being configured poorly: 

\noindent $\bullet$ \underline{\em SLO violations: } If SLO violations for a key are observed for more than a threshold duration or during a window containing a  threshold number of requests, \legostore~chooses to  reconfigure it. 
Section~\ref{sec:reconf-load}, \ref{sec:recon-dcfail} show that LEGOStore's reconfiguration occurs within 1 sec; the threshold should be  set sufficiently larger than this to avoid harmful  oscillatory behavior over-optimizing for transient phenomena. 
If, in addition to SLO violations, some quorum members are suspected of being slow or having failed, these nodes are removed from consideration when determining the next configuration.  

\noindent $\bullet$ \underline{\em Cost sub-optimality:}  Alternatively, a key's configuration might meet the SLO but the estimated running cost might exceed the expected cost. We consider such sub-optimality to have occurred if the observed cost exceeds modeled cost by more than a threshold percentage as assessed over a window of a certain duration.\footnote{ Detailed exploration of the impact of the thresholds on performance is future work.} 
Having determined the need for a change, \legostore~ reconfigures the key based on our {\em cost-benefit analysis} described below. 

\noindent {\bf Should Such a Key be Reconfigured?}   For the case of SLO violations, \legostore~will reconfigure as we consider SLO maintenance to be sacrosanct. 
In the rest of the discussion, we focus on the case of cost sub-optimality. We  assume that such a key's workload features can be predicted for the near term. In the absence of such predictability---and this applies more generally to any similar system---\legostore's options are limited to using a state-of-the-art latency-oriented optimization (see our baselines ABD Nearest and CAS Nearest in Section~\ref{sec:setup}) that is likely to be able to meet the SLO; the impact of such a heuristic on cost is examined in Section~\ref{sec:opt-ins}.

Assuming predictability, \legostore~computes the new configuration with the updated workload characteristics using the  optimization framework from Section~\ref{sec:opt}. For an illustrative instance of such decision-making, denote this newly computed configuration as $c_{new}$ and the existing configuration as $c_{exist}$. Let us denote by {\tt Cost}$(c)$ the per time unit cost incurred when using configuration $c$. We assume an additional predicted feature $T_{new}$, the minimum duration for which the predicted workload properties will endure before changing. \legostore~ compares the cost involved in reconfiguring with potential cost savings arising due to it.  Reconfiguring a key entails (a) {\em explicit costs} arising from the addtional data transfer; and (b) {\em implicit costs}  resulting from requests that are slowed down or rejected. An evaluation (see Section~\ref{sec:reconf-load} for details) of our reconfiguration protocol suggests that the number of operations experiencing slowdown is small. 
Therefore, we consider only (a). 

    
\legostore's cost-benefit analysis is simple.  A reconfiguration to $c_{new}$ is carried out if the potential (minimum) cost savings $T_{new} \cdot  \big(${\tt Cost}$(c_{exist})-${\tt Cost}$(c_{new})\big)$ significantly outweigh the explicit cost of reconfiguration as captured by {\tt ReCost}$(c_{old},c_{exist})\cdot (1+\alpha)$. {\tt ReCost}$(.,.)$ is the cost of network transfer induced by our reconfiguration and its calculation involves ideas similar to those presented in our optimizer (see  Appendix~\ref{app:optimization} of~\cite{legostore-arxiv}). The $(1+\alpha)$ multiplier ($\alpha>0$) serves to capture how aggressive or conservative \legostore~ wishes to be.  The efficacy of our heuristics depends on the predictability in workload features and the parameter $T_{new}$. 

Algorithms~\ref{algo:recon_cont}, \ref{algo:recon_server_client} of \cite{legostore-arxiv} show that the time to complete a reconfiguration (call it $T_{re}$) is largely dictated by the RTTs between the controller and the servers farthest from it in the quorums involved in various phases of the reconfiguration protocol.
Figure~\ref{fig:recon_timeline} of \cite{legostore-arxiv} highlights this period when both the old and the new configurations use ABD. 


%% file: 7evaluation.tex
\section{Evaluation}
\label{sec:eval}

We implement a prototype LEGOStore on the Google Compute Engine (GCE) public cloud and make the code for our prototype as well as the optimizer available  under the Apache license 2.0 at \href{https://github.com/shahrooz1997/LEGOstore}{github.com/shahrooz1997/LEGOstore}. Details of our prototype implementation may be found in Appendix~\ref{app:impl} of \cite{legostore-arxiv}. 
We evaluate LEGOStore in terms of its ability to (i)  lower costs compared to the state-of-the-art; and  (ii) meet latency SLOs. We use  Porcupine~\cite{athalye2017porcupine} for verifying linearizability of execution histories of our prototype.

\subsection{Experimental Setup}
\label{sec:setup}

\noindent {\bf Prototype Setup:} We deploy our LEGOStore  prototype across 9 Google Cloud Platform (GCP) DCs with locations, pairwise RTTs, and resource pricing shown in Tables~\ref{tab:dcs} and~\ref{tab:net+rtt}. We locate our users within these data centers as well for the experiments. 
We conduct extensive validation of the efficacy of the latency and cost modeling underlying our optimizer (Appendix~\ref{app:valid} of~\cite{legostore-arxiv}).

\noindent \textbf{Workloads:}  We employ a custom-built workload\footnote{We do not use an existing workload generator such as YCSB~\cite{ycsb} because we wish to explore a wider workload feature space than covered by available tools. } generator which emulates a user application {with an assumption that it sends requests as per a Poisson process.} We explore a large workload space by systematically varying our workload parameters as follows. 
\squishlist
\item
    3 per-key sizes in KB: (i) 1, (ii) 10, and (iii) 100;
\item 
    3 per-key read ratios for high-read (HR), read-write (RW), and high-write (HW) workloads, resp.: (i) 30:1, (ii) 1:1, and (iii) 1:30; 
\item 
    3 per-key arrival rates in requests/sec: (i) 50, (ii) 200, and (iii) 500; 
\item
    3 sizes for the overall data: (i) 100 GB, (ii) 1 TB, and (iii) 10 TB. 
\item 
    7 different client distributions: (i) Oregon, (ii) Los Angeles, (iii) Tokyo, (iv) Sydney, (v) Los Angeles and Oregon, (vi) Sydney and Singapore, and (vii) Sydney and Tokyo. 
\squishend

This gives us a total of 567 
diverse \emph{basic} workloads for a given availability target and latency SLO. Finally, we also vary the availability target ($f$=1 in this section and $f$=2 in Appendix~\ref{app:addnresults} of~\cite{legostore-arxiv}) 
and latency SLO (in the range 200 ms---1 sec) in our experiments. 
Additionally, we use the following customized workloads to explore  particular performance related phenomena: (i) a uniform client distribution across all 9 locations; (ii) workloads related to  Figures~\ref{fig:k}-\ref{fig:recon-lat}; we describe these in the text accompanying these figures. Finally, while our exact metadata size varies slightly between ABD and CAS, we round it up to an overestimated  100 B. 


\noindent \textbf{Baselines:} We would like to compare LEGOStore's efficacy to the most important state-of-the-art approaches. To enable such comparison, we construct the following baselines:
    
\noindent $\bullet$    \textbf{ABD Fixed and CAS Fixed:} These baselines use only ABD or only CAS, respectively. The baseline employs either a fixed degree of replication or a fixed set of CAS parameters.  These  parameter values ($3$ for ABD and $(5,3)$ for CAS) are the ones chosen  most frequently by our optimizer across our large set of experiments described in Section~\ref{sec:opt-ins}. For these fixed parameters, these baselines pick the DCs  with the smallest average network prices for their quorums, where the average for a DC $i$ is calculated over the price of transferring data to all user locations. 
    A comparison with these baselines demonstrates that merely knowing the right parameters does not suffice---one must pick the actual DCs judiciously. 

\noindent $\bullet$    \textbf{ABD Nearest and CAS Nearest:} These baselines also use only ABD or only CAS. 
However, they do not a priori fix the degree of replication or the EC parameters. Instead, we pick the optimized value for each parameter and choose quorums that result in the smallest latencies for the GET/PUT operations ignoring cost concerns. They solve a variant of our optimizer where the objective is latency minimization, expressions involving costs are not considered, and all other constraints are the same.  
These baselines serve as representatives of existing works (e.g., Volley~\cite{volley} and  \vc{\cite{Sharov:2015:TMY:2824032.2824047}}) that primarily focus on latency reduction. While the baselines are admittedly not as sophisticated as Volley, our results demonstrate that unbridled focus on latency  can lead to high costs.

\noindent $\bullet$    \textbf{ABD Only Optimal and CAS Only Optimal:} These are our most sophisticated baselines meant to represent state-of-the-art approaches. ABD Only Optimal and CAS Only Optimal are representative of works that optimize replication-based systems such as SPANStore~\cite{spanstore} or EC-based systems such as Pando~\cite{pando}.

It is instructive to note that our baselines are quite powerful. \vc{In fact, our optimizer picks the lower cost feasible solution among ABD Only Optimal and CAS Only Optimal, so we expect at least one of these baselines to be competitive for any given workload. Yet, we will demonstrate that: (i) these baselines \emph{individually} perform poorly for many of our  workloads; and (ii) the choice of which of the two is better for a particular workload is highly non-trivial.} \

\begin{figure*}[t]
\centering
\resizebox{0.9\textwidth}{!}{
\begin{subfigure}{\columnwidth}
  \centering
  \includegraphics[width=0.95\columnwidth]{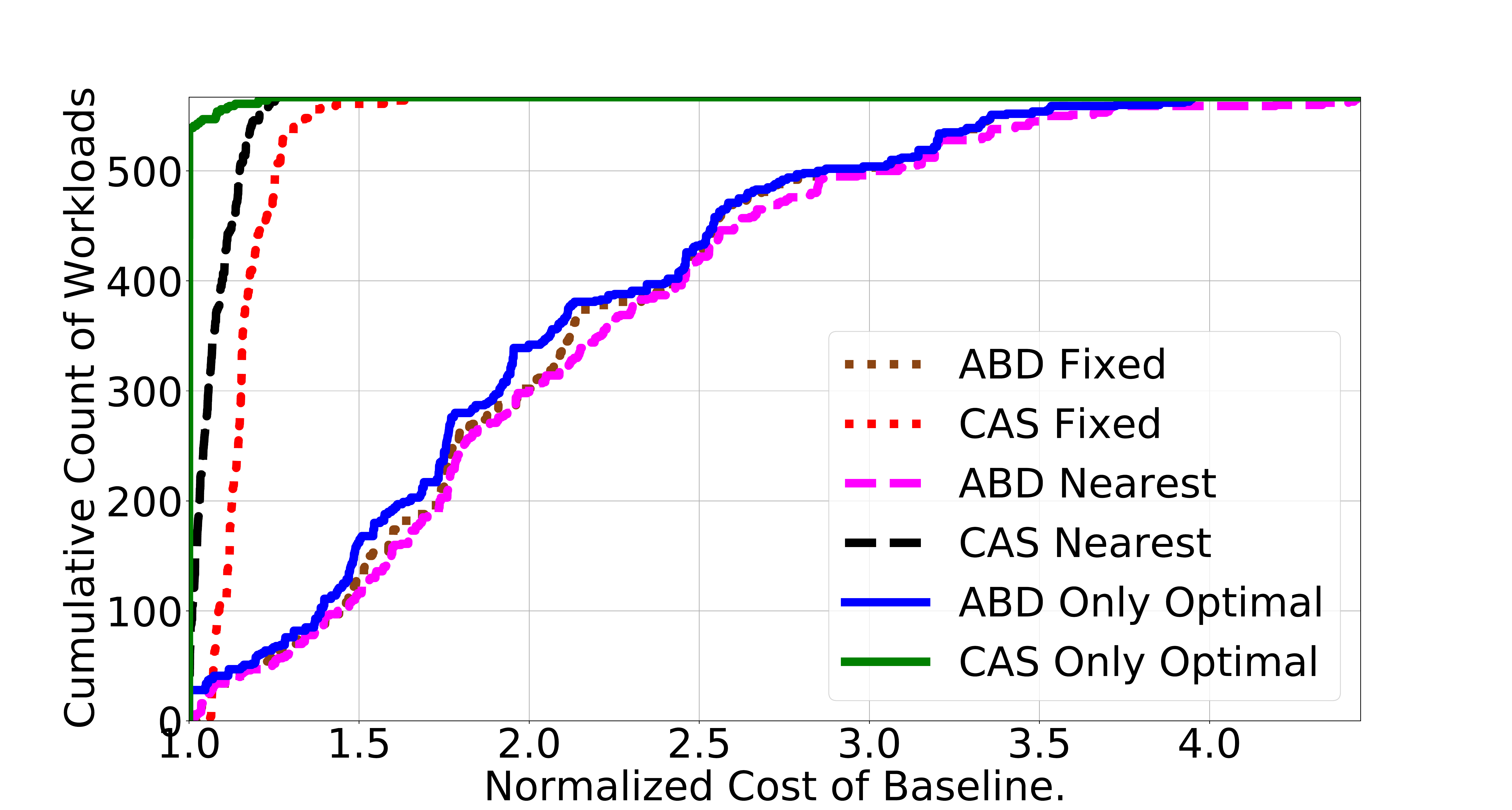}
  \caption{Latency SLO=1 sec.}
\end{subfigure}%
\begin{subfigure}{\columnwidth}
  \centering
  \includegraphics[width=0.95\columnwidth]{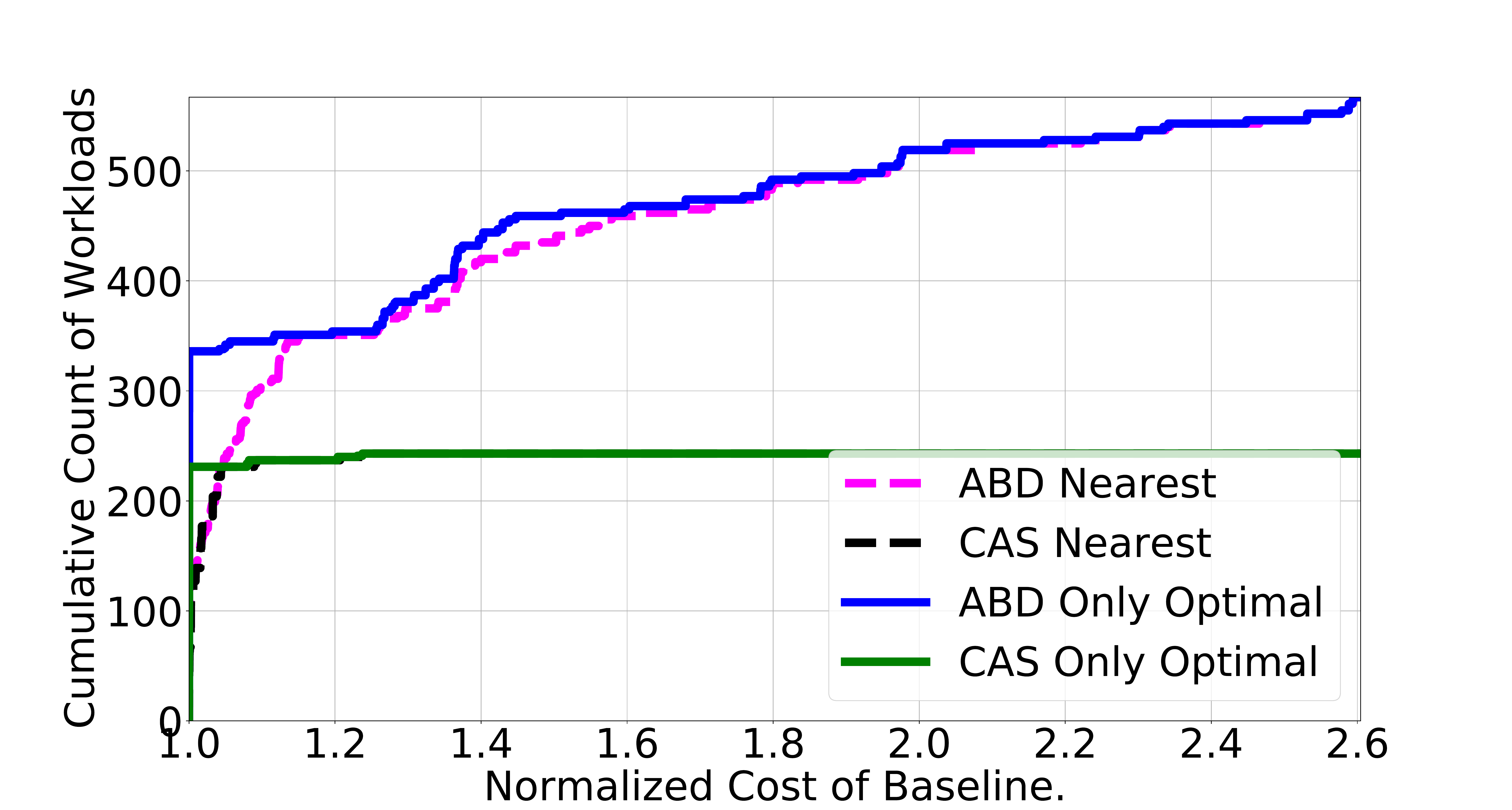}
  \caption{Latency SLO=200 msec.}
\end{subfigure}}
\caption{ Cumulative count of the normalized cost of our baselines (for our 567 basic workloads) with $f=1$ and two extreme latency SLOs.}
\label{fig:normalized-costs}
\end{figure*}

\subsection{Insights from Our Optimization}
\label{sec:opt-ins}

\begin{figure*}[ht]
    \includegraphics[width=\textwidth]{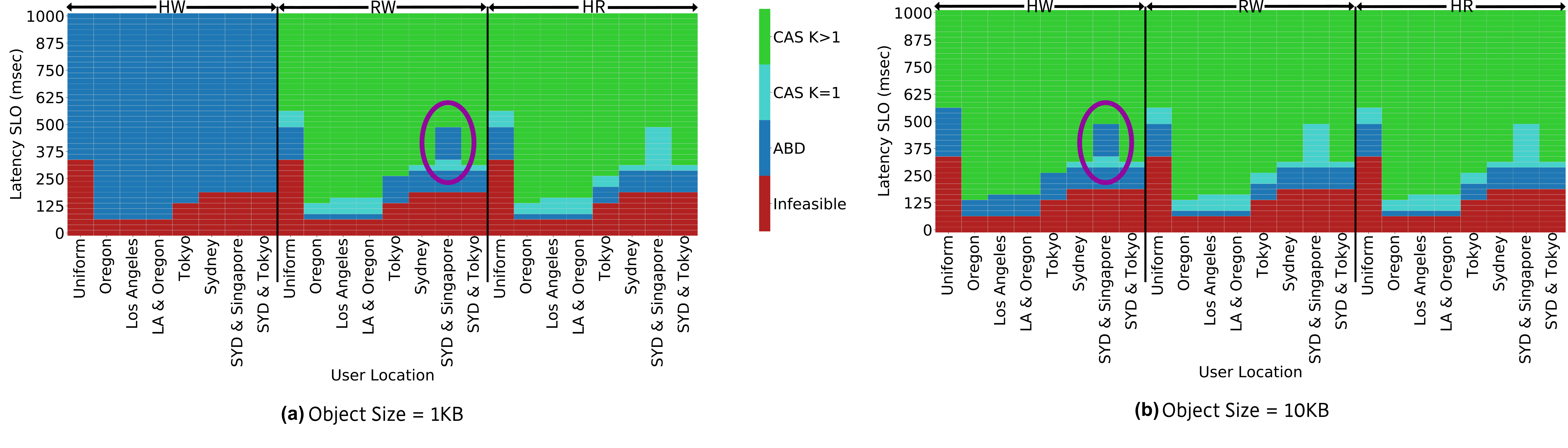}
    \caption{Sensitivity of the optimizer's choice to the latency SLO. We consider 2 object sizes (1KB and 10KB), 8 different client distributions, arrival rate=500 req/sec, and $f$=1. We consider 3 different read ratios (HW, RW, HR defined in Section~\ref{sec:setup}). 
    }
    \label{fig:slo-latency}
\end{figure*}

\subsubsection{The Extent and Nature of Cost Savings}
\label{subsec:nature}

In Figure~\ref{fig:normalized-costs}, we express our optimizer's  cost savings over our baselines  via each  baseline's {\em normalized cost} (cost offered by baseline / cost offered by our optimizer); \vc{note that our optimizer has the lowest cost among all the baselines, so this ratio is at least $1$}. We consider  our collection of 567 basic workloads with $f$=1 and (a) a relaxed SLO of 1 sec and (b) a more stringent SLO of 200 msec. \vc{At least one of the baselines - and consequently, our optimizer - meet the SLOs for all the $567$ workloads chosen}.  We begin by contrasting the first-order strengths and weaknesses of ABD and CAS as they are understood in conventional wisdom. For the relatively relaxed latency SLO of 1 sec in Figure~\ref{fig:normalized-costs}(a), we find that ABD Only Optimal (and other ABD-based variants) have more than twice the cost of our optimizer for more than 300 (i.e, more than half) of our workloads. On the other hand, CAS Only Optimal  closely tracks our optimizer's cost.
That is, as widely held, if high latencies are tolerable, EC can save storage and networking costs. Figure~\ref{fig:normalized-costs}(b), with its far more stringent SLO of 200 msec, confirms another aspect of conventional wisdom.  \vc{CAS only optimal is now simply unable to meet the SLO for many workloads (324 out of 567).} This is expected given its 3-phase PUT operations and larger quorums. {(See similar results with $f$=2 in Figure~\ref{fig:normalized-costs2} of~\cite{legostore-arxiv})).} What is  surprising, however, is that when we focus on the subset of 243 workloads for which CAS Only Optimal is feasible, it proves to be the cost-effective choice---nearly all workloads for which CAS Only Optimal is feasible have a normalized cost of 1. So, even for stringent SLOs, EC does hold the potential of saving costs. Unlike in Figure~\ref{fig:normalized-costs}(a), however, CAS Fixed or CAS Nearest are nowhere close to being as effective as CAS Only Optimal. That is, while EC can be cost-effective for these workloads, its 
quorums need to be chosen carefully rather than via  simple greedy heuristics.\footnote{As a further nuance, only 3 workloads out of 243 use CAS with $k$=1.} 
Replication tends to be the less preferred choice for more relaxed  SLOs but, again, there are exceptions.

\subsubsection{Sensitivity to Latency SLO} 
\label{subsec:latency}

We focus on how the cost-efficacy of ABD vs. CAS depends on the latency SLO by examining the entire range of latencies from 50 msec to 1 sec. Furthermore, we separate out this dependence based on read ratio, availability target and object size. Our selected results are shown in 
Figure~\ref{fig:slo-latency}. 
As expected, as one moves towards more relaxed SLOs,  the optimizer's choice tends to shift from ABD to CAS (recall the 3-phase PUTs and larger quorums in CAS). 
The complexity that the figures bring out is {\em when} this transition from replication to EC occurs---we see that, depending on workload features, this transition may never occur (e.g., HW in Figure~\ref{fig:slo-latency}(a)) or may occur at a relatively high latency (e.g., at 575 msec for the uniform user  distribution for RW/HR).\footnote{The reader might be intrigued by the portions of Figure~\ref{fig:slo-latency} highlighted using ovals. Here,  our optimizer's choice shifts from ABD to CAS as the latency SLO is relaxed (as expected) but then it shifts back to ABD! We consider this to be a  quirk of the heuristics embedded in our optimizer rather than a fundamental property of the optimal solution. } In particular, more spatially distributed workloads correspond to a tendency to choose replication over EC; for instance, for workloads with uniformly distributed users,  SLOs smaller than 300 msec are infeasible due to a natural lower bound implied by the inter-DC latencies. 
We find that $f$ also has a complex impact on the optimizer's choice; see results with $f$=2 in Figure~\ref{fig:slo-latencyf2},  Appendix~\ref{app:addnresults} in \cite{legostore-arxiv}. 

\subsubsection{Read- vs. Write-Intensive Workloads}
\label{subsec:asymmetry}

One phenomenon that visibly stands out in Figure~\ref{fig:slo-latency} is how write-intensive workloads for the  relatively small object sizes (HW in Figure~\ref{fig:slo-latency}(a)) prefer ABD  even for the more relaxed SLOs. This preference of ABD over CAS becomes less  pronounced when we increase the object size to 10KB in  Figure~\ref{fig:slo-latency}(b). Finally, we also observe that read-intensive/moderate workloads tend to prefer CAS ($K$=1) over ABD, even when replication is used.  To understand this asymmetry,  note the following:

\noindent $\bullet$ \textbf{Reads:}     
    Whereas both ABD and CAS have a ``write-back'' phase for read operations, ABD's write-back phase carries data, while CAS's only carries metadata, and thereby incurs much lower network cost. Thus, our optimizer tends to prefer CAS for HR workloads.

\noindent $\bullet$ \textbf{Writes:} 
    For writes,  CAS  involves 3 phases whereas ABD only requires 2. Since each phase incurs an additional overhead in terms of metadata, the metadata costs for write operations are higher for CAS. Therefore, especially for small object sizes (Figure~\ref{fig:slo-latency}(a)) and write-heavy workloads, our optimizer will tend to  prefer  ABD. 

Collectively, the results in Figure~\ref{fig:slo-latency} convey the significant complexity of choosing between ABD and CAS. Within CAS-friendly workloads, there is further substantial complexity in how the parameter $K$ depends on workload features.

\begin{figure*}[t]
\centering
\begin{subfigure}{.33\textwidth}
  \centering
  \includegraphics[width=\columnwidth]{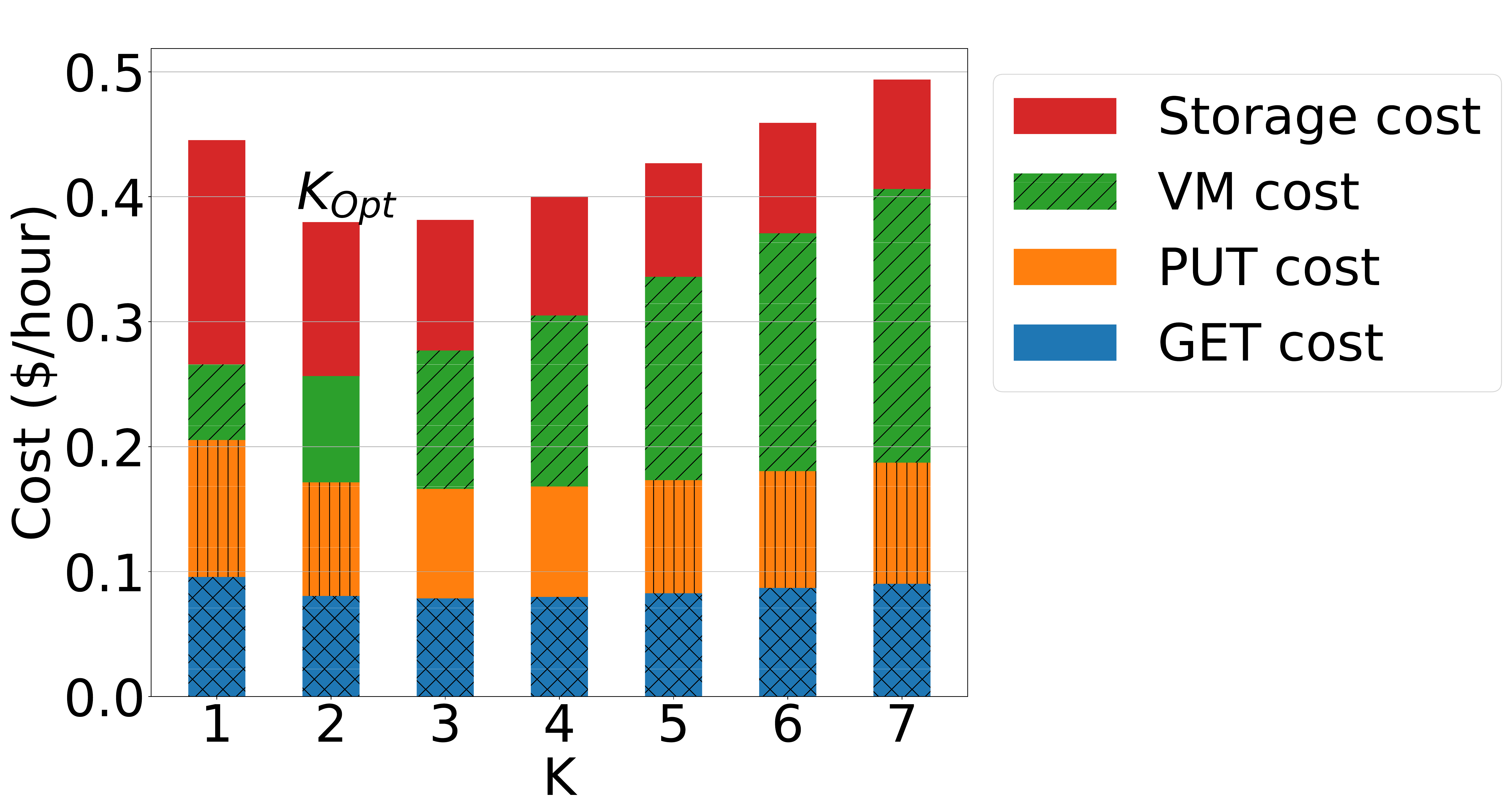}
  \caption{}
\end{subfigure}%
\begin{subfigure}{.33\textwidth}
  \centering
  \includegraphics[width=\columnwidth]{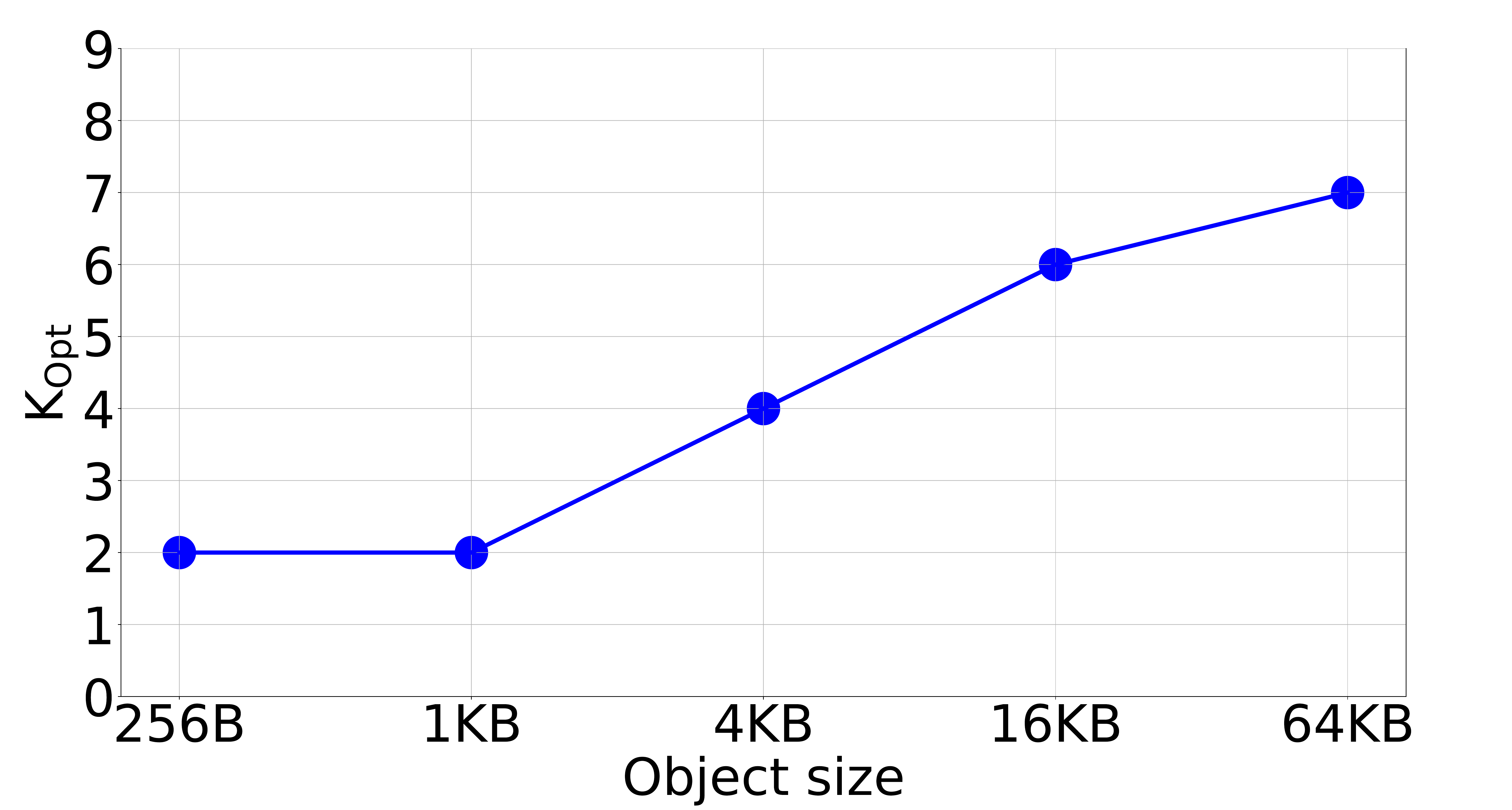}
  \caption{}
\end{subfigure}
\begin{subfigure}{.33\textwidth}
  \centering
  \includegraphics[width=\columnwidth]{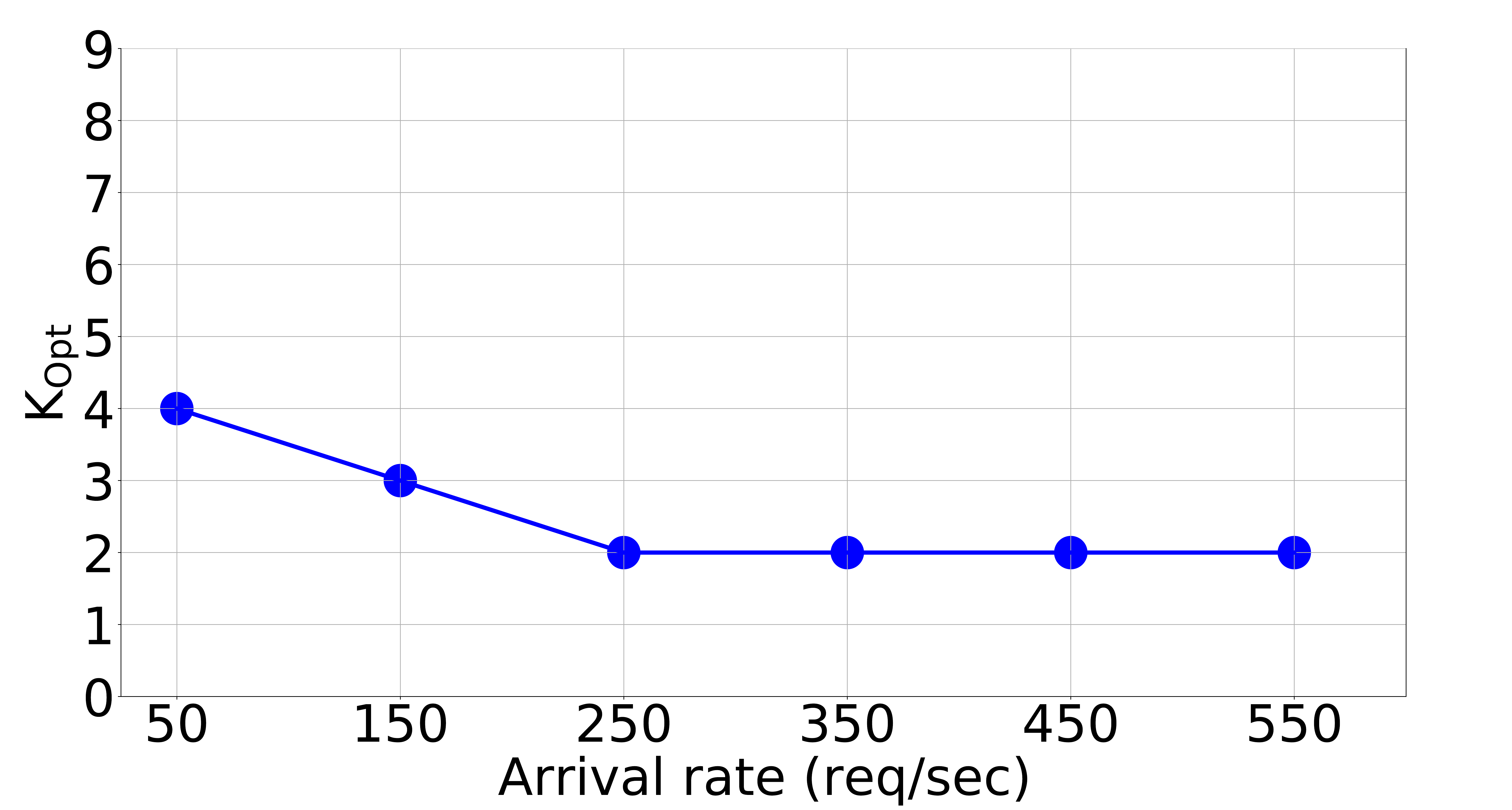}
  \caption{}
\end{subfigure}
\caption{For CAS-based solutions, cost is non-monotonic in $K$ and $K_{opt}$ has a complex relation with object size and arrival rate. Latency SLO is 1 sec.} 
\label{fig:k}
\end{figure*}

\subsubsection{Factors Affecting Optimal Code Dimension \texorpdfstring{$K$}{K}} \label{subsubsec:opt-k-dep} 

We illustrate our findings using the representative results in Figure~\ref{fig:k}(a)-(c) based on a workload with the following features for which CAS is the cost-effective choice:  object size=1KB; datastore size=1TB; arrival rate=200 req/sec; read ratio= RW (50\%); user locations are Sydney and Tokyo; latency SLO=1 sec.  
To understand the effects in Figure~\ref{fig:k}(a)-(c), we develop a simple analytical model based on the empirical results (details in Appendix~\ref{app:k}, \cite{legostore-arxiv}). Our model relates cost to $K$, object size ($o$), arrival rate ($\lambda$), and $f$ as follows:
\begin{equation}
    cost = \Big(c_1 \cdot \lambda \cdot K + c_2 \cdot o \cdot \lambda  \cdot \frac{f}{K} +c_3 \cdot o \cdot \frac{2f}{K} + \overline{c}_{4}\Big). 
\end{equation}
Here, $c_1,c_2,c_3$  are system-specific constants VM cost, network cost, and storage cost, respectively\footnote{$\overline{c}_{4}$ is a constant and does not affect $K_{opt}$.}. Our model captures and helps understand the {\em non-monotonicity of cost  in $K$} seen in Figure~\ref{fig:k}(a). 
This behavior emerges because the following cost components move in opposite directions with growing $K$: network and storage costs decrease due to reduction in object size, while VM costs increase due to increase in quorum sizes. Fundamentally, this implies that even under very relaxed latency constraints, the highest value of $K$ is not necessarily optimal \vc{(contrary to the coarse analysis of Table \ref{table:comparison})}. Our model yields the following optimal value of $K$:
$K_{opt} = \sqrt{\frac{ o\cdot f \cdot (c_2 \cdot \lambda + 2 c_3)}{c_1\cdot \lambda}}.
$
Observe that $K_{opt}$ increases with object size\footnote{A qualification to note is that the phenomenon is connected to our modeling choice of having VM cost independent of the object size $o$. E.g., if the VM cost were chosen as an affine function of $o$, then the dependence of $K_{opt}$ on $o$ would diminish.}, which is in agreement with  Figure~\ref{fig:k}(b). 
We observe a similar qualitative match  between our model-predicted dependence of $K_{opt}$ on arrival rate and that in Figure~\ref{fig:k}(c). $K_{opt}$ is a decreasing function of the arrival rate {$\lambda$}, and saturates to a constant $K^{*}$ when   {$\lambda \to \infty,$ i.e.,} when the storage cost becomes a negligible component of the overall cost. {Interestingly, even for $\lambda \to \infty$, the system does not revert to replication, i.e., $K^{*}$ is not necessarily $1$.}

\subsubsection{Does EC Necessarily Have Higher Latency Than Replication?}
\label{subsec:ECvsrep}

Conventional wisdom dictates that EC has lower costs than replication but suffers from higher latency.  
We show that perhaps surprisingly, this insight does not always lead to the right choices in the geo-distributed setting. Note that for a linearizable store, requests cannot be local~\cite{attiya1994sequential}, and so even with replication, requests need to contact multiple DCs  and the overall latency corresponds to the response time of the farthest DC. Thus, in a geo-distributed scenario where there are multiple DCs at similar distances as the farthest DC in a replication-based system, EC can offer comparable latency at a lower cost. 
Our optimizer corroborates this insight. Consider a  workload where requests to a million objects of 1 KB come from users  in Tokyo. The workload is HR (read ratio of 97\%) with an arrival rate of 500 req/sec. To tolerate $f$=1 failure, the lowest GET latency achievable via ABD is 139 msec at a cost of $\$1.057\text{ per hour}$, whereas using CAS achieves a GET latency of 160 msec at a cost of $\$0.704\text{ per hour}$ - a cost saving of $33$\% for a mere 21 msec of latency gap. To tolerate $f$=2 failures for the same workload, the lowest GET latency with ABD is 180 msec at a cost of $\$1.254\text{ per hour}$, whereas CAS offers a GET latency of 190 msec at  a cost of $\$0.773\text{ per hour}$ - 38\% lower cost for a mere 10 msec  latency increase.

\subsubsection{Are Nearest DCs Always the Right Choice?} 
\label{sec:nearest}

Our optimizer reveals that, perhaps surprisingly, the naturally appealing approach of using DCs nearest to user locations~\cite{volley} 
can lead to wasted costs. We describe one such  finding in Appendix~\ref{app:nearest} of~\cite{legostore-arxiv}.

\begin{figure}[t]
\centering
\resizebox{0.9\textwidth}{!}{
\begin{subfigure}{.5\columnwidth}
  \centering
  \includegraphics[width=\columnwidth]{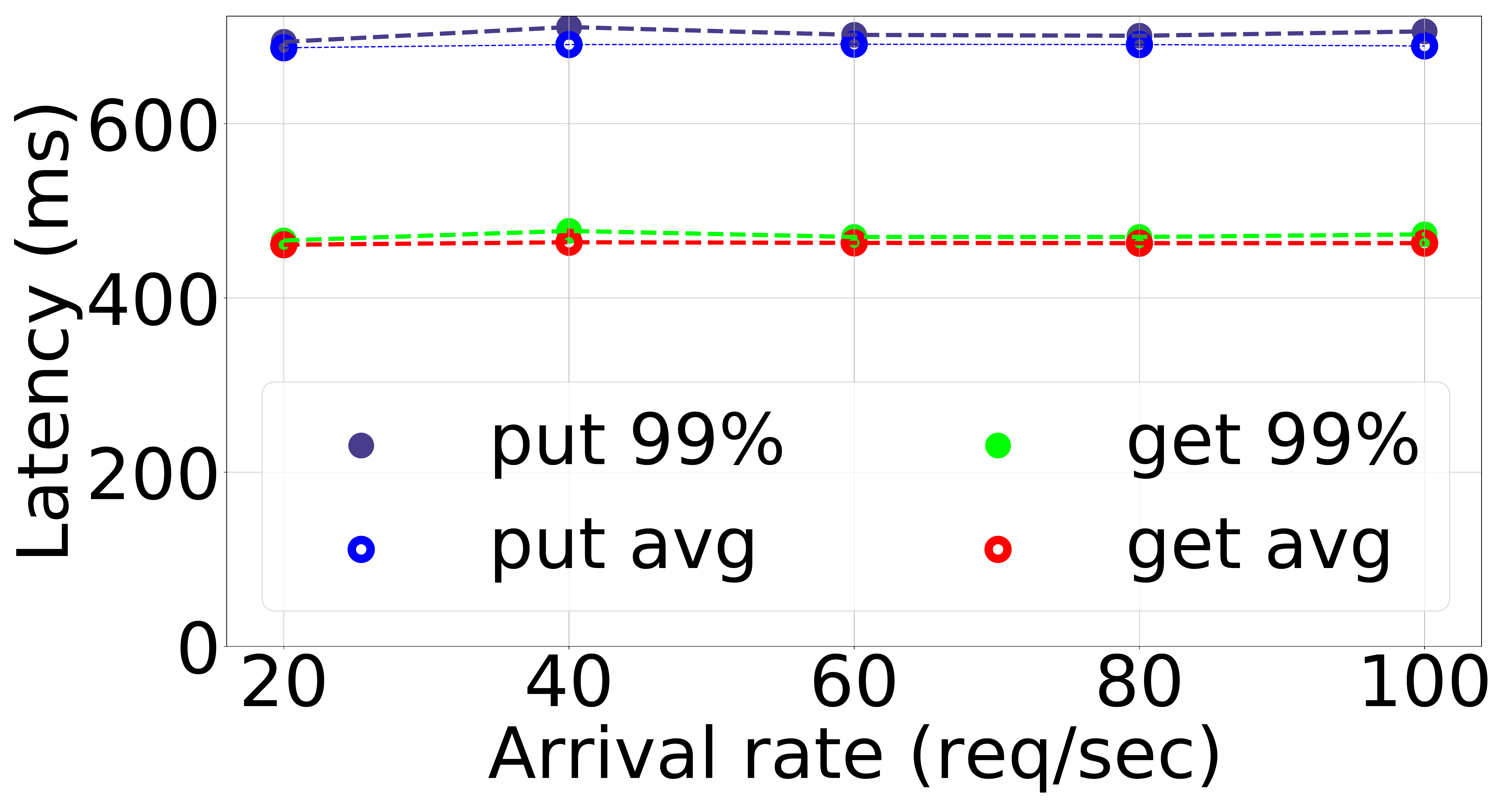}
  \caption{Read ratio=50\%.}
  \label{fig:slo-latency1}
\end{subfigure}%
\begin{subfigure}{.5\columnwidth}
  \centering
  \includegraphics[width=\columnwidth]{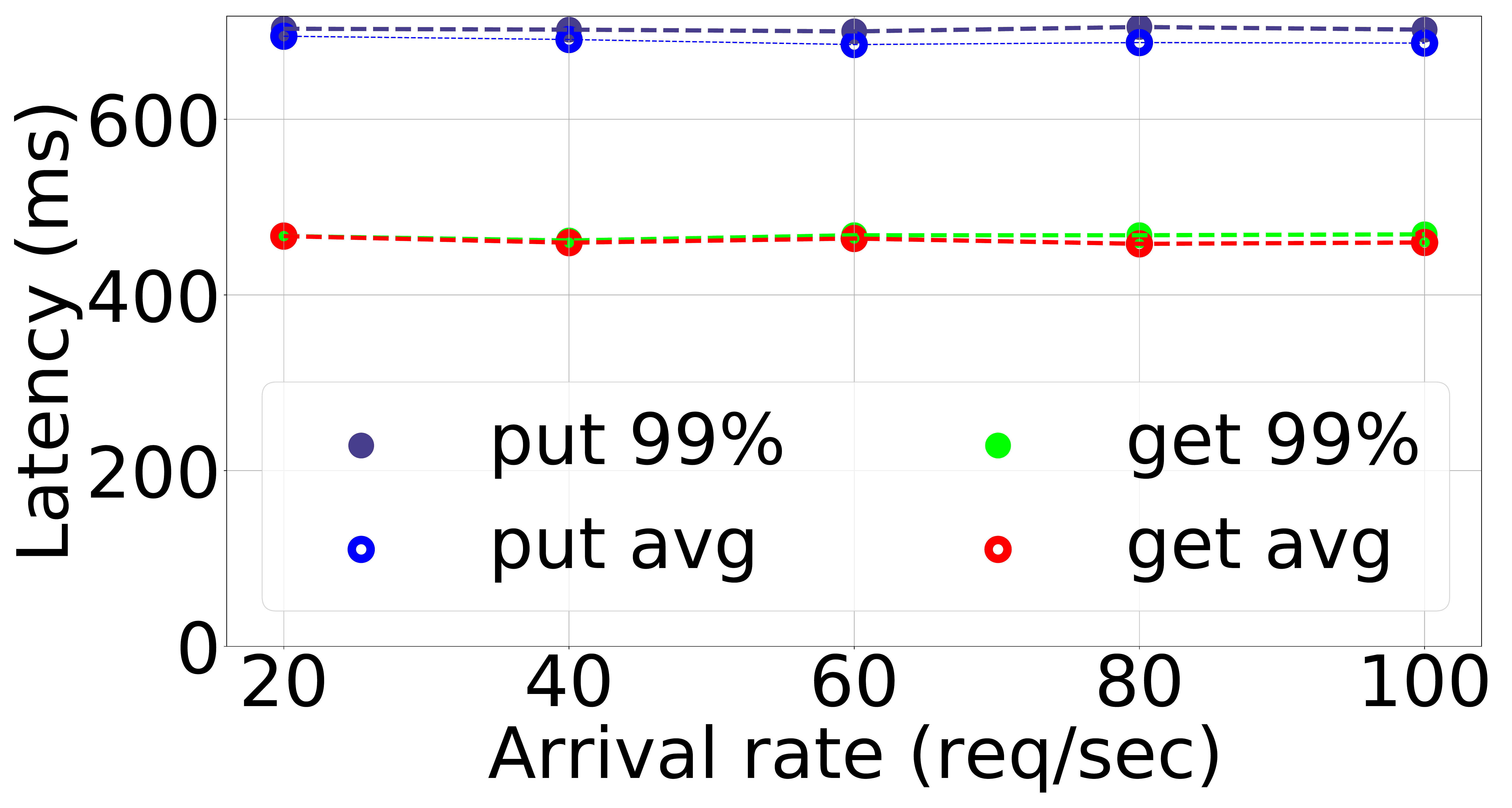}
  \caption{Read ratio=3.2\%.}
  \label{fig:slo-latency2}
\end{subfigure}}
\caption{LEGOStore is able to ensure that latency offered to a key is robust even at for highly concurrent accesses. Here we plot the latency experienced by clients at the Tokyo location for arrival rates in [20-100] req/sec. 
}
\label{fig:concurrency}
\end{figure}

\subsection{Scalable Concurrency Handling}

\label{sec:concurrency}

A distinguishing feature of LEGOStore is that it is designed to provide reliable tail latency even in the face of highly concurrent access to a key. 
For consensus-based protocols that apply operations sequentially to a replicated state machines one after another, even in the optimistic case where all operations are issued at a leader that does not fail, the latency is expected to grow linearly with the amount of concurrency. Furthermore, in distributed consensus, due to FLP impossibility \cite{FLP}, concurrent operations may endure several (in theory, unbounded) rounds of communication. 

To validate our expectation of robust tail latency even under high concurrency, we increase the arrival rate for the \emph{same key} with object size 1 KB. The object is configured to use CAS(5, 3) with DCs in Singapore, Frankfurt, Virginia, California, and Oregon. In particular, requests from uniformly-distributed user locations come to the single key. We run the experiments for both HW and RW for a period of 1 minute for each arrival rate. We plot the latency experienced by clients at the Tokyo location against arrival rate in  Figure~\ref{fig:concurrency}. LEGOStore demonstrates a remarkable robustness of the latency of operations. Even for an arrival rate of 100 req/sec to the same key, every operation completes and we see no degradation in performance for the average and tail latencies. We recorded a maximum concurrency of 142 write operations on one key for an arrival rate of 100 req/sec and 30:1 write ratio. Little's law suggests an average concurrency of around 60 operations for this experiment. Note the contrast with consensus-based protocols, where the tail latency is crucially dependent on limited concurrency for a given key. {E.g., in \cite{pando} Figure~13, even with somewhat limited concurrency, the  latency of only ``successful'' writes can grow up to 30s without leader fallback,  and at least doubles with leader fallback.}

The similarity of the latency in Figure~\ref{fig:concurrency}(b) which has a HW workload as compared  with the RW workload in  Figure~\ref{fig:concurrency}(a) indicates that our latencies remain robust even if the workload is write heavy. It is also worth noting that Figure~\ref{fig:concurrency} is a further corroboration of the robustness of our modeling in Section~\ref{sec:opt}. Specifically, our model ignores intra-DC phenomena such as queuing, and the robustness of latency despite a high arrival rate shows the overwhelming  significance of the inter-DC RTTs in determining response times.

\subsection{Reconfiguration to Handle Load Change}
\label{sec:reconf-load}

In this subsection and the next, we explore \lego's ability to  perform fast reconfiguration in line with the expectations set in Section~\ref{sec:reconfig-proto}.~
We consider a set of 20 keys with similar workloads, each with an object size of 1 KB and $f$=1 to which RW (i.e., read ratio of 50\%) requests  arrive  from 4  locations with the following distribution: Tokyo (30\%), Sydney (30\%), Singapore (30\%), and Frankfurt (10\%). Each user issues one request every 2 seconds on average. Our latency SLOs are 700 msec and 800 msec for GETs and PUTs, respectively. As seen in  Figure~\ref{fig:recon-lat}, till $t$=200 sec, requests arrive at a total rate of 100 req/sec (i.e., 200 users) from the 4 locations. LEGOStore employs configurations with CAS(5,3) for our keys with DCs in Tokyo, Sydney, Singapore, Virginia, and Oregon. The figure plots the latency experienced by users at Sydney and Frankfurt;  users at Singapore and Tokyo experience similar  SLO adherence.  LEGOStore successfully meets SLOs. In fact, a small number of GET requests (shown using a right-facing arrow) see superior performance as they are "optimized" GETs (recall Section~\ref{sec:back}).  At $t$=200 sec, the collective request arrival rate increases 4-fold to 400 req/sec (i.e., 800 users) while all other workload features remain   unchanged. 
We assume that the 
controller located at LA issues a reconfiguration without delay on detecting this workload change. For the new workload, LEGOStore's optimizer recommends a new configuration performing ABD with replication factor of $3$ over DCs in Tokyo, Sydney, and Singapore. Across multiple measurements, we find that reconfiguration concludes in less than 1 sec. The breakdown of overall reconfiguration for a sample instance that takes 717 msec is: (i) reconfig query=68 msec; (ii) reconfig finalize=208 msec; (iii) reconfig write=139 msec; (iv) updating metadata=163; and (v) reconfig finish=139 msec.

We examine user experience during and in the immediate aftermath of reconfiguration. We show the latencies experienced by all the users each at the Sydney and Frankfurt locations to isolate the performance degradation experienced at each user location more clearly.
 A user request experiences one of two types of degradation which mainly depends on when it arrives in relation to the reconfiguration. {\bf Type (i)} A small number of requests (small due to how quick the reconfiguration is) is blocked at the old configuration servers with the possibility of either getting eventually serviced by these old servers or having to restart in the new configuration (see Section~\ref{sec:reconfig-proto}). These are the requests experiencing latencies in the 750 msec - 1 sec range and highlighted using boxes for the GET requests. {\bf Type (ii)} A second possibility applies to all other requests that do not get blocked at the old configuration servers. These requests incur an additional delay of about 200 msec (users in Sydney) and 250 msec (Frankfurt) to acquire the new configuration from LA and are shown using an ovals for the GET requests. This increase in latency happens because the users do not know that a reconfiguration has occurred and try to do an operation with the old configuration, e.g., see requests at $t \sim$ 200 sec experiencing a slight degradation among GET operations from Sydney users.

\subsection{Reconfiguration to Handle DC Failure}
\label{sec:recon-dcfail}

\begin{figure*}[t]
\centering
{
\begin{subfigure}{0.49\columnwidth} \label{fig:recon-lat1}
  \centering
  \includegraphics[width=\columnwidth]{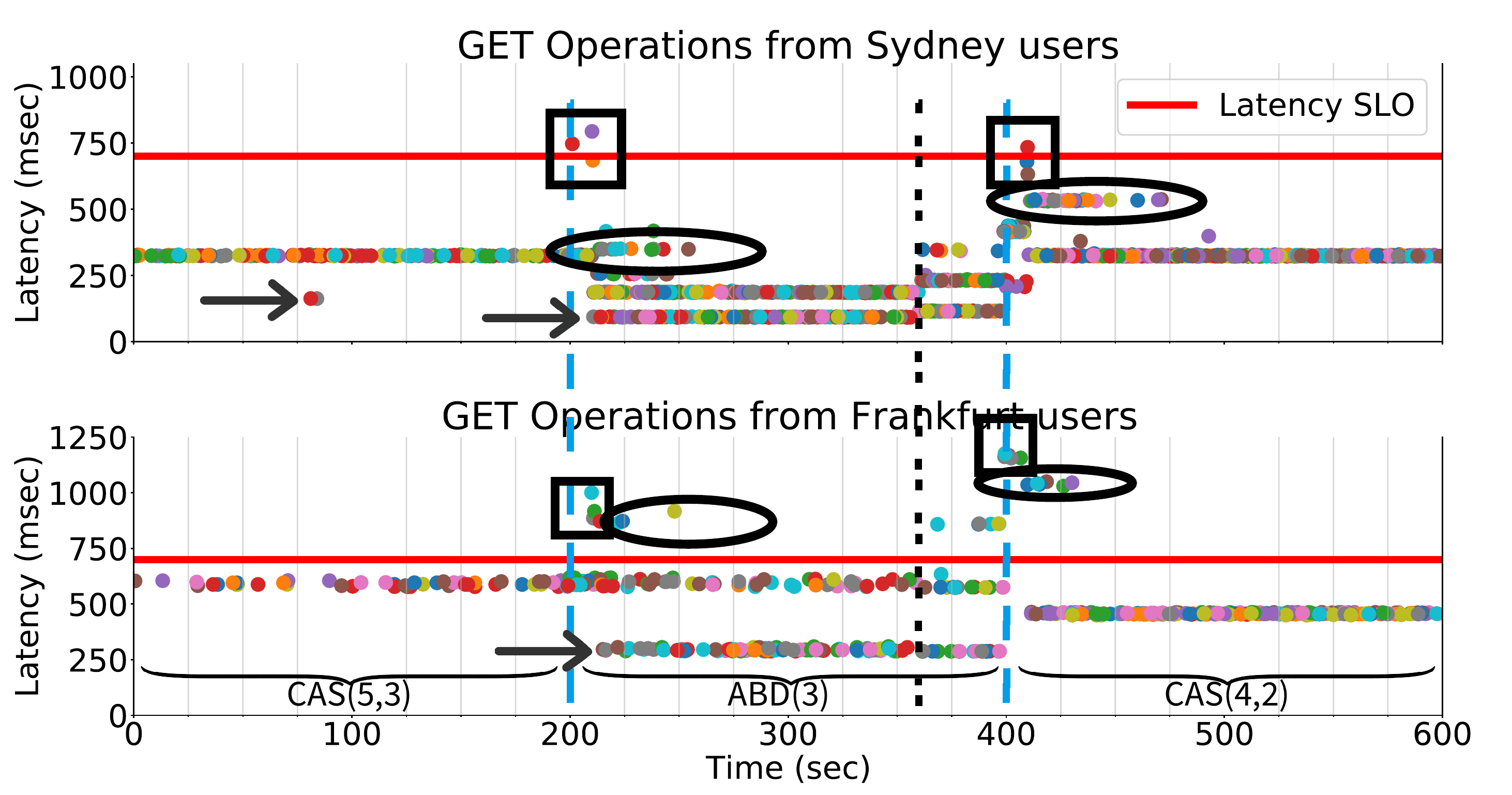}
  \caption{GET operations.}
\end{subfigure}%
\begin{subfigure}{0.49\columnwidth} \label{fig:recon-lat2}
  \centering
  \includegraphics[width=\columnwidth]{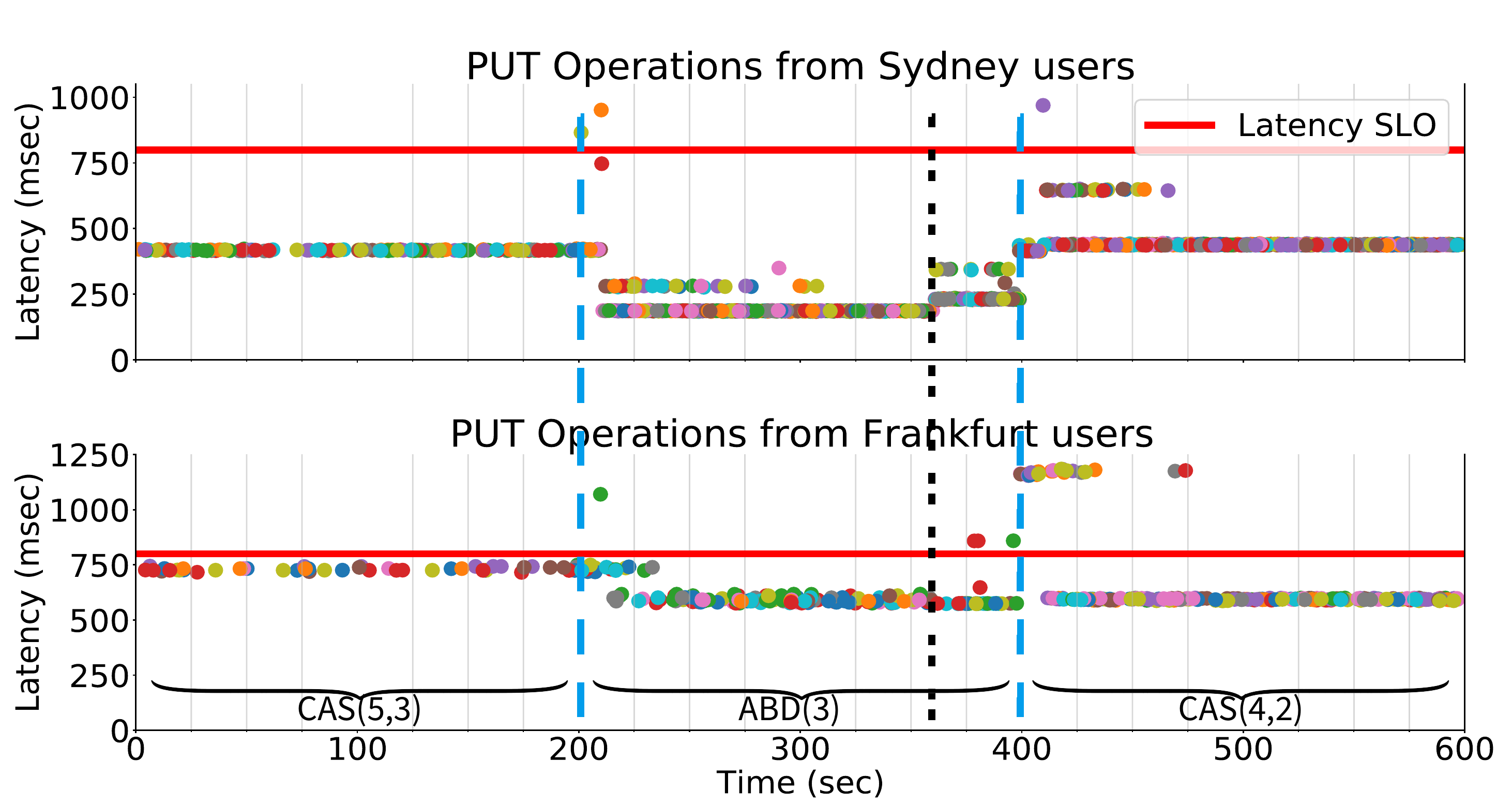}
  \caption{ PUT operations.}
\end{subfigure}}
\caption{The efficacy and performance impact of two reconfigurations is shown for one of 20 keys with similar workloads. The first reconfiguration occurs in response to a 4-fold increase in request arrival rate at $t$=200 sec. The second reconfiguration occurs at $t$=400 sec in response to the Singapore DC failing at $t$=360 sec. The arrows show the optimized GET operations while the squares and ovals respectively highlight two types of performance degradation associated with reconfiguration: (i) requests blocked in the old configuration, and (ii) first request issued by a user after the reconfiguration which needs to acquire the new configuration from the controller at LA. The different colors for the latency dots represent different users. }
\label{fig:recon-lat}
\end{figure*}

When a DC in one of the quorums fails, LEGOStore will send the request to all other DCs participating in the configuration that are not in the quorum. \vc{This will in general be sub-optimal cost-wise and may also fail to meet the SLO. In Appendix~\ref{app:perf-failure}   Figure~\ref{fig:latency-verif} in \cite{legostore-arxiv}, we show a sample result where a DC failure results in such SLO violation.} To alleviate this, upon detecting a failure,\footnote{LEGOStore can work with any existing approach for failure detection. } LEGOStore invokes its optimizer to determine a new cost-effective configuration that discounts the failed DC and then transitions to this new configuration.
Figure~\ref{fig:recon-lat} depicts a scenario where the Singapore DC, a member of both ABD quorums,  fails at $t$=360 sec. We assume that this failure is detected and remediated via a transition to a new configuration using CAS(4,2) at $t$=400 sec.
Again, we find that the transition occurs within a second and has a small adverse impact on request latency--- most requests whose latency exceeds the SLO are of the unavoidable Type (ii). 

\begin{figure}[t]
\centering
\includegraphics[width=0.7\columnwidth]{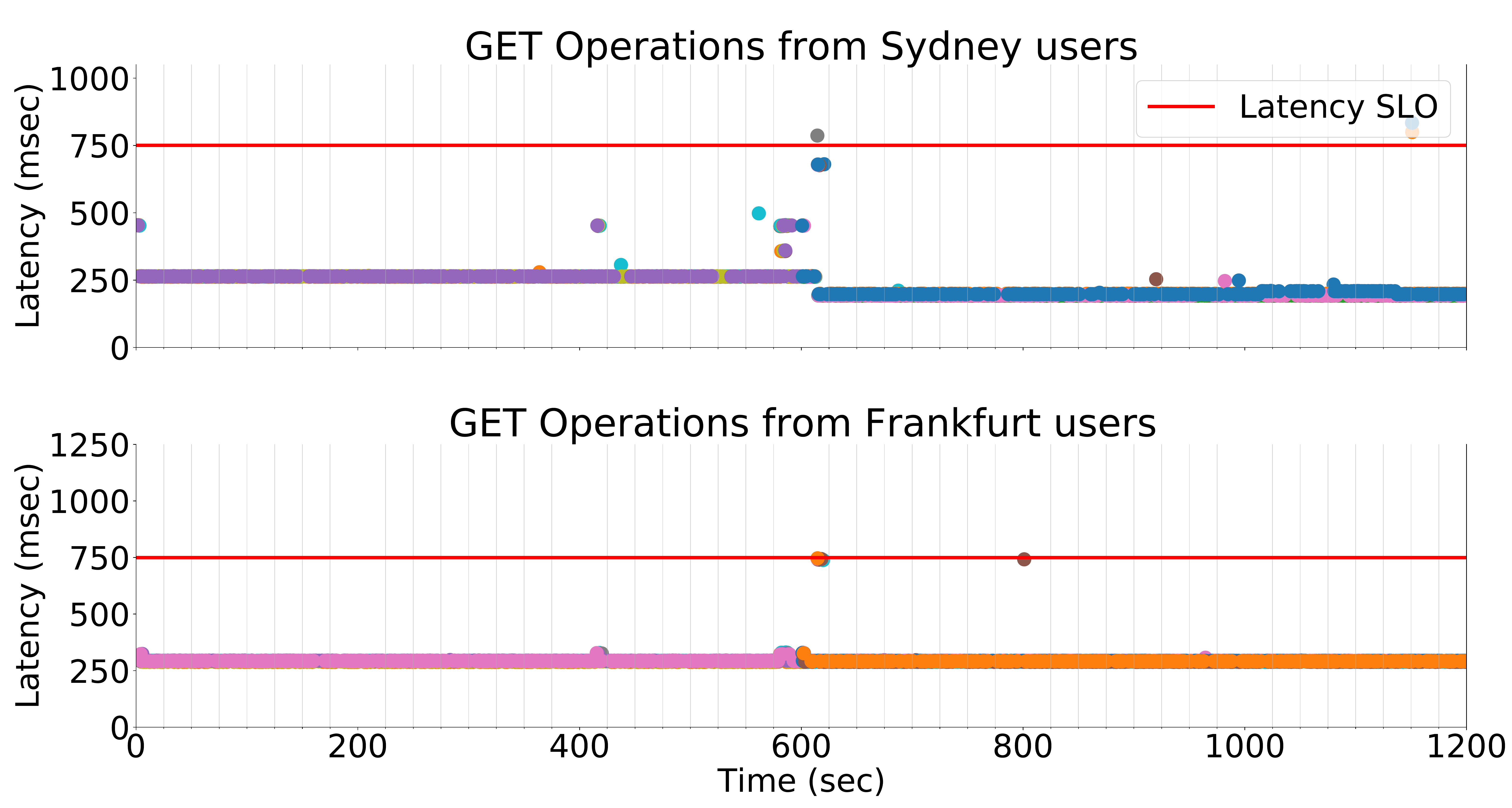}
\caption{The efficacy and performance impact of a reconfiguration (at t=10 min) is shown for a  key that we derive from the Wikipedia dataset.}
\label{fig:realreconf}
\end{figure}

\bu{
\subsection{LEGOStore for a Real-World Workload}
\label{sec:realreconf}

In this subsection, we construct our workload using a publicly available dataset collected from Wikipedia's web server~\cite{wikitrace}. This is a read-mostly workload with a highly skewed popularity distribution.  We extract arrival time and request size information from the dataset and interpret each request as a GET or a PUT based on its type. We sample a set of 1550 distinct objects (each interpreted as a key) from the dataset whose aggregate arrival rate can be accommodated by our prototype. We consider workload features for this set of keys over two 1-hour long periods (call these $T_1$ and $T_2$). Since the workload itself does not reveal a distribution of clients, we assume a uniform distribution of clients among 5 of our DCs (Tokyo, Sydney, Singapore, Frankfurt, London) for $T_1$ and a uniform distribution among all 9 DCs for $T_2$. We use our optimizer to determine cost-effective configurations for each of these keys for both of our 1-hour periods. We choose a latency SLO of 750 msec.

Our findings demonstrate that \lego~offers cost savings over baselines for the Wikipedia workload. We compare the cost offered by the optimizer against our various baselines for all of the 1550 keys in Figure~\ref{fig:wikicomparison} of \cite{legostore-arxiv}. Even for a fixed duration, the results highlight the importance of the optimizer as a variety of different configurations are chosen for different objects - this includes both replication and CAS and different parameters for CAS. Further, with change in client distribution for a given key, \lego's reconfiguration and optimizer couple to ensure sustained cost effectiveness and improvement over baselines. In Figure~\ref{fig:realreconf},  we highlight an illustrative key's performance in our \lego~ prototype over a 20 minute period with the first 10 minutes from $T_1$ and the second 10 minutes from $T_2$. The arrival rate to this key changes from 16 to 35 req/sec.
Our optimizer chooses CAS (m=5, k=1) for $T_1$ and CAS (m=8, k=1) for $T_2$. The latter yields a 20\% cost reduction over the former  and triggers a reconfiguration. Figure~\ref{fig:realreconf} is centered around the reconfiguration LEGOStore carries out for this key at t=10 min. Similar to earlier experiments, our prototype accomplishes the reconfiguration within 1.96 seconds with an increase in response times for a small number of requests during the reconfiguration. 

}

\bu{
We find that our optimizer, itself executed on public cloud VMs, contributes negligibly to the operational cost of LEGOStore. As an illustrative calculation, for the workload in this section we consider extremely frequent reconfigurations occurring once every 5 minutes for the key with the highest arrival rate of 20.16 req/sec. Each invocation of our optimizer costs about \$0.0001 on average (the average optimizer execution time is 18 sec, see Appendix~\ref{app:impl}). This turns out to be a mere 0.48\% the overall costs for this key.
}

%% file: 8related_work.tex
\section{Related Work}
\label{sec:related}




\noindent {\textbf{EC Based Data Storage:}} 
Several papers have studied the design of in-memory KV stores ~\cite{rashmi2016ec, shankar2017high, lu2017scalable, zhao2016toward, xiang2016joint, chen2017efficient, wang2017wps, Abebe2018ECStoreBT, duan2018scalable}. A significant body of work focuses on minimizing repair costs and encoding/decoding \cite{ClayCodes,wang2019xorinc, bai2019fast, xie2019az,li2019openec, mitra2016partial, Taranov_ECKV, xia2017revisiting, yiu2017erasure, li2017parix, li2017bcstore, duan2018scalable}.
The cost savings offered by 
EC have motivated its use particularly in production archival (i.e., write-once/rarely)  systems~\cite{huang2012erasure,FacebookF4}. These papers do not focus on consistency aspects that are relevant to workloads with both reads and writes, nor do they study the geo-distributed setting; therefore, the key factors governing their performance are different from us. 
Strongly consistent EC-based algorithms and KV stores are developed in \cite{GWGR, Hendricks, Dutta, dobre_powerstore, FAB}; however, none of these works study the geo-distributed setting or the public cloud.


\noindent{\bf  Strongly Consistent Geo-Distributed Storage:} 
There are several strongly consistent geo-distributed KV stores \cite{Spanner,spanstore,giza,pando,gupta2018fogstore,zhang2013transaction}.
SpanStore~\cite{spanstore} develops an optimization  to minimize costs while satisfying latencies for a strongly consistent geo-distributed store on the public cloud.  While there are several technical differences (e.g., SpanStore uses a blocking protocol via locks), the most important advance made by  \lego~ is its  integration of EC into the picture. Besides tuning EC parameters, \lego~ integrates the constraints of structurally more complex EC-based protocols to enable cost savings. Most closely related to our work are  Giza ~\cite{giza} and Pando~\cite{pando}, which are both strongly consistent EC-based geo-distributed data stores. Both data stores modify consensus protocols (Paxos and Fast Paxos) to utilize EC and minimize latency. The most notable difference between these works and \lego~ is that \lego~ is designed to keep  \emph{tail} latency predictable and robust and keep costs low in the face of dynamism.  Since Giza and Pando are based on consensus, they will tend to have  higher latency under  concurrent writes, e.g., for hot objects with high arrival rates. Furthermore, neither Pando nor Giza have an explicit reconfiguration algorithm. 
On the other hand, since Giza and Pando use consensus, they offer more complex primitives such as Read-Modify-Writes and versioned objects. A noteworthy comparison point vs. Giza is that it  does not operate in the public cloud and does not contain an optimization framework for cost minimization. 


\noindent{\bf Reconfiguration:} There is a growing body of work that develops \emph{non-blocking} algorithms for reconfiguration~\cite{nicolaou2019ares,DYNASTORE}. Algorithms in~\cite{nicolaou2019ares,DYNASTORE} require an additional phase of a client to contact a controller/configuration service in the critical path of \emph{every} operation. In LEGOStore, for the common case of operations that are not concurrent with a reconfiguration, the number of phases (and therefore the latency, costs) are identical to the baseline static protocol. \vc{Our algorithm has a resemblance to an adaptation of \cite{RAMBO} in the tutorial \cite{aguilera2010reconfiguring}. That algorithm works mainly for replication and requires clients to propagate values to the new configuration rather than the controller, which can incur larger costs. Our reconfiguration algorithm utilizes concepts/structures that appear in previous algorithms; our main contribution is to adapt the existing algorithms specifically to ABD and CAS in order to keep the reconfiguration latency/costs low and predictable.  In particular, our algorithm piggybacks read/write requests for the reconfiguration along with messages that are sent to block ongoing operations, and makes careful choices on operations that can be completed in older configurations to provably ensure linearizability.} Several works \cite{Sharov:2015:TMY:2824032.2824047,volley,ardekani2014self} design heuristics to determine when and which objects to reconfigure. Sharov et. al~\cite{Sharov:2015:TMY:2824032.2824047} give a method for optimizing the configuration of quorum-based replication schemes, including the placement of the leaders and replica locations for read and write operations as well as transactions. \vc{The paper shares conceptual similarities with \lego's optimization, but was limited to replication-based schemes and focused solely on minimizing latency for placement, whereas we focus  on erasure-coded schemes and include costs in our placement decisions; on the other hand, our methods do not readily apply to systems that support transactions. A similar comparison applies to the replication-oriented optimizers described in~\cite{zakhary18,abede2020, glasbergen18}. }  Volley~\cite{volley} describes techniques for dynamically migrating data among Microsoft's geo-distributed data centers to keep content closer to users and keeping server loads well-balanced. 
 
\noindent {\bf Data Placement and Optimization for Public Cloud:} There is a rich area of data placement and tuning of consistency parameters for replication based geo-distributed stores \cite{terry2013consistency, ardekani2014self, spanstore, replication-placement, shankaranarayanan2014performance, Abebe2018ECStoreBT, su2016systematic, ec-geo-placement}. These works expose the role of diverse workloads and costs in system design,  and our optimization framework is  inspired by this body of work. However, most of these references \cite{terry2013consistency, ardekani2014self,spanstore, replication-placement, shankaranarayanan2014performance} only consider replication.  Reference \cite{Abebe2018ECStoreBT} studied placement and parameter optimization for EC \emph{within} a DC; while some insights are qualitatively similar, our geo-distributed setting along with its diversity makes the salient factors that govern performance different. \vc{From an optimization viewpoint, closest are \cite{ec-geo-placement,su2016systematic} which study EC over geo-distributed public clouds; however, they do not consider consistency and related quorums constraints and costs.}



%% file: 9conclusion.tex
\section{Conclusion}
\label{sec:future}

We developed LEGOStore, a linearizable geo-distributed key-value store which procured resources from  a public cloud provider. LEGOStore's goal was to offer tail latency SLOs that were predictable and robust in the face of dynamism. 
We focused on salient aspects of EC's benefits for LEGOStore. 
Several additional aspects of key-value store design  
constitute interesting future directions. 
For instance, LEGOStore's effectiveness depends on a module that detects workload change and then reconfigures based on the detected changes. 
Additionally,
we focused on read/write operations and have not implemented read/modify/write (RMW) operations~\cite{hermes}, which inevitably suffer from less robust tail-latency due to FLP impossibility. The recent paper Gryff \cite{burke2020gryff} designs a provably strongly consistent data store with RMW operations and yet provides the favorable tail-latency properties of  ABD for read and write operations. Gryff is based on replication, and the development of a similar system that uses EC is an interesting area of future research.

\section*{Acknowledgements}

This work was supported in part by a Google Faculty Award, and the NSF under grants CCF-
1553248 and CNS-1717571. We thank  Raj Pandey for his help in Section~\ref{sec:realreconf}. 

%% file: appendix/appendix.tex
\appendix

\input{appendix/abd_algorithm}
\input{appendix/cas_algorithm}

\input{appendix/constraints_for_opt}
\input{appendix/reconfiguration_lin_proof}

\input{appendix/k-analysis.tex}

\input{appendix/additionalresults.tex}

%% file: appendix/abd_algorithm.tex
\section{The ABD algorithm}
\label{app:ABD}

\begin{figure}
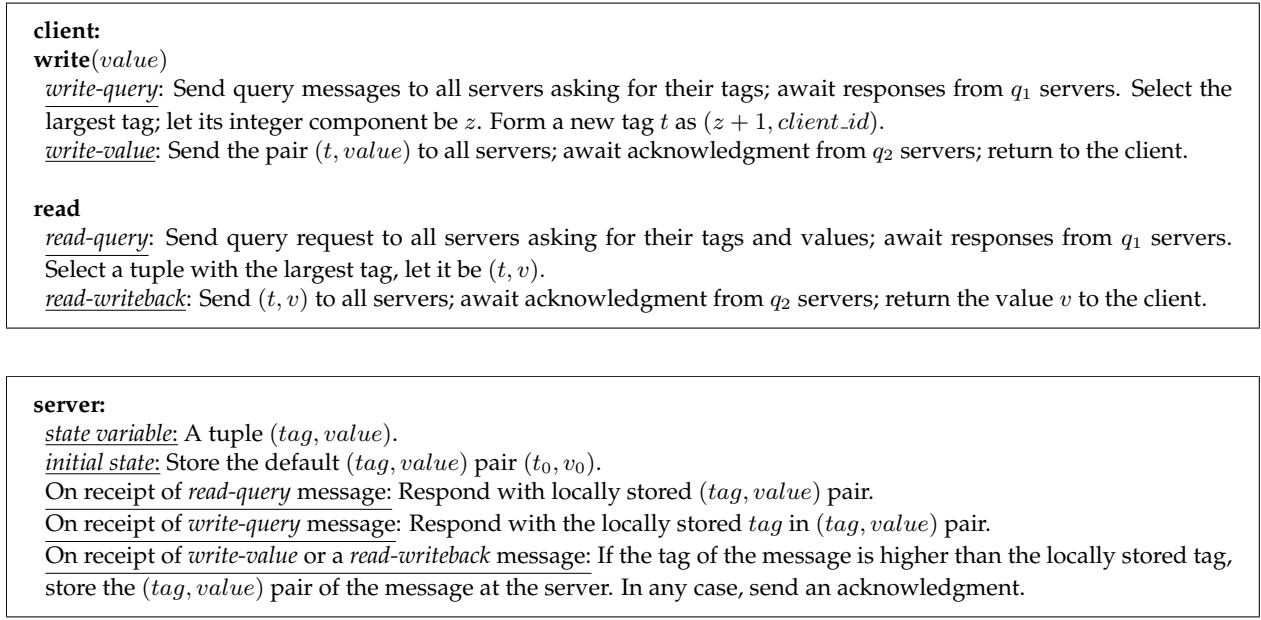

\begin{mdframed}
\footnotesize
\begin{tabbing}
    {\bf client: } \\
    {\bf write$(value)$} \\
    \ \ \begin{minipage}[t]{\textwidth}%
        {\underline{\emph{write-query}}: Send query messages to all servers asking for their tags; await responses from  $q_{1}$ servers. Select the largest tag; let its integer component be $z$. Form a new tag $t$ as $(z+1, client\_id).$}
        
        {\underline{\emph{write-value}}: Send the pair $(t, value)$ to all servers; await acknowledgment from $q_{2}$ servers; return to the client.}
    \end{minipage}
\end{tabbing}

\footnotesize
\begin{tabbing}
    {\bf read} \\
    \ \ \begin{minipage}[t]{\textwidth}%
        {\underline{\emph{read-query}}: Send query request to all servers asking for their tags and values; await responses from $q_{1}$ servers. Select a tuple with the largest tag, let it be $(t, v).$}

        {\underline{\emph{read-writeback}}: Send $(t, v)$ to all servers; await acknowledgment from $q_{2}$ servers; return the value $v$ to the client.}
    \end{minipage}
\end{tabbing}
\end{mdframed}

\begin{mdframed}
\footnotesize
\begin{tabbing}
    {\bf server:} \\
    \ \
   \begin{minipage}[t]{\textwidth}%
        {\underline{\emph{state variable}:} A tuple $(tag, value)$.}
    
        {\underline{\emph{initial state}:} Store the default $(tag, value)$ pair $(t_{0}, v_{0})$.}
    
        {\underline{On receipt of \emph{read-query}  message:} Respond with locally stored $(tag, value)$ pair.}
    
        {\underline{On receipt of \emph{write-query} message}: Respond with the locally stored $tag$ in $(tag, value)$ pair.}
    
        {\underline{On receipt of \emph{write-value} or a \emph{read-writeback} message:} If the tag of the message is higher than the locally stored tag, store the $(tag, value)$ pair of the message at the server. In any case, send an acknowledgment.}
    \end{minipage}
\end{tabbing}
\end{mdframed}
\caption{Client and server protocols for ABD. Note that the key is implicit.} \label{fig:ABD}
\end{figure}

We describe the ABD algorithm in Figure~\ref{fig:ABD}.  The algorithm here actually refers to a multi-writer variant presented in ~\cite{Lynch1996}. In the algorithm description, we assume that every client has a unique identifier, and every tag is of the form $(integer,$ $client\_id).$
The $client\_id$ field is used to break ties. ABD employs 2 quorums of servers $Q_1$ and $Q_2$. It uses $Q_1$ to gather timestamps and data for GET operations and only timestamps for PUT operations. $Q_2$ is used to write the data into. We use the notation: $q_{1} = |Q_{1}|$ and $q_{2} = |Q_{2}|$. The algorithm implements a linearizable object when the quorums satisfy $q_{1}+q_{2} \geq N,$ and $q_{1},q_{2} \leq N-f$ where $N$ is the number of DCs and $f$ is the number of failures can be tolerated. 
To save costs, we only send requests to the servers in the set $Q_{1}$ in phase 1 (or $Q_{2}$ in phase 2) and only approach additional servers if one or more servers within this set does not respond within a timeout. 

We also consider an optimized version of ABD, called $ABD$-$Opt$ wherein the \emph{read-writeback} of read operations can sometimes be omitted. The bigger the $read\_ratio$ is, the more useful this modification is. To enable the $ABD$-$Opt$ algorithm, the \emph{read-query} phase of read operations in Figure~~\ref{fig:ABD} should be replaced with: \\
{\underline{\emph{read-query-opt}}: Send query request to all servers asking for their tags and values; await responses from $max(q_{1}, q_{2})$ servers. Let $(t, v)$ be the pair with the maximum tag. If at least $q_{2}$ number of responses have the tag $t$, return the value $v$ to the client. Otherwise, go to the \emph{read-writeback} phase.}

%% file: appendix/cas_algorithm.tex
\section{The CAS Algorithm}
\label{app:cas}

We present the CAS protocol in  Figures~\ref{fig:CASwrite} and~\ref{fig:CASserver}. The \emph{tag} refers to the logical timestamp of the form (integer,client-id), where each client has a unique client-id. The set of tags is denoted by $\mathcal{T},$ the set of codeword symbols is denoted by $\mathcal{W}$, and  the set of nodes is denoted by $\mathcal{N}$. CAS employs 4 quorums: $Q_1$ to query for the highest timestamp; $Q_2$ to send the coded elements to; $Q_3$ to send the finalized tag to; and, finally, $Q_4$ to request  the coded elements. We use the notation $q_{i} = |Q_{i}|, \forall i$, and the protocol satisfies safety and liveness despite $f$ failures if the following constraints hold:
\label{app:safety_constraints}
\begin{gather}
\label{eq:20}
    q_1 + q_3 > N \\
    q_1 + q_4 > N\\
    q_2 + q_4 \geq N+K\\
    q_4 \geq K\\
    \forall i~~~q_i \leq N - f \label{cas-f-constraint}
\end{gather}

\noindent {\bf Relation between $N$, $K$, and $f$:} By substituting $q_2$ and $q_4$ with $N-f$ from~(\ref{cas-f-constraint}), we have:

\begin{gather}
\label{eq:n-k-f}
    2N+2f \geq N+k \Rightarrow N-k \geq 2f
\end{gather}

Like ABD, we present an optimized version of CAS, called CAS-Opt where the \emph{read-finalize} phase of read operations can sometimes be omitted. To enable such optimization, we replace \emph{read-query} phase with the following:\\
{\underline{\emph{read-query-opt}}: Send query request to all servers asking for their tags and values; await responses from $max(q_{1}, q_{2})$ servers. Let $(t, v)$ be the pair with the maximum tag. If at least $q_{4}$ number of responses have the tag $t$, return the value $v$ to the client. Otherwise, go to the \emph{read-finalize} phase.}

Unlike several other erasure coding based protocols~\cite{Dutta, Hendricks}, CAS only sends one codeword symbol per operation, therefore, the number of bits sent across DCs will be significantly less than replication (and other erasure coding based algorithms). A `fin' label indicates that the value has been transferred to a sufficient number of nodes, and can be exposed to reads. Propagation of the `fin' label is indeed the reason for the additional phase of write operations. 

\begin{figure}[!t]
  \begin{mdframed}
  \footnotesize
  \begin{tabbing}
    {\bf client: } \\
  	{\bf write$(value)$} \\
	\ \ \begin{minipage}[t]{\textwidth}%
	{\em \ul{query}:}  Send query messages to all nodes asking for the 
	highest tag with label `fin'; {await responses from  $q_1$ servers.} 

	{\em \ul{pre-write}:} {Select the largest tag from the \emph{query} phase; let its integer component be $z$. Form a new tag $t$ as $(z+1,\text{`}\mathrm{id}\text{'})$, where `$\mathrm{id}$' is the identifier of the client performing the operation.} Apply the $(n,k)$ MDS code to the value to obtain coded elements ${w_1, w_2, \ldots, w_n}$. Send $(t, w_s, \text{`}\mathrm{pre}\text{'})$ to node $s$ for every $ \in \mathcal{N}.$ Await response{s} from $q_2$ servers. 

{\em \ul{finalize}:} Send a \emph{finalize} message {$(t,\text{`}\mathrm{null}\text{'},\text{`}\mathrm{fin}\text{'})$ to all
	nodes.}  Terminate after receiving responses from $q_3$ servers.\smallskip

	\end{minipage} 
\end{tabbing}

\begin{tabbing}
		{\bf read} \\
	\ \ \begin{minipage}[t]{\textwidth}%
	{\em \ul{query}:} As in the write protocol. 

	{\em \ul{finalize}:} Send a \emph{finalize} message with tag $t$ to all the nodes requesting 
	the associated coded elements. Await responses from $q_4$ servers. {If at least $k$ nodes include their locally stored coded elements in their  
	responses, then obtain the $value$ from these coded elements by decoding and terminate by returning $value$.} \smallskip
	\end{minipage}  
\end{tabbing}
\end{mdframed}
\caption{Client protocol for CAS. The key is implicit.}\label{fig:CASwrite}
\end{figure}

\begin{figure}
    \begin{mdframed}
    
      \footnotesize

\begin{tabbing}
	{\bf server:} \\
	\ \ \begin{minipage}[t]{\textwidth}%
\emph{\ul{state variable}:} A variable that is a subset of $\mathcal{T} \times \left(\mathcal{W} \cup \{\text{`}\mathrm{null}\text{'}\}\right) \times \{\text{`}\mathrm{pre}\text{'}, \text{`}\mathrm{fin}\text{'}\}.$ 

		{{\em \ul{initial state}:} Store {$(t_0, w_{0,s}, \text{`}\mathrm{fin}\text{'})$ where $s$ denotes the server and $w_{0,s}$ is the coded element corresponding to server node $s$ obtained by encoding the initial value $v_0$.}} 

		\ul{On receipt of \emph{query} message}: Respond with the highest locally known tag that has a label $\text{`}\mathrm{fin}\text{'}$, i.e., the highest tag $t$ such that the triple $(t, *, \text{`}\mathrm{fin}\text{'})$ is at the node, where $*$ can be a coded element or `$\mathrm{null}$'. 

	\ul{On receipt of \emph{pre-write} message}: If there is no record of the tag of the message in the list of triples stored at the node, then {add }the incoming triple {in the message }to the list of stored triples; otherwise ignore. {Send acknowledgment.} 

	\ul{On receipt of \emph{finalize} from a write}: Let $t$ be the tag of the message. If a triple of the form $(t, w_s, \text{`}\mathrm{pre}\text{'})$ exists in the list of stored triples, then update it to $(t, w_s, \text{`}\mathrm{fin}\text{'})$. Otherwise add $(t, \text{`}\mathrm{null}\text{'}, \text{`}\mathrm{fin}\text{'})$ to list of stored triples\footnotemark[16]. Send {acknowledgment}. 

	\ul{On receipt of \emph{finalize} from a read}: Let $t$ be the tag of the message. If a triple of the form $(t, w_s, *)$ exists in the list of stored triples where $*$ can be $\text{`}\mathrm{pre}\text{'}$ or $\text{`}\mathrm{fin}\text{'}$, then update it to $(t, w_s, \text{`}\mathrm{fin}\text{'})$ and send $(t, w_s)$ to the client. Otherwise add $(t, \text{`}\mathrm{null}\text{'}, \text{`}\mathrm{fin}\text{'})$ to the list of triples at the node and send an {acknowledgment.}   

     \smallskip 
	\end{minipage}
	\end{tabbing}

  \end{mdframed}
  \caption{Server protocol for CAS. The key is implicit.}
  \label{fig:CASserver}
\end{figure}

%% file: appendix/constraints_for_opt.tex
\section{LEGOStore's Optimizer}
\label{app:optimization}


\noindent {\bf A Detailed Look at our Objective:}
The networking cost per unit time of PUTs for key $g$ must be represented differently based on whether ABD or CAS is used for $g$. We use the boolean variable $e_g$ to capture this (0 for ABD and 1 for CAS). 

\begin{equation}
\label{eq:put-cost}
    C_{g,put} = \underbrace{e_g \cdot C_{g,put,CAS}}_{\textrm{n/w cost if CAS chosen}} + \underbrace{(1-e_g) \cdot C_{g,put,ABD}}_{\textrm{n/w cost if ABD chosen}}.
\end{equation}

The terms $C_{g,put,ABD}$ and $C_{g,put,CAS}$ are designed to capture the intensity of network traffic exchanged between clients and the various quorums of servers they need to interact with per the concerned protocol's idiosyncrasies. Note the role played by the key boolean decision variable $iq^k_{ijg}$ whose interpretation is: $iq^k_{ijg}$=1 iff data center $j$ is in the $k^{th}$ quorum for clients in/near data center $i$.  We have: 

\begin{equation}
\begin{split}
\label{eq:put-abd-cost}
C_{g,put,ABD} = (1-\rho_g) \cdot \lambda_g \sum\limits_{i=1}^D \alpha_{ig}\Big(\underbrace{o_m\sum\limits_{j=1}^D {p^n_{ji} \cdot iq^1_{ijg}}}_{\textrm{n/w cost for phase 1}} + \\
        \underbrace{o_g\sum\limits_{k=1}^D {p^n_{ik} \cdot iq^2_{ikg}}}_{\textrm{n/w cost for phase 2}}\Big).
\end{split}
\end{equation}

Here, $(1-\rho_g) \cdot \lambda_g \cdot \alpha_{ig}$ captures the PUT request rate arising at/near data center $i$ and the $o_m$ and $o_g$ multipliers convert this into bytes per unit time. The terms within the braces model the per-byte network transfer prices and are worthy of some elaboration. The first term represents network transfer prices that apply to the first phase of the ABD PUT protocol whereas the second term does the same for ABD PUT's second phase. The  term $p^n_{ji} \cdot iq^1_{ijg}$ should be understood as follows: since ABD's first phase involves clients in/near data center $i$ sending relatively small-sized {\em write-query} messages to all servers in their "read quorum" (i.e., quorum 1, hence the 1 in the superscript of $iq$) followed by these servers responding with their (tag, value) pairs, the subscript in $p^n_{ji}$ is selected to denote the price of data transfer from $j$ (for the server at data center $j$) to $i$ (for clients located in/near data center $i$); the multiplication with $iq^1_{ijg}$ achieves the effect of considering these prices only for data centers that are in the read quorum for the clients being considered.  The networking cost for CAS PUT is as follows:

\begin{equation}
\begin{split}
\label{eq:put-cas-cost}
 C_{g,put,CAS} = (1-\rho_g) \cdot \lambda_g  \sum\limits_{i=1}^D
 \alpha_{ig}\Big\{o_m\Big(\underbrace{\sum\limits_{j=1}^D{p^n_{ji}} \cdot iq^1_{ijg}}_{\textrm{phase 1}} + \\ 
  \underbrace{\sum\limits_{k=1}^D{p^n_{ik} \cdot iq^3_{ikg}}}_{phase 3}\Big) +  \underbrace{\frac{o_g}{k_g}\sum\limits_{m=1}^D{p^n_{im} \cdot iq^2_{img}}}_{phase 2}\Big\}.
\end{split}
\end{equation}





It is instructive to compare the network cost per unit time for ABD in (\ref{eq:put-abd-cost}) with that for CAS in (\ref{eq:put-cas-cost}). First, notice how the number of terms within braces correspond to the number of phases involved in the PUT operation for each protocol (2 for ABD and 3 for CAS). Second, notice the nature of network cost savings offered by CAS: terms in (\ref{eq:put-cas-cost}) involve either  meta-data transfer ($o_m$ in the first 2 terms)  or only a fraction of the data size ($\frac{o_g}{k_g}$  in the 3rd term) as opposed to 
(\ref{eq:put-abd-cost}) both of whose terms involve the entire data size  $o_g$. 

Modeling the cost per unit time incurred towards storage for key $g$ is relatively straightforward (recall from Table~\ref{tbl:inputsandvars} that we use $m_g$ for length of our code which, for ABD, is the degree of replication): 
\begin{equation}
\label{eq:storage-cost}
C_{g,store} = p^s \cdot \big(e_g \cdot m_g \cdot \frac{o_g}{k_g} + (1 - e_g) \cdot m_g \cdot o_g \big).
\end{equation}  

Finally, we consider the  VM costs per unit time for key $g$ (to meet the computational need of our protocol processing). We make the following simplifying yet reasonable assumptions. First, VM capacity may be procured at a relatively fine granularity. We consider this reasonable given that cloud providers offer small-sized VMs with fractional CPU capacities and a few 100 MBs of DRAM (e.g., f1-micro in GCP~\cite{small-vm}). Consequently, our decision variable $v_i$ (number of VMs at data center $i$) is a non-negative real number. Second, across diverse VM sizes (within a VM "class"), VM price tends to be proportional to CPU capacity and DRAM sizes~\cite{small-vm}.\footnote{How price relates to CPU/DRAM capacity may vary across classes, e.g., CPU- vs memory- vs. IO-optimized VMs or on-demand vs. reserved vs. spot vs. burstable VMs. We keep \legostore's VM procurement simple in the sense of confining it to a single such VM class. } Third, we assume that the extensive existing work on VM autoscaling~\cite{burscale, guo-autoscale} can be leveraged to ensure satisfactory provisioning of VM capacity at each data center (i.e., neither excessive and wasteful  over-provisioning nor under-provisioning that may degrade performance). Furthermore, we assume that this suitable VM capacity chosen by such an autoscaling policy is proportional to the total request arrival rate at data center $i$ for key $g$. With these assumptions, the VM cost for key $g$ at data center $i$ is: 

\begin{equation}
\begin{split}
\label{eq:vm-cost}
 C_{g,VM} = \theta^v \cdot \sum\limits_{j=1}^D p_j^v \cdot \lambda_g \sum\limits_{i=1}^D \alpha_{ig} \sum\limits_{k=1}^4 iq^k_{ijg}
\end{split}
\end{equation}

where $\theta^v$ is an empirically determined multiplier that estimates VM capacity needed to serve the computational needs of the request rate arriving at data center $j$ for $g$.  



\noindent {\bf Constraints:} Our optimization needs to capture the 3 types of constraints related to: (i) ensuring linearizability; (ii) meeting availability guarantees corresponding to the parameter $f$; and (iii) meeting latency SLOs. 
Recall that our SLOs are based on tail latency. The key modeling choices we make are: (i) to use worst-case latency as a "proxy" for tail latency; and (ii) ignore latency contributors within a data center other than data transfer time (e.g., various sources of queuing in the storage or compute layers, encoding and decoding time). 
We make a few observations related to these modeling choices.  For our particular problem setting, worst-case latency turns out to be an effective proxy for tail latency because inter-DC latencies---by far the largest components of GET/PUT latencies---tend to be fairly stable over long timescales. Latency contributors within a data center tend to exhibit higher variation but tend to be much smaller than inter-DC latencies. Our choice to ignore queuing effects aligns with our earlier assumption of a well-designed autoscaling policy. Coincidentally,  constraints based on worst-case latencies enable our decomposition of datastore-wide optimization into per-key optimization formulations (recall that the other enabler of such decomposition was the composability property of linearizability). As a counterpoint, SLOs based on cross-key average latencies would require constraints that combine latencies of various keys. 

With these assumptions, our latency constraints take the following form. 
For GET operations, we have:

\begin{equation}
\label{eq:lat-cas-get}
\begin{split}
\forall {i, j, k \in \mathcal{D},}&\\
&\underbrace{iq^1_{ijg} \cdot \Big(l_{ij} +  l_{ji}+\frac{o_m}{B_{ji}}\Big)}_{\textrm{Latency of first phase of GET}} + \\ &\underbrace{iq^4_{ikg}\cdot \Big(l_{ik}+\frac{o_m}{B_{ik}}+l_{ki}+\frac{o_g/k_g}{B_{ki}}\Big)}_{\textrm{Latency of second phase of GET}} \leq l_{get}.     
\end{split}
\end{equation}


\begin{equation}
\label{eq:lat-cas-put}
\begin{split}
\forall {i, j, k \in \mathcal{D},}&\\
&\underbrace{iq^1_{ijg} \cdot \Big(l_{ij} + l_{ji} + \frac{o_m}{B_{ji}}\Big)}_{\textrm{Latency of first phase of PUT}} + \underbrace{iq^2_{img}\cdot \Big(l_{im} + {\frac{o_g/k_g}{B_{im}} + l_{mi}\Big)}}_{\textrm{Latency of second phase of PUT}} \\
& + \underbrace{iq^3_{ikg}\cdot \Big(l_{ik} + \frac{o_m}{B_{ik}} + l_{ki}\Big)}_{\textrm{Latency of third phase of PUT}} \leq l_{put}.
\end{split}
\end{equation}

    
While the LHS expressions of (\ref{eq:lat-cas-get}) and (\ref{eq:lat-cas-put}) capture worst-case latencies, they do not employ an explicit {\tt max} operator---forcing the latencies of {\em all} servers in the relevant quorums accomplishes this automatically. It is useful to compare the LHS expressions of (\ref{eq:lat-cas-get}) and (\ref{eq:lat-cas-put}) alongside the CAS protocol in Appendix~\ref{app:cas}. GET operations require 2 phases with the first involving quorum $Q_1$ and the second involving quorum $Q_4$; PUT operations require 3 phases involving the quorums 
$Q_1, Q_2,$ and $Q_3$, respectively.

The two following equations show the same constraint for GET and PUT in ABD respectively.
\begin{equation}
\label{eq:lat-abd-get}
\begin{split}
\forall {i, j, k \in \mathcal{D},}&\\ &\underbrace{iq^1_{ijg} \cdot \Big(l_{ij} +  l_{ji}+\frac{o_m}{B_{ji}}+\frac{o_g}{B_{ji}}\Big)}_{\textrm{Latency of first phase of GET}} +\\ &\underbrace{iq^2_{ikg}\cdot \Big(l_{ik}+l_{ki}+\frac{o_m}{B_{ik}}+\frac{o_g}{B_{ik}}\Big)}_{\textrm{Latency of second phase of GET}} \leq l_{get}.     
\end{split}
\end{equation}


\begin{equation}
\label{eq:lat-abd-put}
\begin{split}
\forall {i, j, k, m \in \mathcal{D},}&\\
&\underbrace{iq^1_{ijg} \cdot \Big(l_{ij} + l_{ji} + \frac{o_m}{B_{ji}}\Big)}_{\textrm{Latency of first phase of PUT}} +\\ &\underbrace{iq^2_{ikg}\cdot \Big(l_{ik} + l_{ki} + {\frac{o_g}{B_{ik}}\Big)}}_{\textrm{Latency of second phase of PUT}}
\leq l_{put}.
\end{split}
\end{equation}

Please compare the LHS expressions of (\ref{eq:lat-abd-get}) and (\ref{eq:lat-abd-put}) alongside the ABD protocol in Appendix~\ref{app:ABD}, and recall GET and PUT operations need 2 phases involving both $Q_1$ and $Q_2$ in each phase.

\vc{
Finally, we have the following constraints for the quorum sizes.
\begin{eqnarray}
\sum_{j}iq^{\ell}_{ijg} &=& q_{\ell,g}, \forall i \in \mathcal{D},\forall g\\
 q_{i,g} &\leq& N-f,\forall i=1,2,3,4\\
    1-e_{g} + e_{g}\left(q_{1,g} + q_{3,g}-N\right) &>& 0 \label{eq:cas1}\\
    1-e_{g} + e_{g}\left(q_{1,g} + q_{4,g}-N\right) &>& 0\label{eq:cas2}\\
  e_{g}\left(q_{2,g} + q_{4,g} -N-K\right) &\geq& 0\label{eq:cas3}\\
    e_g(q_{4,g}-K) &\geq &0\label{eq:cas4}\\
    e_g+(1-e_{g})(q_{1,g}+q_{2g}-N) &>& 0
\end{eqnarray}
The final constraint above pertains to ABD, whereas (\ref{eq:cas1})-(\ref{eq:cas4}) come from CAS.}


\noindent {\bf The Case of "Optimized" GETs:}
\label{subsec:optimized-get}
While our modeling assumes that GET operations have two phases, the ABD protocol can complete some GET operations in one phase; we refer to such a scenario as an ``Optimized GET.''  A GET operation becomes an  Optimized GET  if all the servers return the same (tag,value) pair in the first phase. This scenario occurs if there is no concurrent operation, and some previous operation to the same key has propagated the version to the servers in the quorum that responds to the client. This optimization mirrors consensus protocols such as Paxos, where operations can complete in one phase in optimistic circumstances. Our paper also involves an optimization for CAS that enables some GET operations to complete in one phase. Since our optimization focuses on worst-case latencies, our modeling makes a conservative estimate that no operation performs an Optimized GET. A minor extension of our work can allow nodes to gossip the versions in the background (after the operation is complete) to increase the fraction of Optimized GETs for ABD protocol at an increased communication cost (See, e.g., \cite{pando} that performs such an optimization).


\noindent {\bf Discussion:} We conclude with a discussion of a few important aspects of our optimization, especially paying attention to their implications for \legostore's operational feasibility and efficacy. A key concern for a practitioner would be the obviously high computational complexity of our formulation---notice the presence of several discrete variables and  quadratic terms in our constraints. Instead of searching for the optimal configuration in all possible combinations of data centers in each quorum, we use a heuristic to reduce the search scope for the optimal configuration and, consequently, time to find one. Our heuristic will sort the data centers based on their network price coming into the client. For example, to select quorums for client $i$, we pick the data centers with the lowest network price to client $i$. (See Table~\ref{tab:net+rtt} for price discrepancies.)

In our evaluation, individual runs of the optimizer for selecting among 9 data centers concluded within 6.5 minutes on 8 \emph{e2-standard-4} machines from GCP which we consider adequate for the scale of our experiments. Papers describing commercial-grade systems (e.g., Volley~\cite{volley}) exploit techniques (including parallelization) to scale such optimization formulations to the much larger input sizes these systems must deal with. Our optimizer is also highly parallelizable---parallelization can happen in granularity of keys as well as computing the cost within each key.

While our work only considers two protocols (ABD and CAS) to pick from for a key, it is an easy conceptual enhancement to incorporate other protocols (e.g., underlying protocols used in Cassandra~\cite{CassandradB} or Google Spanner~\cite{Spanner}). The key enhancements needed would consist of (i) generalizing $e_g$ to a vector of booleans of size equal to the number of considered protocols; (ii) generalizing its use to express the exclusive use of one of the considered protocols; and (ii) expressing cost and latency for each of the considered protocols similarly to what was described above.  

An important concern would be the data granularity at which to perform reconfigurations. We restrict our attention to a key granularity. 
Generally, one could identify groups of keys with similar predicted workload properties and compute the new configuration for such a group using a single execution of the optimizer. We consider these concerns as representing relatively minor difficulty beyond our current work and leave these to future work. 

The GETs are modeled similar to PUTs for both ABD and CAS:

\begin{equation}
\label{eq:get-cost}
    C_{g,get} = \underbrace{ e_g \cdot C_{g,get,CAS}}_{\textrm{n/w cost if CAS chosen}} + \underbrace{(1-e_g) \cdot C_{g,get,ABD}}_{\textrm{n/w cost if ABD chosen}}.
\end{equation}

\begin{equation}
\label{eq:get-abd-cost}
\begin{split}
C_{g,get,ABD} = \rho_g \cdot \lambda_g \cdot o_g \sum\limits_{i=1}^D \alpha_{ig}\Big(\underbrace{\Sigma_{j=1}^D p^n_{ji} \cdot iq^1_{ijg}}_{\textrm{ABD phase 1}} +\\
\underbrace{\Sigma_{k=1}^D p^n_{ik} \cdot iq^2_{ikg}}_{\textrm{ABD phase 2}}\Big). 
\end{split}
\end{equation}

\begin{equation}
\label{eq:get-cas-cost}
\begin{split}
C_{g,get,CAS} = \rho_g \cdot \lambda_g  \sum\limits_{i=1}^D \alpha_{ig}\Big\{o_m\Big(\sum\limits_{j=1}^D p^n_{ji} \cdot iq^1_{ijg} + \\  \sum\limits_{k=1}^D p^n_{ik} \cdot iq^4_{ikg}\Big) + \frac{o_g}{k_g}\sum\limits_{k=1}^D p^n_{ki} \cdot iq^4_{ikg}\Big\}.
\end{split}
\end{equation}

\noindent {\bf Discussion:} We conclude with a discussion of a few important aspects of our optimization, especially paying attention to their implications for \legostore's operational feasibility and efficacy. A key concern for a practitioner would be the obviously high computational complexity of our formulation---notice the presence of several discrete variables and  quadratic terms in our constraints. Instead of searching for the optimal configuration in all possible combinations of data centers in each quorum, we use a heuristic to reduce the search scope for the optimal configuration and, consequently, time to find one. Our heuristic will sort the data centers based on their network price coming into the client. For example, to select quorums for client $i$, we pick the data centers with the lowest network price to client $i$. (See Table~\ref{tab:net+rtt} for price discrepancies.)

In our evaluation, individual runs of the optimizer for selecting among 9 data centers concluded within 6.5 minutes on 8 \emph{e2-standard-4} machines from GCP which we consider adequate for the scale of our experiments. Papers describing commercial-grade systems (e.g., Volley~\cite{volley}) exploit techniques (including parallelization) to scale such optimization formulations to the much larger input sizes these systems must deal with. Our optimizer is also highly parallelizable---parallelization can happen in granularity of keys as well as computing the cost within each key.

While our work only considers two protocols (ABD and CAS) to pick from for a key, it is an easy conceptual enhancement to incorporate other protocols (e.g., underlying protocols used in Cassandra~\cite{CassandradB} or Google Spanner~\cite{Spanner}). The key enhancements needed would consist of (i) generalizing $e_g$ to a vector of booleans of size equal to the number of considered protocols; (ii) generalizing its use to express the exclusive use of one of the considered protocols; and (ii) expressing cost and latency for each of the considered protocols similarly to what was described above.  

An important concern would be the data granularity at which to perform reconfigurations. We restrict our attention to a key granularity. 
More generally, one could identify groups of keys with similar predicted workload properties and compute the new configuration for such a group using a single execution of the optimizer. We consider these concerns as representing relatively minor difficulty beyond our current work and leave these to future work. 

%% file: appendix/reconfiguration_lin_proof.tex
\section{Description and Proof Sketch of the  Linearizability of Our Reconfiguration Protocol}
\label{app:reconfig}

We provide a pseudocode for the reconfiguration algorithm (described in Section~\ref{sec:reconfig-proto}) in Figures \ref{algo:recon_cont}, \ref{algo:recon_server_client}. We prove that algorithm satisfies the desired safety properties, that is, the system is linearizable despite reconfigurations. Our proof assumes that reconfigurations are issued sequentially by the controller (i.e., reconfiguration requests are not concurrent).

\input{algorithms/reconfig}

The same proof works for both CAS and ABD algorithms; we focus on the CAS algorithm here for brevity. Consider any execution of the algorithm where every operation terminates\footnote{For linearizability, it suffices to consider executions where every operation terminates~\cite{Lynch1996}}. With our reconfiguration protocol, note that an operation can span multiple configurations, in particular, if it starts in one configuration and then has to restart on receiving a $\texttt{operation\_fail, new\_configuration}$ message. For any operation that terminates in the execution, we refer to the configuration of that operation as the configuration in which it \emph{terminates}. The \emph{tag} of an operation is defined as the tag associated with it in the configuration where it terminates; for a read, it is the tag acquired in its query phase, and for a write, it is the tag sent along with the value/codeword symbol.

To ensure linearizability, the main invariant we require to show (See \cite{Lynch1996}) is the following: if operation $\pi$ begins after operation $\phi$ ends, then, in the configuration where it terminates, $\pi$ gets a tag at least as large as the tag of $\phi$ in its query phase. To show this for our protocol, we consider three cases. 
\squishlist
    \item Case (1), $\phi$ and $\pi$ are in the same configuration;
    
    \item Case (2), $\phi$ completes in an old configuration and $\pi$ in a new configuration;
    
    \item Case (3) $\phi$ completes in a new configuration and $\pi$ in an old configuration. 
\squishend

For Case (1), the identical proofs as \cite{CadambeCoded_NCA} apply.

We show that Case (3) is impossible. For a configuration, we define the point it is \emph{enabled} as the first point where the controller has received the acknowledgment of all the servers in the configuration. Suppose $\phi$ completes in configuration $c_{new}$ and $\pi$ completes in configuration $c_{old}.$ Since $\pi$ completes in the old configuration $c_{old}$, we infer that the point of the enabling of $c_{new}$ is after the point of invocation of $\pi$. This is because, based on the protocol, an operation that begins after the enabling of a new configuration does not contact an old configuration.   Since $\phi$ is an operation in $c_{new}$, the point of completion of $\phi$ is after the point of enabling of $c_{new}$ and therefore after the point of invocation of $\pi.$ This contradicts the assumption that $\phi$ completes before $\pi$ begins. Therefore Case (3) is impossible. 

Now, we consider Case (2). In this case, it suffices to argue that the controller read a value with a tag at least as large as the tag of $\phi.$ This is because $\pi$ acquires a tag at least as large as the tag acquired by the controller. Since $\phi$ is in an old configuration, the controller issues a read $r$ as a part of the reconfiguration. It suffices to show  that the read obtains a value with a tag that is as large as the tag of $\phi$. There are two possibilities, (a) at least one server receives a message from the last phase of $\phi$ and then responds to the first message of $r$, and (b) there is no server that responds to the first message of $r$ after the last message of $\phi$. In case (a) the proof is similar to the proofs of \cite{CadambeCoded_NCA}; effectively, since $\phi$ completes, at least one server that received the $fin$ label from $\phi$ responds to $r$. Therefore, $r$ obtains a tag at least as large as the tag of $\pi.$

Case $(b)$ is the critical case. Now, since $\phi$ completes, that means there are $n-f$ servers that responded to the final phase of $\phi$. Furthermore, because we are operating in case $(b),$ each of these servers responded to the first message of $r$ before sending the response. From the protocol note that these servers halt responding to operations after receiving the query message from $r$ until they get a $finalize\_reconfig$. Since these $n-f$ servers responded to $\phi,$ and they responded after the query message from $r$, we conclude from the protocol that they responded after receiving the $finalize\_reconfig$ message. The fact that they responded implies that the tag of  $\phi$ is at most as large as the tag acquired by the read $r$ of the controller. This completes the proof.

%% file: algorithms/reconfig.tex
\begin{figure}
\centering
    \includegraphics[width=0.6\columnwidth]{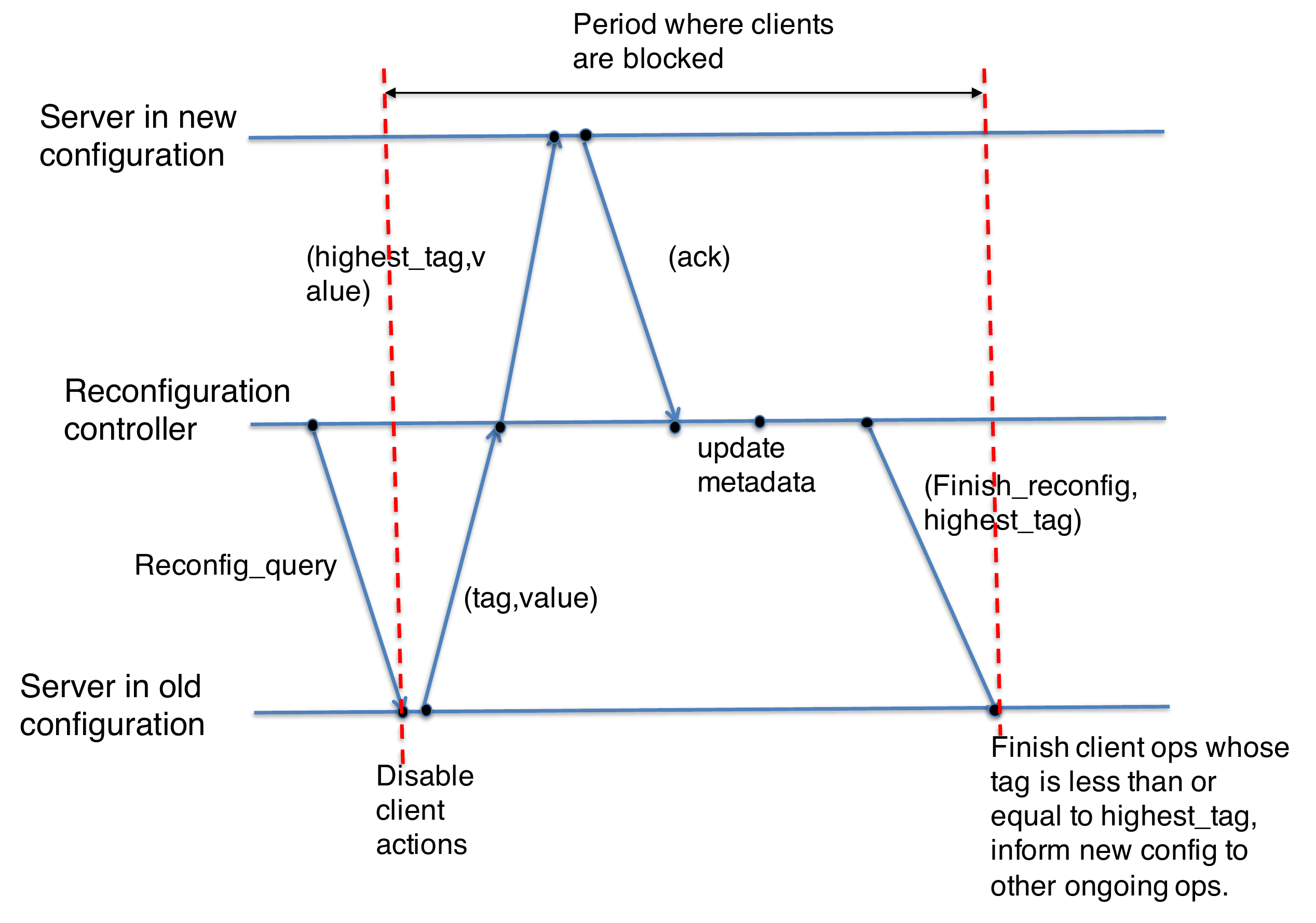}
    \caption{ Reconfiguration protocol timeline. The timeline is depicted assuming both configurations are performing replication-based ABD.}
    \label{fig:recon_timeline}
\end{figure}

{
\begin{algorithm}
\SetAlgoLined
\DontPrintSemicolon

\textbf{Inputs: } old\_config, new\_config

Send \texttt{reconfig\_query} messages to all servers in \texttt{old\_config}

\eIf(\tcc*[f]{CAS}){old\_config.$e_{g}$ == 1}{

    Await responses from $\max(\texttt{old\_config}.N-\texttt{old\_config}.q_3+1, \texttt{old\_config}.N-\texttt{old\_config}.q_4+1)$ servers
    
    Let $t$ denote the highest logical timestamp received
    
    Send \texttt{reconfig\_get}($t$) message to all servers in \texttt{old\_config}
    
    Wait for \texttt{old\_config}$.q_4$ servers to respond
    
    Decode the corresponding value $v$
    
}(\tcc*[f]{ABD}){
    Await responses from $\texttt{old\_config}.N-q_2+1$ servers
    
    Let $t$ denote the received pair with the highest logical timestamp
    
    Let $v$ be the value of pair $t$
    
    
        
        
}

\eIf(\tcc*[f]{CAS}){\texttt{new\_config}$.e_g==1$}{
    
        Encode the value $v$ to $chunks$ using Reed-Solomon code
        
        Send \texttt{reconfig\_write}($t, chunk[i], 'fin'$) to $server_i$ for all servers in \texttt{new\_config}
        
        Wait for $\max(\texttt{new\_config}.q_2,$ $ \texttt{new\_config}.q_3)$ servers to respond
    }(\tcc*[f]{ABD}){
        Send \texttt{reconfig\_write}($t,v$) pair to all servers in \texttt{new\_config}
        
        Wait for $\texttt{new\_config}.q_2$ servers to respond
    }

Update global metadata (probably local to the controller) to reflect new configuration;

Send a \texttt{finish\_reconfig}$(t, v$, \texttt{new\_config}$)$ message to all servers in \texttt{old\_config}.


 
 \caption{Reconfig - Controller}
 \label{algo:recon_cont}
\end{algorithm}
}

{
\begin{algorithm}
\SetAlgoLined



\SetKwBlock{Onrecreque}{On receiving a \texttt{reconfig\_query}}{end}
\Onrecreque(\tcc*[f]{from controller}){

    Disable all actions for clients for $\texttt{old\_config}$
    
    \eIf(\tcc*[f]{CAS}){\texttt{new\_config}$.e_g==1$}{
        
        Send the highest tag labeled fin to controller
        
    }(\tcc*[f]{ABD}){
        Send stored $(t,v)$ pair to controller
    }
}

\SetKwBlock{Onrecrefin}{On receiving a \texttt{reconfig\_get}($t$)}{end}
\Onrecrefin(\tcc*[f]{from controller}){
    
    \eIf{the server has a locally stored tuple $(t,chunk,*)$}{
        
        If `*' is `pre', change the label to `fin'
        
        Send $(t,chunk)$ to the controller.
        
    }{
        Store $(t,NULL,'fin')$
        
        Send an ack
    }
}

\SetKwBlock{Onrecfinre}{On receiving a \texttt{finish\_reconfig}($t, v$, {\tt new\_config}$)$}{end}
\Onrecfinre(\tcc*[f]{from controller}){
    
    Send an \texttt{(operation\_fail, new\_config)} message to all \texttt{get\_timestamp} queries.
    
    Unblock other queries and respond as described per protocol to every operation with a tag less than or equal to $t$.
    
    Send an \texttt{(operation\_fail, new\_config)} message to all client operations with tag larger than $t$.
    
}

\SetKwBlock{Onrecchunk}{On receiving a \texttt{reconfig\_write}$(t,v)$ or \texttt{reconfig\_write}$(t,chunk,'fin')$ tuple}{end}
\Onrecchunk(\tcc*[f]{from controller}){
    
    Store the tuple
    
    Send an ack \tcc*[f]{To controller}
}

\SetKwBlock{Onrecopfail}{On receiving \texttt{\texttt{operation\_fail}, \texttt{new\_config})} during an operation}{end}
\Onrecopfail(\tcc*[f]{in client}){
    
    Restart the operation with \texttt{new\_config}.
}
 
\caption{Reconfig - Server and Client}
\label{algo:recon_server_client}
\end{algorithm}
}

%% file: appendix/k-analysis.tex
\section{An Analytical Model Relating Cost to \texorpdfstring{$K$} for CAS-Based Configurations}
\label{app:k}

As $K$ increases, the network and storage costs decrease inversely proportional as $K$.  However, the VM costs increase because the quorum sizes increase to accommodate a larger quorum intersection,  indirectly increasing the effective arrival rate for each participating DC. Further the object size increases, network and storage costs increase, whereas, in our model, the VM costs remains constant. Similarly, as the arrival-rate increases, the network and VM costs increase, whereas the storage size remains constant. Thus, overall, the total cost can be modeled as: 
\begin{align*}c_1 \cdot \lambda \cdot  \frac{N+K}{2} + c_2 \cdot \lambda \cdot o\frac{N+K}{2K }+c_3 \cdot o \cdot \frac{N}{K}\end{align*}


if we neglect the metadata costs, the spatial variations in the pricing and assume that each quorum is of size $\frac{N+K}{2}$. Note here that $c_1,c_2,c3$ respectively capture the dependence on vm cost, network and storage costs. If we fix the failure tolerance as $f$ and choose $N=K+2f$ - the most frequent choice of the optimizer, then the overall cost  can be written as:
 \begin{align*}c_1 \cdot \lambda \cdot K + c_2 \cdot o \cdot \lambda  \cdot \frac{f}{K} +c_3 \cdot o \cdot \frac{2f}{K} + \overline{c}_{4}\end{align*}
 
 where $\overline{c}_{4}$ is a constant that does not depend on $K.$

\section{Garbage Collection}
\label{app:gc}

To ensure termination of certain concurrent operations, CAS   requires each server to store  codeword symbols corresponding to older versions, not just the latest version. A garbage collection (GC) procedure is required to remove older values and keep storage costs close to our models. GC does not affect safety, and can only affect termination of concurrent operations to the same key \cite{CadambeCoded_NCA}. We use a simple GC heuristic that deletes codeword symbols corresponding to older versions after a threshold time set larger than the maximum predicted latency of all operations. In our prototype we set this parameter to 5 minutes, an order of magnitude larger than operation latencies (10s-100s of msec). We empirically confirm that this heuristic keeps the storage overheads negligibly small. 

%% file: appendix/additionalresults.tex
\section{Additional Results}
\label{app:addnresults}

\begin{figure*}[ht]
\centering
\begin{subfigure}{0.48\columnwidth}
  \centering
  \includegraphics[width=\columnwidth]{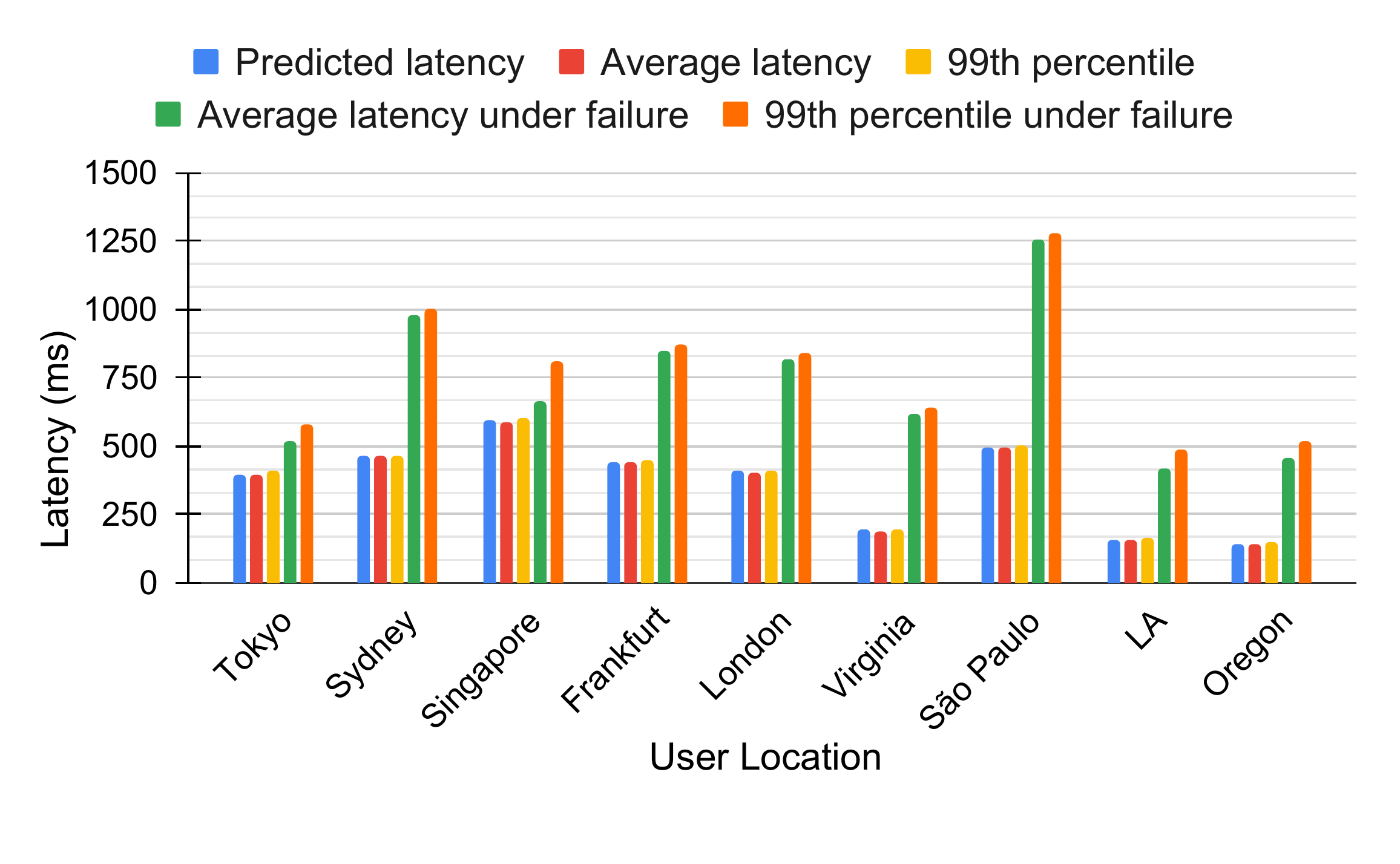}
  \caption{Latency of PUT operations}
  \label{fig:latency-verif-put}
\end{subfigure}%
\begin{subfigure}{0.48\columnwidth}
  \centering
  \includegraphics[width=\columnwidth]{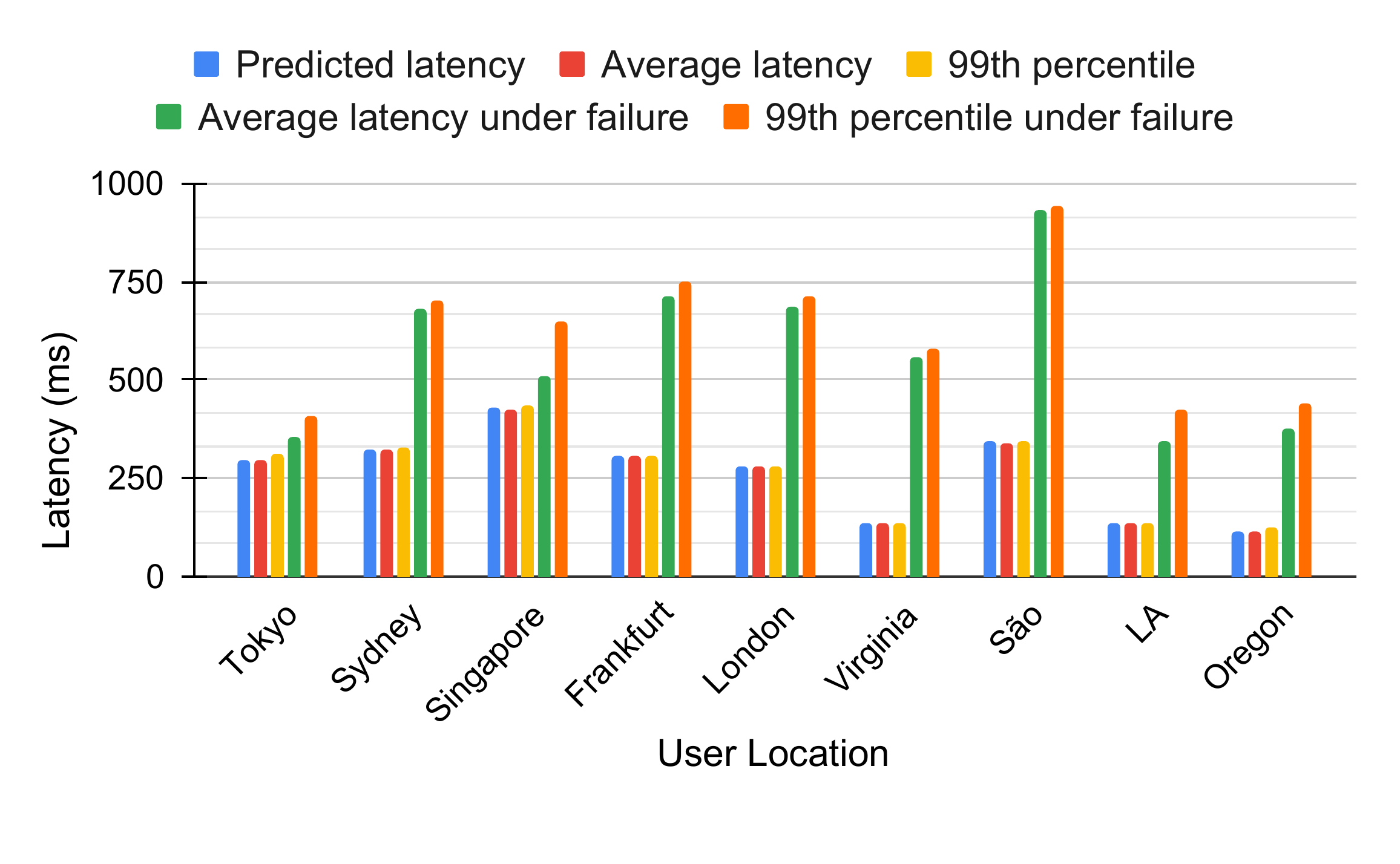}
  \caption{Latency of GET operations}
  \label{fig:latency-verif-get}
\end{subfigure}
\caption{\footnotesize A sample comparison of our models for PUT and GET latencies vs. observations on our prototype. The workload chosen has a uniform user distribution and we show the latency numbers for requests arising at all 9 locations.  The two rightmost bars for each location depict observed latencies when the server in LA has failed. Please note that this server is available in all quorums, so it has the worst effect on the latency. If a server that does not participate in any quorums fail, there will be no change in the latencies.}
\label{fig:latency-verif}
\end{figure*}

\subsection{Prototype-Based Validation}
\label{app:valid}

We conduct extensive validation of the efficacy of the latency and cost modeling underlying our optimizer. As a representative result, in Figure~\ref{fig:latency-verif} we compare the modeled latencies against those observed on our prototype during a 10 minute long run of the following workload for which our optimizer chooses CAS(4,2): (i) Uniform client distribution; object size=1KB; datastore size=1 TB; 
request rate=200 req/s; read ratio=HW; latency SLO=1 sec. Besides the excellent match between modeled and observed latencies, we also note that the observed tail latency is not much higher than the average. This is because the latency's dominant component are inter-DC RTTs that we have found to be largely stable over long periods. This dominance of inter-DC RTTs also implies that the effect of stragglers on latency and the benefits of EC over replication in mitigating it as recently noted in~\cite{rashmi2016ec} in the context of a datastore confined to a single data center is not an important concern for us. In several experiments, we have also observed that for some workloads, depending on the arrival rate and read ratio, our average GET latencies could be much smaller than the predicted latency because a fraction of the reads return in one phase due to optimized GET operations (See Section~\ref{app:ABD} and~\ref{app:cas}).

\begin{figure*}
\centering
\begin{subfigure}{0.48\columnwidth}
  \centering
  \includegraphics[width=\columnwidth]{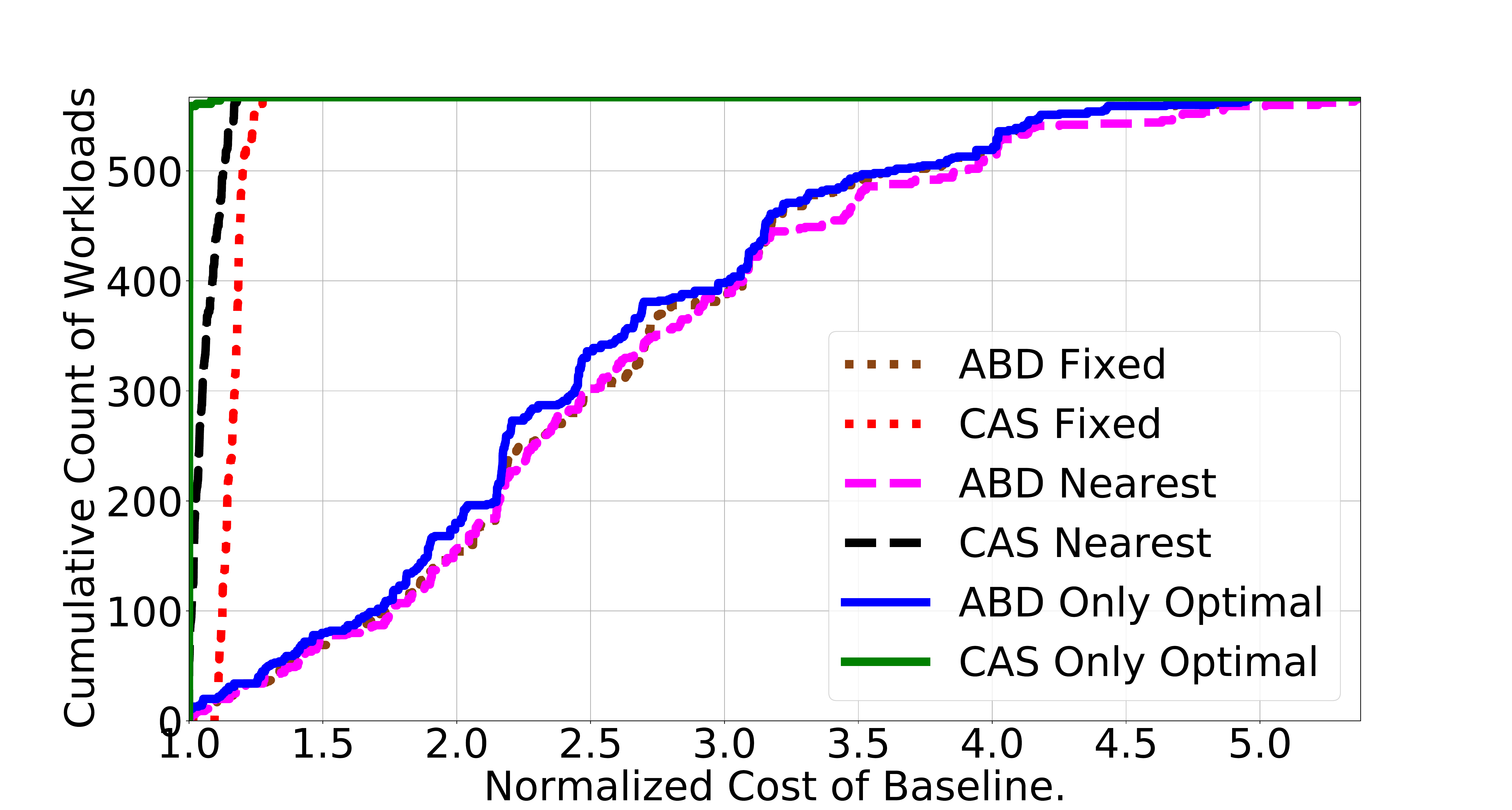}
  \caption{Latency SLO=1 sec.}
\end{subfigure}%
\begin{subfigure}{0.48\columnwidth}
  \centering
  \includegraphics[width=\columnwidth]{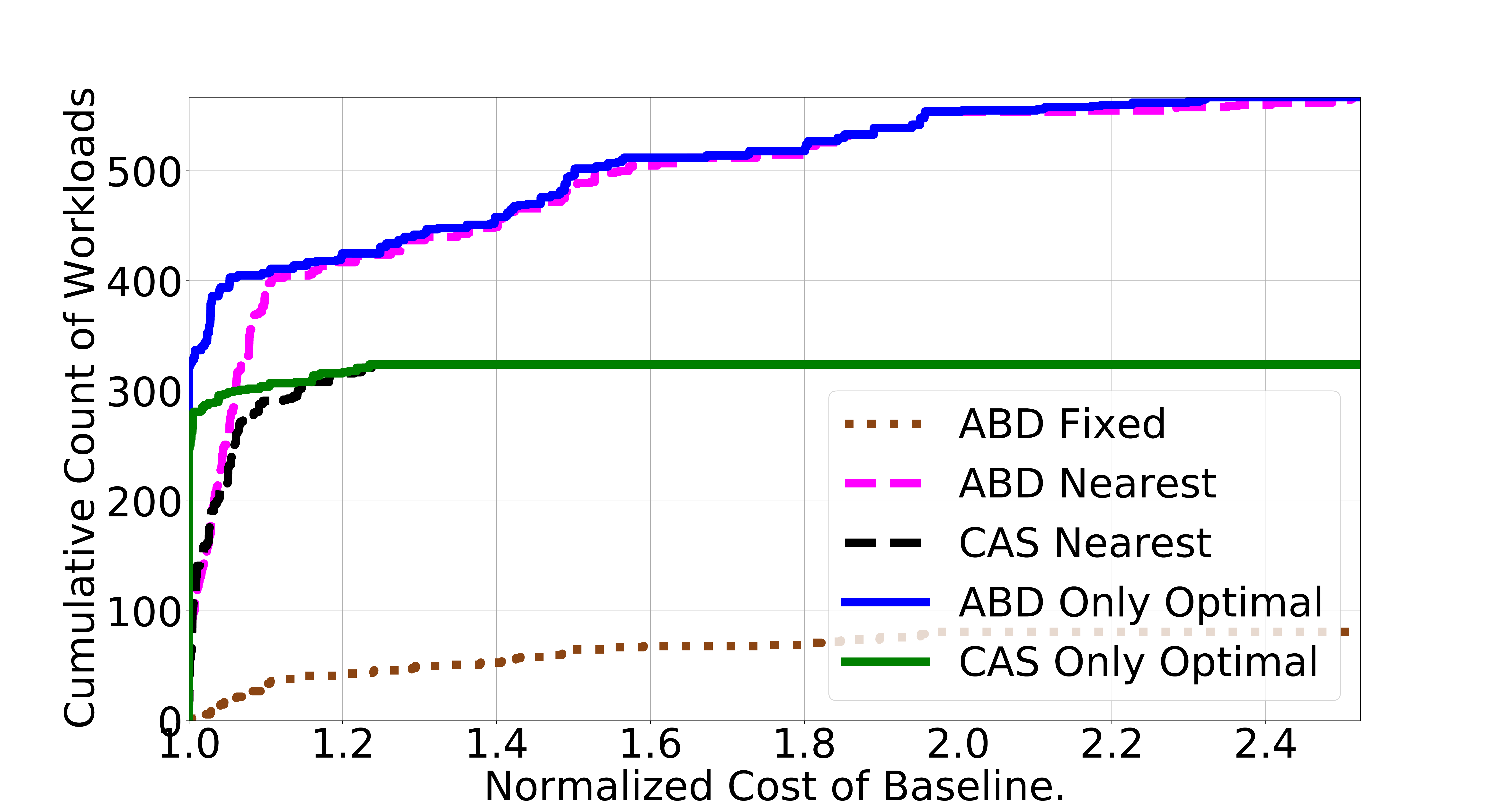}
  \caption{Latency SLO=300 msec.}
\end{subfigure}
\caption{Cumulative count of the normalized cost of our baselines (for our 567 basic workloads) with $f=2$ and two extreme latency SLOs. Note that no configuration was feasible for CAS Fixed when the latency SLO is 300 msec.}
\label{fig:normalized-costs2}
\end{figure*}

\begin{figure*}
    \includegraphics[width=\textwidth]{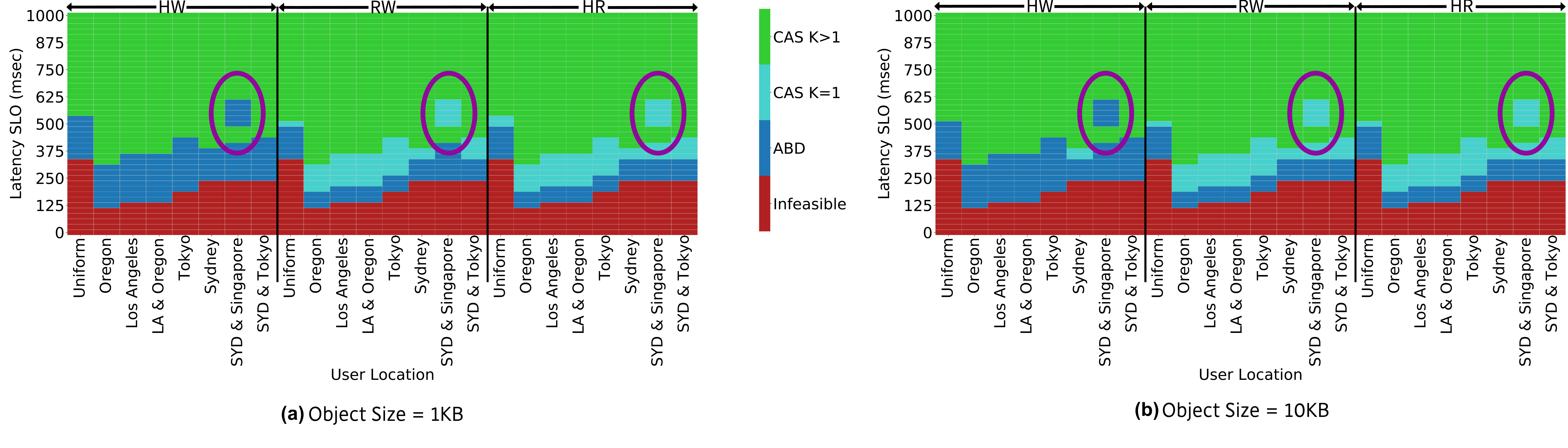}
    \caption{\footnotesize Sensitivity of the optimizer's choice to the latency SLO. We consider 2 object sizes (1KB and 10KB), 8 different client distributions, arrival rate=500 req/sec, and $f$=2. We consider 3 different read ratios (HW, RW, HR defined in Section~\ref{sec:setup}).}
    \label{fig:slo-latencyf2}
\end{figure*}

\subsection{Are Nearest DCs Always the Right Choice?} 
\label{app:nearest}

\begin{figure*}
\centering
\begin{subfigure}{0.48\columnwidth}
  \centering
  \includegraphics[width=\textwidth]{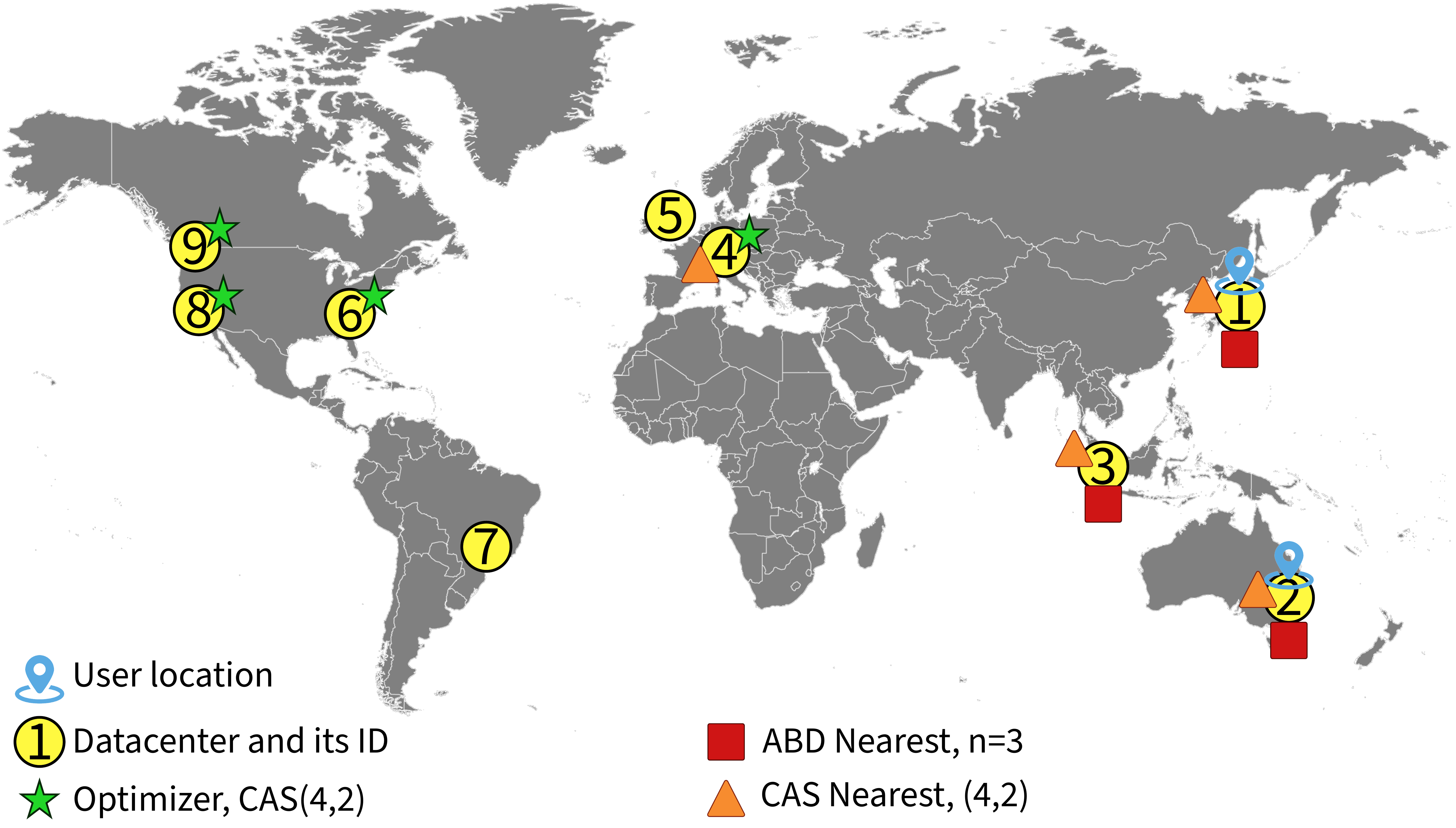}
  \caption{DC and user locations;  DCs chosen by different solutions}
  \label{fig:nearest-cost1}
\end{subfigure}%
\begin{subfigure}{0.48\columnwidth}
  \centering
  \includegraphics[width=\textwidth]{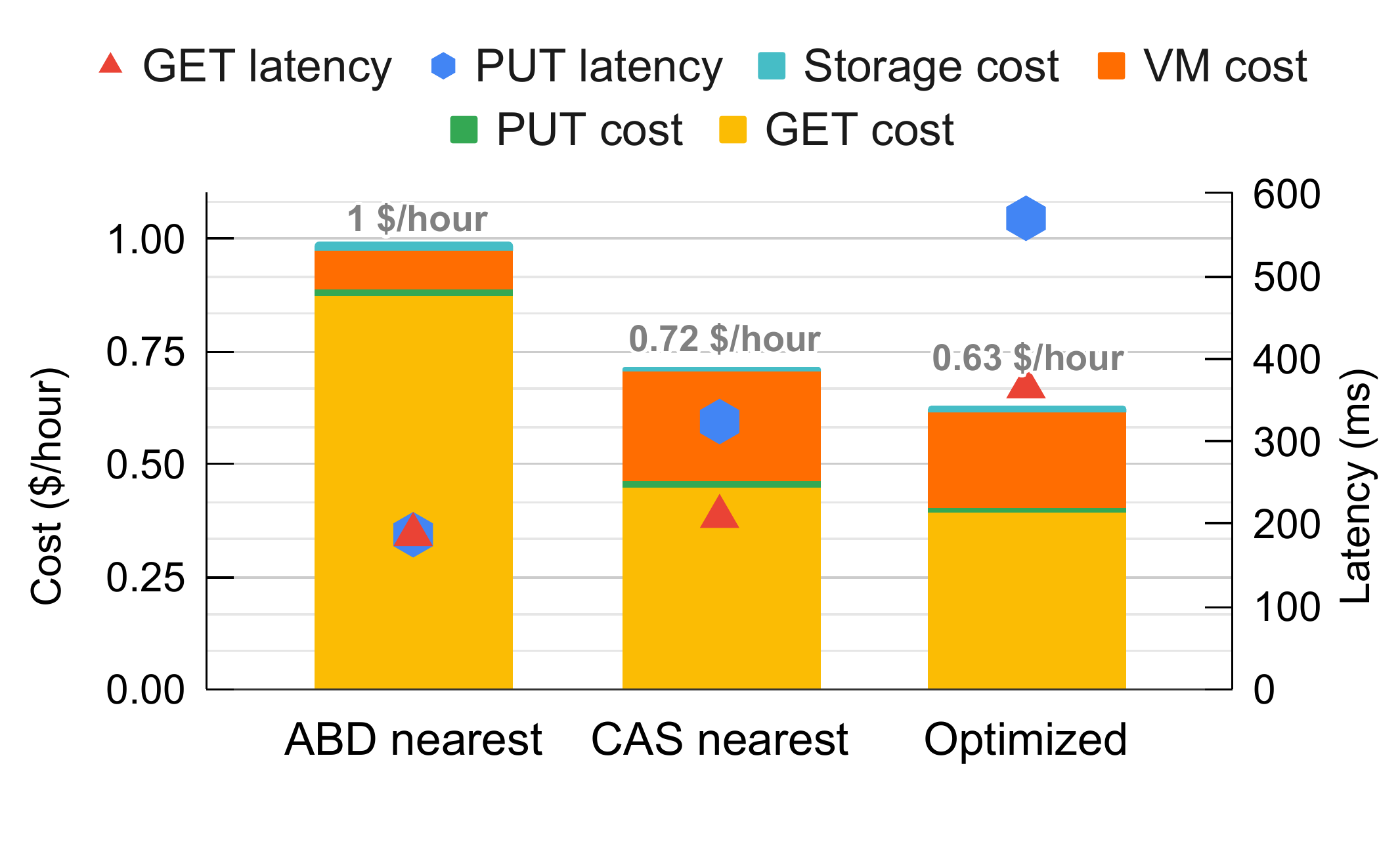}
  \caption{Cost and latency}
  \label{fig:nearest-cost2}
\end{subfigure}
\caption{\footnotesize An experiment demonstrating that choosing the nearest data centers may be sub-optimal. In (a), we show our client locations and the data enters chosen by ABD Nearest, CAS Nearest, and our optimizer. In (b), we compare the component-wise costs (\$/hour) and GET/PUT latency offered by these three approaches.
}
\label{fig:nearest-cost}
\end{figure*}

Our optimizer reveals that, perhaps surprisingly, the naturally appealing approach of using DCs nearest to user locations~\cite{volley} 
can lead to wasted costs. To demonstrate an extreme case of this, we consider an HR workload 
with 50\% of the requests each coming from Sydney and Tokyo with a latency SLO of 1 sec, $f$=1, and object size of 1 KB. The set of DCs that our optimizer chooses includes neither Tokyo nor Sydney---see Figure~\ref{fig:nearest-cost1}. Our optimizer chooses CAS with parameters ($N$=4,$K$=2) with four DCs spread across North America and Europe. This is because the network prices for the Singapore and Sydney data centers are nearly twice as expensive (recall Table~\ref{tab:net+rtt}) as the others. We compare the component-wise costs (in \$/hour) as well as latencies offered by our nearness-oriented baselines with our optimizer in Figure~\ref{fig:nearest-cost}(b). Notice that GET networking costs are the dominant component (since the read ratio is 30:1). The use of EC by the optimizer as well as CAS Nearest significantly reduces this component while incurring higher VM costs because EC needs to use more data centers to maintain $N - K > 2f$ (see Equation~\ref{eq:n-k-f}). In the balance, however, EC offers better costs. Although CAS Nearest uses the \emph{right} EC parameters, its choices of the specific data centers is poorer causing its cost to be about 14\% higher than our optimizer's. 

\subsection{Performance Under Failure}
\label{app:perf-failure}



LEGOStore continues servicing user requests as long as the number of DC failures does not exceed $f$. During DC failures, users may experience a degradation in latency or cost depending on the specific DCs that need to be relied upon to compensate for the unavailable DCs; \vc{in such cases, SLO violation can also occur until the node is failed node replaced}. If more than $f$ DCs become unavailable, operations timeout at the client and the user is returned an error. The reconfiguration is used to replace failed nodes with non-failed nodes.
If the reconfiguration controller for a key fails in the midst of an ongoing reconfiguration, then all operations to that key can get stalled indefinitely, whereas the functioning of the rest of the data store remains unaffected. LEGOStore's design allows for the reconfiguration controller to be made fault tolerant via state machine replication (e.g., using Paxos), without much change in performance of operations in the common case where there is no concurrent reconfiguration. In our implementation we do not replicate the reconfiguration controller because (i) this is orthogonal to our main interest in this paper and (ii) it does not affect our results in any meaningful way. If the controller fails when no reconfiguration is underway, it can be safely relaunched at an available DC. 

We test the performance of our prototype under failure. We consider workloads that tolerate $f=1$ failure and study the effect of a node failure on performance. Specifically, in Figure~\ref{fig:latency-verif}, we show the change in latency when the server participating in all the quorums fails. Essentially, if the failed server is not in the quorum, there will be no degradation and we can simply replace the failed server. The degradation in Figure~\ref{fig:latency-verif} happens because once a client does not receive a response from at least one server in its optimized quorum, it will retry the operation by sending the request to all the servers, and once it has enough number of response to maintain linearizability, it will return with success.

\bu{
\subsection{Cost Savings Offered by Our Optimizer for the Wikipedia Workload}
\label{app:costsavingswiki}

\begin{figure}
\centering
\includegraphics[width=0.55\columnwidth]{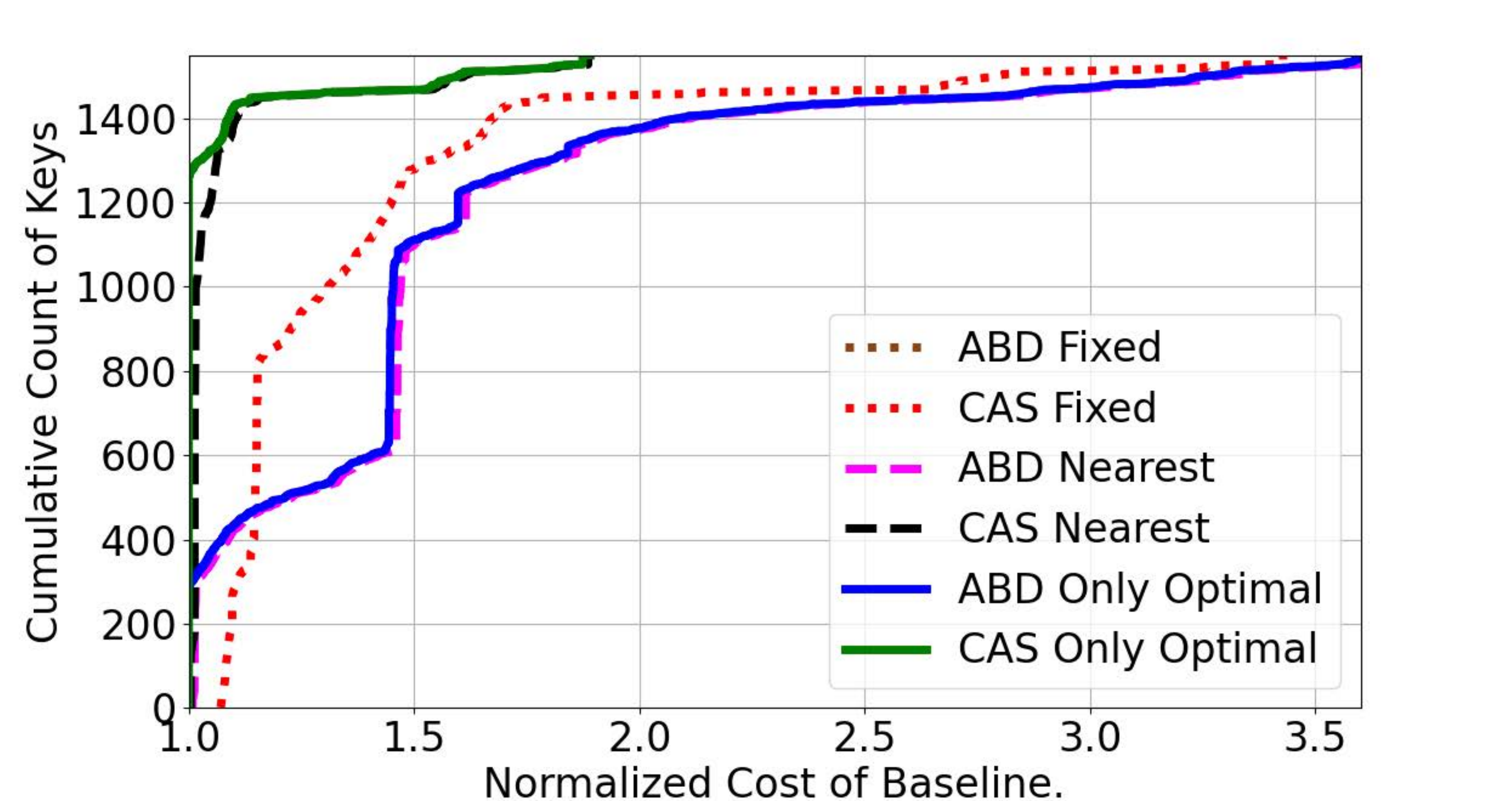}
\caption{\bu{Cumulative count of the normalized cost of our baselines (for our 1550 keys derived from the Wikipedia dataset) with $f=1$ and latency SLO of 750 msec.}} 
\label{fig:wikicomparison}
\end{figure}

In Figure~\ref{fig:wikicomparison}, we depict the cost savings offered by our optimizer over our various baselines for the 1550 keys derived from the Wikipedia dataset as described in Section~\ref{sec:realreconf}. 

}


\section{Salient Implementation Aspects}
\label{app:impl}



We implement our  optimizer 
using Python3.
Since finding the configurations can be done in parallel with each other, we execute up to 48 instances of the optimizer in parallel spanning 8 \emph{e2-standard-4} GCP VMs.\footnote{Our optimizer takes 10 sec for each workload on average for $f$=1.}  We implement our LEGOStore prototype in C++11 using about 5.5K LOC. Our prototype uses the {\tt liberasurecode} library's~\cite{liberasurecode}  Reed-Solomon backend for encoding and decoding data in CAS. 
We refer to LEGOStore's users and clients \& servers of our protocols as its {\em data plane}; we call the per-DC metadata servers, the optimizer and reconfiguration controllers its {\em control plane}. \lego's source code is available under the Apache license 2.0 at \href{https://github.com/shahrooz1997/LEGOstore}{github.com/shahrooz1997/LEGOstore}.

\noindent {\bf Data Plane:} A user application (simply user) needs to link the \lego~library 
and ensure it uses \lego's API. The library implements logic for the user to interact with the appropriate (ABD or CAS) client.
LEGOStore maintains a pool of {\em compute nodes} (VMs) (GCP `custom-1-1024' instance types) that run the client and server protocols and store keys using the RocksDB~\cite{rocksdb} database engine.

Each DC maintains a meta-data server (MDS) locally 
that contains configurations for keys that the clients within it have served and that have not yet been deleted. {\em LEGOStore does not have to have a centralized MDS} with corresponding freedom from concerns related to the failure of such an MDS. When a client does not find a key in the local MDS, it issues queries to other DCs (e.g., in increasing order of RTTs) to learn the key's configuration.

\noindent {\bf Control Plane:}   LEGOStore's control plane comprises the two types of elements for each key described in Section~\ref{sec:interface}: (i) the optimizer and (ii) the reconfiguration controller. Neither of these components is tied to a particular DC location (e.g., See Section~\ref{sec:reconfig-whenwhat} for discussion regarding controller placement).